\documentclass[pra,twocolumn,reprint,amsmath,amssymb,showpacs,floatfix,superscriptaddress]{revtex4-1}
\usepackage{color}
\usepackage{graphicx}

\usepackage[sort&compress]{natbib}
\usepackage{color}
\usepackage{soul}

\usepackage{hyperref}
\hypersetup{
bookmarks=false,
pdftitle={Weak and Strong coupling regimes in plasmonic-QED}
anchorcolor=blue,
colorlinks=true,
citecolor=blue,
urlcolor=blue,
linkcolor=blue}
\usepackage{xcolor}

\usepackage{units}

\begin{document}

\title{Weak and Strong coupling regimes in plasmonic-QED}

\author{T. H\"ummer}
\affiliation{Instituto de Ciencia de Materiales de Aragon y Departamento de Fisica
de la Materia Condensada, CSIC-Universidad de Zaragoza, E-50012 Zaragoza,
Spain}
\affiliation{Ludwig-Maximilians-Universit\"at M\"unchen, D-80799 Munich, Germany}
\affiliation{Max-Planck-Institut f\"ur Quantenoptik, D-85748 Garching, Germany}

\author{F. J. Garc\' {\i}a-Vidal}
\affiliation{Departamento de Fisica Teorica de la Materia Condensada, Universidad
Autonoma de Madrid, E-28049 Madrid, Spain}

\author{L. Mart\'{\i}n-Moreno}
\affiliation{Instituto de Ciencia de Materiales de Aragon y Departamento de Fisica
de la Materia Condensada, CSIC-Universidad de Zaragoza, E-50012 Zaragoza,
Spain}


\author{D. Zueco}
\affiliation{Instituto de Ciencia de Materiales de Aragon y Departamento de Fisica
de la Materia Condensada, CSIC-Universidad de Zaragoza, E-50012 Zaragoza,
Spain}
\affiliation{Fundacion ARAID, Paseo Maria Agustin 36, E-50004 Zaragoza, Spain.}

\begin{abstract}
We present a quantum theory for the interaction of a two level emitter
with surface plasmon polaritons confined in single-mode waveguide
resonators. Based on the Green's function approach, we develop the conditions
for the weak and strong coupling regimes by taking into account the
sources of dissipation and decoherence: radiative and non-radiative
decays, internal loss processes in the emitter, as well as propagation
and leakage losses of the plasmons in the resonator. The theory is
supported by numerical calculations for several quantum emitters, GaAs and CdSe quantum dots and
NV centers together with different types of resonators constructed
of hybrid, cylindrical or wedge waveguides.
We further study the role of temperature and resonator length. 
Assuming realistic leakage rates, we find the existence of an optimal length at
which strong coupling is possible. Our calculations show that the strong coupling regime in plasmonic resonators 
is accessible within current technology
when working at very low temperatures ($\lesssim 4K$). In the weak coupling  regime
our theory accounts for recent experimental results.
By further optimization we find highly enhanced spontaneous emission with Purcell factors
over 1000 at room temperature for NV-centers. We finally discuss more applications for quantum
nonlinear optics and plasmon-plasmon interactions. 

\end{abstract}
\maketitle

\section{Introduction}

Cavity quantum electrodynamics (cavity QED) was invented to study
and control the simplest light-matter interaction: a two level emitter
(called TLS or emitter throughout this paper) coupled to a light monomode\cite{Haroche1993}.
At first associated with quantum optics, the emitter was an atom or
a collection of them, while the electromagnetic (EM) field was confined
in a high-finesse cavity\cite{Raimond2001}. Nowadays cavity QED experiments
cover quite a lot of implementations. Atoms may be replaced by other two level systems,
artificial or not, such as quantum dots or superconducting qubits. The
light mode can be any single bosonic mode quantized in \textit{e.g.}
superconducting cavities\cite{Wallraff2004}, nanomechanical resonators\cite{LaHaye2009},
carbon nanotubes\cite{Steele2009}, photonic cavities\cite{hennessy2007quantum}
or (collective) spin waves in molecular crystals\cite{Zhu2011}.

Cavity QED relies on the comparison between the ``light-matter''
coupling strength per boson and the irreversible losses from both
emitter and bosonic mode. Depending on their ratio two main regimes
appear: \textit{weak} and \textit{strong} coupling. In the weak regime
losses dominate and the emission spectrum consists of a single
peak around the dressed TLS resonant transition while the lifetime
is modified because of the field confinement inside the cavity. This
modification is nothing but the Purcell effect. In the strong coupling (SC) regime the
coupling dominates the losses. In this case a double peak emerges in the emission spectrum, 
arising from the emitter-resonator
level anticrossing. Cavity QED is interesting
per se: it demonstrates the quantum nature of both light and matter,
and serves \emph{e.g.} for testing quantumness in bigger and complex
systems\cite{Romero-Isart2010}. But, cavity QED is also a resource,
e.g. for optimizing single photon emission \cite{deleon2012tailoring}
or lasing\cite{mu1992oneatom}. Besides, systems in the SC regime
may behave as non linear media \cite{Yin2012}, generate photon-photon
interactions \cite{Englund2012} and are the building blocks in quantum
information processing architectures \cite{Ladd2010}.

Though the weak coupling (WC) regime is relevant, the ultimate goal is to reach the
SC regime. The former can be easily reached if the latter
is set. Being in the SC regime is determined by the EM
field lifetime, confinement and dipole moment. Usually, when working
with macroscopic mirrors and atoms having small dipole moments, the
field confinement is not optimized but the cavity has extremely long
lifetimes, \textit{i.e.}, very high finesse or quality factors. In
other setups like superconducting circuits all parameters -- quality
factor, dipole moment and field confinement -- are optimized such that the
so called 
ultra-strong coupling regime has been demonstrated \cite{Niemczyk2010,
Forn-Diaz2010}.
Circuits are promising on-chip setups but have to be operated
at microwave frequencies and mK temperatures.

A possible alternative at optical or telecom frequencies, with their plethora
of applications in quantum communication, is provided by the subwavelength
confined EM fields of surface plasmon polaritons (SPP)
in \emph{plasmonic waveguides}\cite{chang2007strongcoupling}. By
making resonators out of those waveguides, the energy density and
therefore the coupling is highly enhanced. The payoff is that metals
introduce considerable losses, further increasing with higher confinement.
Therefore, it is not clear under which conditions SC
could be reached with plasmonic resonators. On the other hand, advanced
architectures of plasmonic waveguides present a good trade off between confinement and
losses\cite{oulton2008confinement}, e.g. hybrid \cite{oulton2008ahybrid},
wedge\cite{boltasseva2008triangular} or channel\cite{bozhevolnyi2006channel}
waveguides . Plasmonic waveguides have already shown remarkable properties, such as 
focusing\cite{Stockman04,Volkov09}, lasing \cite{oulton2009plasmon},
superradiance \cite{MartinCano10}, 
mediators for entanglement between quits\cite{Tudela11} 
and single plasmon emission,
\cite{deleon2012tailoring, Russell2012}. Still, a very challenging perspective is their use for achieving
quantum cavity QED with plasmons in the SC regime\cite{gong2007designof}.
Plasmonic QED is not just another layout for repeating what has been
done in other cavity QED implementations but offers interesting advantages.
As shown in this paper SC can be obtained inside nanometric
resonators. It can be mounted \textit{on a chip} in combination with
dielectric waveguides. The latter have minor losses but weakly interact
with quantum emitters.

This paper aims to be self-contained.
We first summarize the 
quantum theory for the coupling between
quantum dipoles and resonators made out of one-dimensional (1D) plasmonic
waveguides.  Within this theoretical framework, we properly include
the losses and map to
a Jaynes-Cummings model and, therefore,  to the physics and applications
of traditional cavity QED. 
We present extensive finite-element
simulations for a  variety of resonator layouts and several quantum
emitters.  Our simulations allow to set the conditions to reach the SC
regime. We also motivate the study of these systems in the less demanding
WC  regime because of very high achievable Purcell factors
into the plasmon channel
of more than 1000.
We use our simulations for explaining the numbers
provided in a recent experiment for quantum resonators in the WC
regime \cite{deleon2012tailoring}.

The paper is organized as follows. We first develop, in section \ref{sec:II} the light-matter
interaction in plasmon resonators within the Green's function approach.
In section \ref{sec:Realization} the different realizations for plasmonic
resonators and emitters are discussed. We continue in section \ref{sec:ws}
with numerical results setting the parameter landscape for WC and
SC regimes. Section \ref{sec:applications} is devoted
to emphasize different applications. Some technical details are discussed
in the appendices.

\section{Interaction of a plasmonic structure and an emitter\label{sec:Interaction}}
\label{sec:II}

\subsection{Green's function approach for dissipative field quantization}

Surface plasmon polaritons ("SPPs" or just "plasmons") are
surface wave quanta bound to the interface between two media
characterized by
permittivities ($\epsilon(\omega)=\epsilon^{\prime}(\omega)+i\epsilon^{\prime\prime}(\omega)$)
with real parts of different signs and negative sum. Usually, the interface separates
a dielectric ($\epsilon^{\prime}(\omega)>0$) and a metal, which presents $\epsilon^{\prime}(\omega)\ll0$
at optical frequencies, see e.g. Ref~\cite{novotny2006principles}. On the other hand, the
imaginary part, $\epsilon^{\prime\prime}(\omega)$ is responsible
for dissipation in the metal (in order to minimize this dissipation, commonly used metals are silver and gold). Complex permittivities can be easily
incorporated in the macroscopic Maxwell equations. However, a problem
arises when trying to quantize the EM field: Maxwell
equations with a complex permittivity, $\epsilon^{\prime\prime}(\omega)\neq0$,
cannot be obtained from a Lagrangian and consequently, a straightforward
canonical quantization is not possible. On the other hand, (linear)
dissipation can be modeled by coupling the EM field to
an additional bath of harmonic oscillators: the system-bath approach\cite{weiss2008quantum}. 
Importantly, the system and bath can be
cast to a \textit{total} Lagrangian and consequently this allows the
quantization of the EM field in dispersive
media\cite{huttner1992quantization,philbin2010canonical}. For
self-completeness we outline this theory in Appendix \ref{app:A}.
To apply this quantization to complex geometries -- needed for plasmonic
structures -- the theory can be conveniently reformulated by means
of the Green's tensor of the classical problem\cite{knoll2000qedin, dung1998threedimensional,dung2003electromagneticfield,gruner1996greenfunction}.
The usefulness of this approach can be appreciated by looking at a
key result, the quantum expansion of the electric field
\begin{widetext}
\begin{equation}
\vec{E}(\vec{r},\omega)=i\sqrt{\frac{\hbar}{\pi\epsilon_{0}}}\frac{\omega^{2}}{c^{2}}\int{\rm d}^{3}r^{\prime}\sqrt{\epsilon^{\prime\prime}(\vec{r}^{\prime},\omega)}\stackrel{\leftrightarrow}{G}(\vec{r},\vec{r}^{\prime},\omega)\left(f^{\dagger}(\vec{r}^{\prime},\omega)-f(\vec{r}^{\prime},\omega)\right)\label{E}
\end{equation}
\end{widetext}
and an analogous expression for the magnetic field. Here, the electric
field can be expanded in normal modes where the coefficients are given
by the Green's function of the classical field. These normal modes of
the combined\emph{ }EM field \emph{and} the dispersive
media are represented by the bosonic creation (annihilation) operators
$f^{\dagger}(\vec{r},\omega)$ ($f(\vec{r},\omega)$). They obey the
commutation relation $\left[f(\vec{r},\omega),\, f^{\dagger}(\vec{r}^{\prime},\omega^{\prime})\right]=\delta(\omega-\omega^{\prime})\,\delta(\vec{r}-\vec{r}^{\prime})$.
We used $\epsilon_{0}$ to denote the vacuum permittivity and $c$
is the speed of light. Finally, $\stackrel{\leftrightarrow}{G}(\vec{r},\vec{r}^{\prime},\omega)$
is the dyadic Green's function of the classical field defined by\cite{novotny2006principles,dung2003electromagneticfield}
\begin{equation}
\Big[\nabla\times\nabla\times-\frac{\omega^{2}}{c^{2}}\epsilon(\vec{r},\omega)\Big]\stackrel{\leftrightarrow}{G}(\vec{r},\vec{r}^{\prime},\omega)=\stackrel{\leftrightarrow}{I}\,\delta(\vec{r}-\vec{r}^{\prime})\,.\label{equation-defining-G}
\end{equation}
Therefore, within this formalism the quantum fields, Eq.~(\ref{E}),
are determined by the classical Green's function, Eq.~(\ref{equation-defining-G}).

\subsection{Emitter-plasmon interaction\textcolor{red}{\label{sub:Emitter-plasmon-interaction}}}

We are interested in the interaction of these quantum fields with
a TLS. The actual physical implementation of the emitters will be
discussed in detail in section~\ref{sub:Quantum-TLS/Emitters}. In
the dipole approximation the interaction can be represented by the
emitter's dipole transition strength $\vec{d}$ and the electric field
at the position of the emitter ($\vec{r}_{e}$)~\cite[Chap. 14]{schleich2001quantum}
\begin{equation}
H_{{\rm int}}=-\sigma_{x}\vec{d}\cdot\vec{E}(\vec{r}_{e})\,, \label{dd}
\end{equation}
with  $\vec{E}(\vec{r}_{e})=\int_{0}^{\infty}{\rm
  d\omega}\vec{E}(\vec{r}_{e},\omega)$. 

In plasmonic waveguides, there are instances (which require a careful positioning of the emitter) where the emitter  radiates mainly into the plasmon channel \cite{chang2007strongcoupling,MartinCano10}. Then, we can isolate the emitter-plasmon coupling and, as explained in Ref. \cite{Dzsotjan2010} and summarized in Appendix
\ref{app:A}, write the emitter-plasmon coupling in terms of operators that annihilate and create  plasmons with frequency $\omega$, $a(\omega)$ and $a^\dagger(\omega)$, respectively. These operators have bosonic character, satisfying 
$[a(\omega), a^\dagger(\omega^\prime)]
= \delta (\omega- \omega^\prime)$. Then the emitter-plasmon Hamiltonian is 
\begin{align}
\label{eq:H}
\nonumber
H/\hbar=
& 
\; \frac{\omega_{e}}{2}\sigma_{z}+\int_{0}^{\infty}{\rm
  d}\omega\;\omega
a^{\dagger}(\omega)a(\omega)\,
\\
& +\int_{0}^{\infty}{\rm
  d}\omega\,\Big(g(\omega)\sigma^{-}a^{\dagger}(\omega)+{\rm
  h.c.}\Big)
\end{align}
This is the \textit{spin-boson} model where the emitter, with level
spacing  $\omega_{e}$,  is represented
by standard Pauli matrices $\sigma_{x,y,z}$, $\sigma^{\pm}=\sigma_{x}\pm i\sigma_{y}$.
The coupling
of the emitter to the plasmon modes is characterized by $|g(\omega)|^{2}$,
also called the \textit{spectral density}, 
\begin{equation}
\left|g(\omega)\right|^{2}=\frac{1}{\hbar\pi\epsilon_{0}}\frac{\omega^{2}}{c^{2}}\vec{d}^{T}{\rm Im}[\stackrel{\leftrightarrow}{G}_{{\rm spp}}(\omega,\vec{r}_{e},\vec{r}_{e})]\vec{d}\,.\label{gG}
\end{equation}
Here, $\stackrel{\leftrightarrow}{G}_{{\rm spp}}$ is the contribution of the plasmon pole to the 
the electromagnetic Green's tensor in the nano-structure. It
contains all the information about the plasmonic structure:
plasmon propagation length, modal shape, etc., and depends on both emitter position
($\vec{r}_{e}$) and emitter orientation (via $\vec d$).

Eq.(\ref{gG}) considers only the coupling between the emitter and the (lossy) plasmons. All other mechanisms for loss in the system will be introduced as Lindblad terms, which affect the non-unitary evolution of the emitter-plasmon density matrix. The origin and description of these dissipative channels will be discussed in Sect
\ref{sub:JC}.

\subsection{Green's function of the plasmonic structure}

\begin{figure}
\includegraphics[width=0.9\columnwidth]{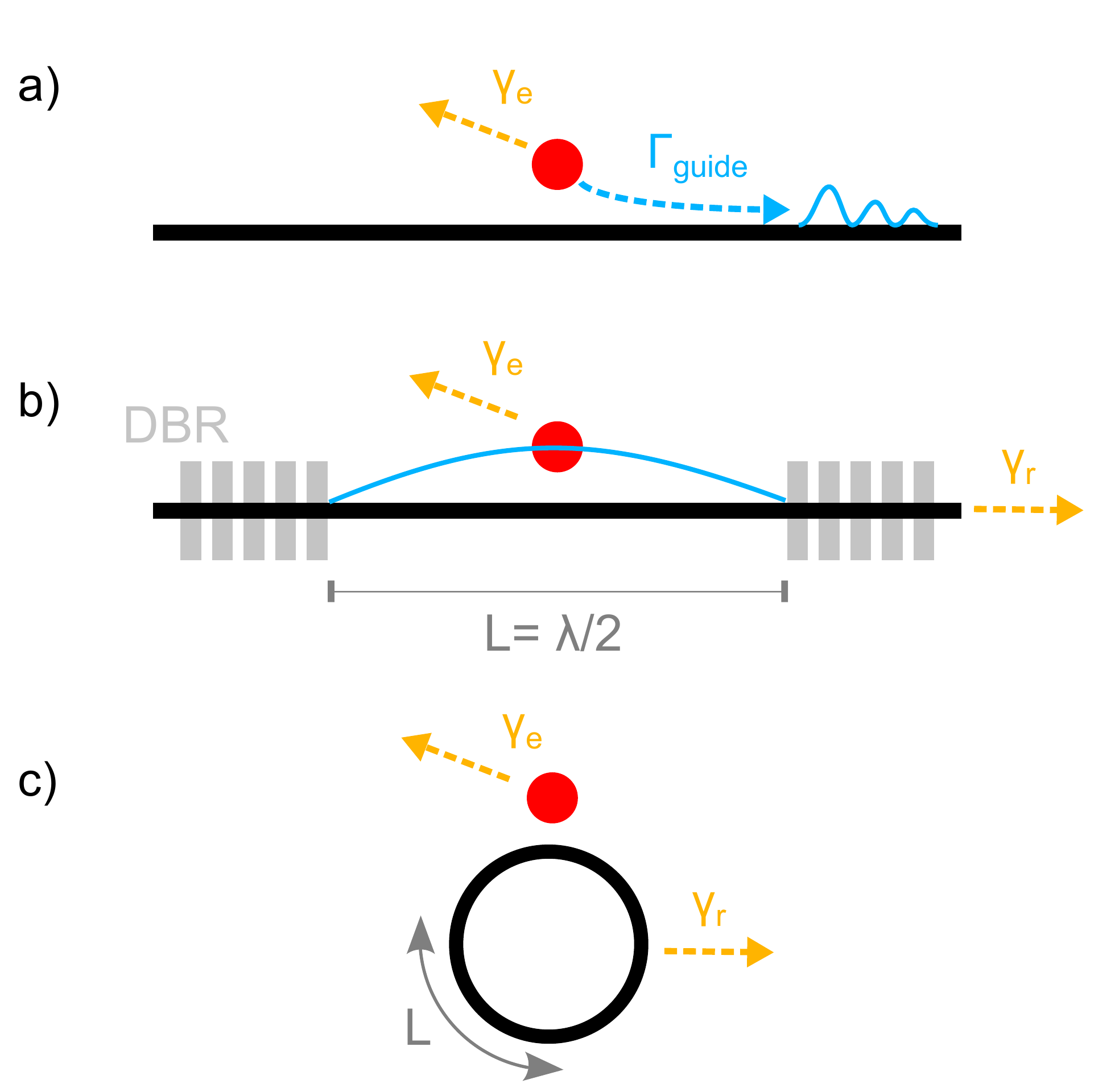}
\caption{(a) 
Sketch of an emitter (red dot) coupled to an open plasmonic wave
guide. It emits with rate $\Gamma_{{\rm guide}}$ into propagating
surface plasmons and with rate $\gamma_{e}$ into other modes. (b)
Linear resonator defined by a waveguide enclosed by mirrors (here
distributed Bragg reflectors, DBR). The excitations from the resonator
are lost with rate $\gamma_{r}$. The length of the resonator has
to be multiples of half the plasmon wavelength for resonances to occur.
(c) A circular resonator configuration. \label{fig:resonators}}
\end{figure}

A strategy for confining the EM field to a small area,
and consequently enhance light-matter interaction, is by using one
dimensional metallic structures (waveguides) that support propagating
plasmons. The plasmonic modes of the waveguide confine the EM
field in two dimensions. Since they are mixed photon-media excitations,
the confinement can exceed the one of free space photons that is limited
by diffraction. To further enhance the interaction the plasmons can
be stored in \emph{resonators}. These can be manufactured out of plasmonic
waveguides by placing two mirrors or building a ring, as sketched in  Fig.~\ref{fig:resonators}.

\subsubsection{Waveguide}

Surface plasmons on an infinite ("\emph{open}") waveguide can be described
by $\vec{E}_{\rm p}(\vec{r})=\vec{E}_{\rm p}(\vec{r}_{t})e^{\pm ik(\omega)z}$, with
the transverse field profile $\vec{E}_{\rm p} (\vec{r}_{t})$, where $z$ is 
the coordinate parallel to the waveguide axis and $\vec{r}_{t}\equiv\left\{ x,y\right\} $
is perpendicular to it. Due to propagation losses of the plasmons,
their propagation constant $k(\omega)=k^{\prime}(\omega)+ik^{\prime\prime}(\omega)$
is a complex quantity. 
The Green's function $G_{{\rm spp}}(\omega,\vec{r},\vec{r}^{\prime})$
can be constructed from the plasmon's electric field \cite{chen2010finiteelement,sondergaard2001general}: 
\begin{equation}
\stackrel{\leftrightarrow}{G}_{{\rm spp}}(\omega,\vec{r},\vec{r}^{\prime})\approx\frac{c^{2}}{\omega v_{g}}\frac{\vec{E}_{\rm p} (\vec{r}_{t})\otimes\vec{E}_{\rm p}^{*}(\vec{r}_{t}^{\prime})}{\int_{A_{\infty}}d^{2}\tilde{r_{t}}\epsilon(\tilde{\vec{r}}_{t})\left|E _{\rm p} (\tilde{\vec{r}}_{t})\right|^{2}}k\, G_{{\rm 1d}}(\omega,z,z^{\prime})\label{eq:green func 3d}
\end{equation}
This is the contribution to the Green's function at the
region outside the metal arising
from the bound surface plasmons. 
In this expression, $c$ is the speed of light and
$v_{g}(\omega)$ is the plasmon group
velocity.

In this work $v_g$ is computed numerically from the dispersion
relation $v_{g}(\omega)=\partial\omega/\partial k$ using frequency dependent values of the permittivity\cite{rodrigo2008influence}. We have found that using approximate expressions which only involve the permittivity at the
working frequency may lead to incorrect results. For instance, the commonly used expression $v_g = \int_A
(\vec{E}_{\rm p} \times \vec{H}_{\rm p} )_z  \, {\rm d}A / \int_A
\epsilon_0 \, \epsilon(\vec r) |E_{\rm p}|^2 {\rm d}A$, exact only for
non-dispersive materials, sometimes strongly overestimates the group velocity.  This is especially important for the case of waveguides with highly confined modes, where the approximate expression may even predict superluminal velocities.

The expression for $G_{{\rm spp}}(\omega,\vec{r},\vec{r}^{\prime})$ is split in a part
perpendicular to the waveguide and a part along the waveguide, $G_{{\rm 1d}}$,
that matches the 1D (scalar) Green's function\cite[Chap. 2]{tai1996dyadicgreen}

\begin{equation}
G_{{\rm 1d}}(\omega,z,z^{\prime})=\frac{i}{2k}e^{ik\left|z-z^{\prime}\right|}\,.\label{eq:G-1D-Open}
\end{equation}
Evaluating the coupling strength $g(\omega)$ at the optimal position
in the waveguide, using Eq.~(\ref{gG})
and Eq.~(\ref{eq:G-1D-Open}) we get
\begin{equation}
\left|g(\omega)\right|^{2}=\frac{1}{2\pi}\Gamma_{0}\frac{3}{\pi\epsilon_{d}^{3/2}}\frac{c}{v_{g}}\frac{A_{0}}{A_{{\rm eff}}}\equiv\frac{1}{2\pi}\Gamma_{{\rm guide}}\,,\label{Open-Coupling}
\end{equation}
where $\Gamma_{0}$ is the ``free space spontaneous emission rate'',
$\Gamma_{0}=\sqrt{\epsilon_{d}}\omega^{3}|d|^{2}/3\pi\hbar\epsilon_{0}c^{3}$,
for an emitter placed in a homogeneous medium with permittivity
$\epsilon_{d}$ (corresponding to the permittivity of the dielectric in which the emitter is placed in). We have defined $\Gamma_{{\rm guide}}$ to
be the emission rate into surface plasmons of the open waveguide. The diffraction limited area, $A_{0}=(\lambda_{0}/2)^{2}$,
is the minimum section in which light of wavelength $\lambda_{0}$ can be confined
to in vacuum. Furthermore, we have introduced the effective mode area of
the plasmon field 
\begin{equation}
A_{{\rm eff}}(r_{t})=\frac{\int_{A_{\infty}}d^{2}\tilde{r_{t}}\epsilon(\tilde{\vec{r}}_{t})\left|E _{\rm p} (\tilde{\vec{r}}_{t})\right|^{2}}{\max\{\epsilon(\vec{r}_{t})\left|E _{\rm p} (\vec{r}_{t})\right|^{2}\}}\,.
\end{equation}

This magnitude is inversely proportional to the maximum energy density and therefore
quantifies the achievable coupling strength.

\subsubsection{Resonator}

The 1D Green's function of a resonator can be obtained by summing over all
the reflected contributions of a wave originating from a $\delta-$source
\cite{tai1996dyadicgreen,hanson2002operator}. The details can be found in Appendix~\ref{app:Green}. 
The change in the Green's function translates into a change in the spectral density, which is related but different to the spectral density in the infinite waveguide.  

Here we will look
at two resonator configurations. Either a linear resonator of length
$L$ terminated by two mirrors with reflectivity $\left|R\right|$
or a circular resonator with circumference $L$.

\begin{figure}
\includegraphics[width=0.9\columnwidth]{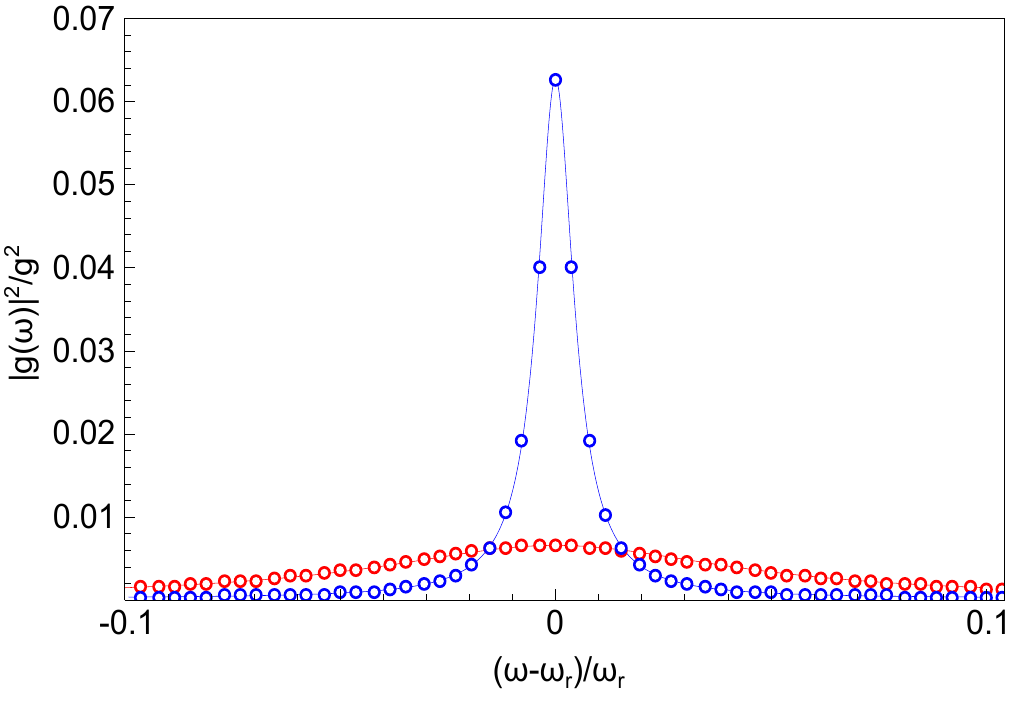}\caption{Spectral density$\left|g(\omega)\right|^{2}$ of the resonator for
different losses (blue $\gamma_{r}=0.01\omega_{r}$ and red $\gamma_{r}=0.1\omega_{r}$).
The solid line is the exact spectral density of a resonator following
Eq.~(\ref{eq:app-1Dgreen-theta}) while the circles are the approximation
with Eq.~(\ref{gwmap}). \label{fig:weak-strong}}
\end{figure}

In both configurations the spectral density $\left|g(\omega)\right|^{2}$
is peaked around the resonance frequencies $\omega_{r}=2\pi v_{p}/\lambda$
as shown in Fig.~\ref{fig:weak-strong}. Here, $v_{p}$ is the phase
velocity and $\lambda=\lambda_{0}v_{p}/c$ the wavelength of the SPPs.
The condition for resonance is
\begin{equation}
L=\frac{\lambda}{2}m=\frac{\pi v_{p}}{\omega_{r}}m\label{eq:L}
\end{equation}
Where $m$ counts the number of the field antinodes in the resonator
and has to be an even integer for circular resonators and any integer
for the linear configuration.

In a real system, the resonator will have losses, with different contributions that can be encapsulated in the coefficient
$\gamma_{{\rm r}}$ 
\begin{equation}
\gamma_{r}\equiv\gamma_{{\rm prop}}+\gamma_{{\rm leak}}=2v_{g}\left(k^{\prime\prime}(\omega_{r})-\frac{1}{L}\ln\left|R\right|\right)\,, \label{eq:gammar}
\end{equation}
where $\gamma_{{\rm prop}}$ are plasmon propagation losses and $\gamma_{{\rm leak}}$
leakage through the mirrors in the linear resonator. For the circular
resonator radiative losses due to bending have to be added but will
not be treated in detail here.

Taking into account losses, the spectral density can be  approximated
close to the resonant frequency
by (see Fig.~\ref{fig:weak-strong}, and Appendix~\ref{app:Green} for a derivation) 
\begin{equation}
|g(\omega)|^{2}\approx g^{2}\frac{2}{\pi}\frac{\gamma_{r}\omega_{r}\omega}{(\omega^{2}-\omega_{r}^{2})^{2}+\gamma_{r}^{2}\omega^{2}}\label{gwmap}\,,
\end{equation}
where we assumed that the resonator linewidth is small compared to
the resonance position, $\gamma_{r}\ll\omega_{r}$, i.e. we have a
well defined resonance. The coupling amplitude is given by 
\begin{equation}
g=\sqrt{\Gamma_{{\rm guide}}\frac{v_{g}}{L}}\,.\label{eq:g-res}
\end{equation}
We assumed that the emitter is positioned at a field antinode in the
linear resonator to yield maximum coupling. In the circular resonator
the emitter can be placed anywhere along the waveguide.

Let us comment on the dependence $\gamma_{{\rm leak}}\sim1/L$. Notice
that $\gamma_{{\rm leak}}$ is the energy loss \emph{per time} through
the resonators mirrors. Therefore the leakage must be proportional
to the energy density at the mirrors which is $\sim1/L$. In a proper
resonator with highly reflective mirrors we can expand $-\ln|R|\equiv-\ln\left(1-\left|T\right|\right)\cong|T|$
for small transmission and absorption coefficients, $\left|T\right|\ll1$.
Thus, the leakage is proportional to the transmittance. In contrast
to the leakage, the propagation losses $\gamma_{{\rm prop}}$ do not
depend on the resonator length since the linewidth (and the loss rate)
quantify the losses \emph{per unit of time} and not per resonator round-trip
of the plasmons. The propagation losses are proportional to the imaginary
part of the plasmon wavevector $k^{\prime\prime}$ or, in other words, inverse to the
plasmon propagation length defined as $\ell\equiv1/(2k^{\prime\prime})$.

\subsection{The JC-model: Plasmonic QED}
\label{sub:JC}
We now use a mathematical result with an enormous physical relevance:
The bosonic bath coupled to a TLS with a peaked spectral density,
$|g(\omega)|^{2}$ like Eq.~(\ref{gwmap}) can be split in a single
boson mode with frequency $\omega_{r}$, coupled to a bath
characterized by a dissipation rate $\gamma_r$, {\it i.e.} the width of
$|g(\omega)|^{2}$ \cite{leggett1984quantum,garg1985effectof,goorden2004entanglement,xian-ting2007decoherence}.
Physically, the $\omega_{r}$-mode is the single resonator mode.
In the end, the plasmonic
resonators discussed here can now be approximated by the Jaynes-Cummings
model 
\begin{equation}
H_{{\rm JC}}/\hbar=\frac{\omega_{e}}{2}\sigma_{z}+\omega_{r}a^{\dagger}a+g\left(\sigma_{-}a^{\dagger}+\sigma_{+}a\right)\label{HJC}
\end{equation}
with additional losses from the emitter (with rate $\gamma_{e}$) and
from the resonator (with rate $\gamma_{r}$). This physics can be encoded
in an \textit{Optical Master Equation} for the density matrix $\rho$,
after tracing out the bath degrees of freedom. It takes the form of
the celebrated \textit{Markovian Lindblad} master equation \cite{breuer2002thetheory,Cohen-Tannoudji1992},
\begin{align}
\dot{\varrho}=  - & \frac{i}{\hbar}[H_{{\rm JC}},\varrho]\label{qme} 
\\ \nonumber
  +& \gamma_{r}\big(a\rho a^{\dagger}-\frac{1}{2}\left\{
    a^{\dagger}a,\varrho\right\} \big)
\\ \nonumber
+&\gamma_{e}\big(\sigma_{-}\varrho\sigma_{+}-\frac{1}{2}\left\{ \sigma_{+}\sigma_{-},\varrho\right\} \big)+\frac{\gamma_{d}}{4}\big(\sigma_{z}\rho\sigma_{z}-\varrho\big)\,.
\end{align}
Here we have introduced two phenomenological rates: $\gamma_e$ and
$\gamma_d$, wich accounts for a dissipative and pure dephasing channel
for the emitter \cite{Tudela11,  martin-cano2011dissipationdriven,  Gonzalez-Tudela2012a}. 

The rate $\gamma_e$ accounts for all processes that provide dissipative transitions between the discrete levels of the qubit. The different contributions to $\gamma_e$ may be written as
\begin{equation}
\gamma_{e}=\gamma_{{\rm rad}}+\gamma_{{\rm nonrad}}
+\gamma_{{\rm int}}.
\end{equation}
Emission into free space radiating EM modes is depicted
by $\gamma_{{\rm rad}}$. Furthermore, another emitter
loss channel specific to plasmonic structures arises: If the emitter
is placed close to a metal surface, it couples to non-propagating,
quickly decaying evanescent modes and the energy is dissipated through
heating of the metal. The associated rate will be called $\gamma_{{\rm nonrad}}$
and can assume very high rates when an emitter is close to a metal
surface. Actually,  $\gamma_{{\rm nonrad}}$ \cite{Chance1978, Barnes1998} is the dominant decay rate at emitter-metal distances below $\sim 10nm$.  To avoid this quenching effect, one may lift the TLS away from the metal
surface, and place it at an intermediate region close enough to still
couple efficiently to plasmons. In all our calculations we assume that the dipole is placed at a 
distance of $10$nm from the metal surface and set  $\gamma_{{\rm nonrad}}=0$. 
This is validated from estimations of  $\gamma_{{\rm nonrad}}$ obtained from the 
fraction of energy radiated by an emitter into waveguide surface plasmons 
(the so called ``$\beta$-factor'') {\cite{chen2010finiteelement,martin-cano2011dissipationdriven}. 
These estimations show that the relevant 
decay rate {\it in plasmonic resonators} arises from both the finite propagation of 
surface plasmons and the transmission at non-perfect mirrors 
(which is another decay route for surface plasmons in the resonator). 
An important point, developed in Appendix~\ref{sub:purcell}, is that in a plasmonic 
resonator the decay into non-radiative channels are penalized with respect to the 
decay into plasmons as the latter can be resonantly enhanced while the former can not. 

Additionally, and in order to present a theory as general as possible (valid for any two level system), we consider in Eq. (16) the possible existence of any other non-electromagnetic dissipative channel, characterized by the phenomenological rate $\gamma_{int}$. 
However, the actual calculations presented in the paper are for either NV-centres or quantum dots, for which the dissipative rates involving transitions between discrete electronic levels are believed to be of electromagnetic origin. Therefore, in all our calculations we set $\gamma_{int}=0$. 

Finally the additional phenomenological pure dephasing
term $\gamma_{d}$ accounts for  transitions between 
the vibro-rotational  manifold of each discrete electronic level of the qubit. Pure dephasing thus models the broadening of the spectral emission
observed in solid state emitters, by \textit{e.g.} coupling to 
phonons\cite{auffeves2009pureemitter,gonzalez-tudela2009effectof,auffeves2010controlling,cui2006emission,majumdar2010linewidth}.

Eq.~(\ref{qme}) enables the mapping between the physics of resonators in  
waveguides and that of cavity-QED systems. As a consequence many
of the results from Jaynes-Cummings physics in cavity QED can be exported
to quantum plasmonic systems. We emphasize that, within the formalism
sketched here, the master equation has been obtained from a \textit{first principles}
theory. Therefore, parameters like the coupling between the single
plasmon resonator mode and the emitter, $g$, and the decoherence
rates, $\gamma_{r}$ and $\gamma_{e}$, can be computed from the emission
spectra of the qubit and the Green's function of the plasmonic structure.

\section{Realization of plasmonic QED\label{sec:Realization} }

In this section we specify the actual emitter and resonator architectures
studied in this work.

\subsection{Waveguides}

The plasmon resonators treated in this work are made out of waveguides.
Therefore, the final resonator properties depend critically on the
specific waveguides used, especially on the achievable field confinement
and plasmon propagation length. Plasmon waveguides are  quasi-1D translational
invariant metal-insulator structures. They possess SPP-eigenmodes that propagate along the waveguide axis, while presenting exponentially decaying evanescent fields in the transverse plane, both in the metal and in the
dielectric. The waveguides we focus at may reach high field confinements
along with low propagation losses. Usually there is a trade-off between
confinement and propagation length, but the actual values are geometrically
dependent. We pick three different waveguide geometries which offer long propagation
lengths along with high field confinements, as well as good fabrication
techniques: The first type are small diameter metal \emph{nanowires}\cite{chang2007strongcoupling}.
The second type are sharp metal \emph{wedges}\cite{moreno2008, boltasseva2008triangular}
offering high field strengths at their tips. Finally, the third class are \emph{hybrid}
waveguides\cite{oulton2008ahybrid,oulton2009plasmon}, formed by
a high refractive index dielectric nanowire (silicon $\epsilon=12.25$)
placed close to a metal surface. There, the SPP of the plane and the
mode of dielectric create a hybrid mode with strong field confinement
in the gap. Transversal cuts through the three waveguides are plotted
in Fig.~\ref{fig:waveguides} along with a sketch of their field
energy distributions. The propagation length and confinement of these
waveguides depend on the concrete geometrical parameters of the waveguides.
In the hybrid waveguide the main parameter is the gap size between
the dielectric and the metal surface. The nanowire properties depend
on the radius and the wedge on the tip angle and the tip radius. For
smaller wires, smaller gaps or tighter angles, respectively, the field
confinement increases and the propagation lengths decrease. Finally,
we consider that the metal waveguides are made of silver, which offers
best propagation lengths at optical and telecom frequencies and are
embedded in PMMA ($\epsilon_d=2$).

\begin{figure}
\includegraphics[width=0.8\columnwidth]{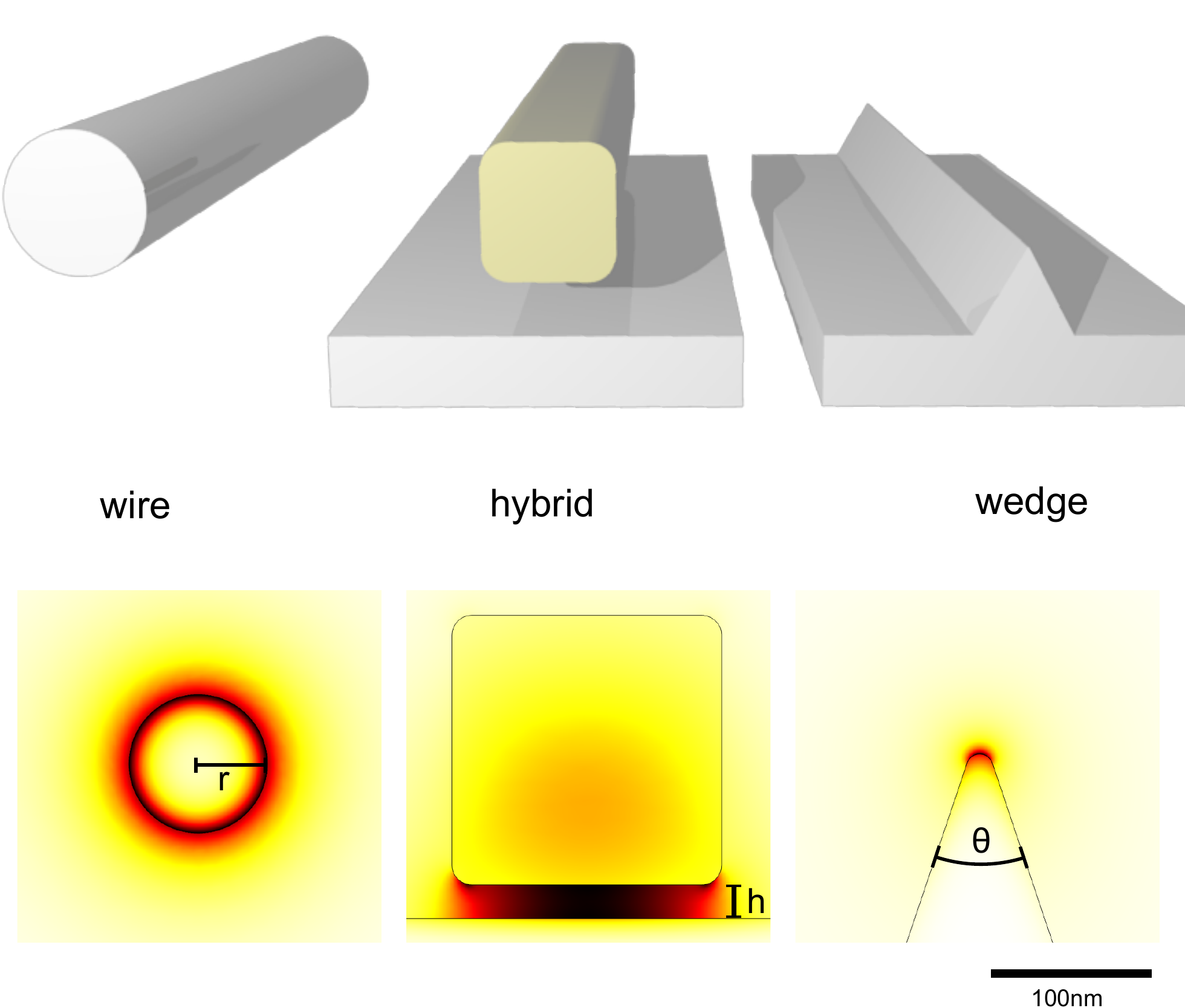}
\caption{Model of the three waveguides treated in this work. Below the relative
energy density of each waveguide is sketched. The parameters used
here are $r=50{\rm nm}$, $h=25{\rm nm}$, $\theta=40^{\circ}$
and the SPP eigenmodes are numerically computed for a frequency corresponding
to a free space wavelength 
\label{fig:waveguides} }
\end{figure}

\subsection{Resonators}

In our calculations we consider two architectures, a \textit{circular}
and a \textit{linear} resonator as sketched in Fig.~\ref{fig:resonators}.

\subsubsection{Circular resonator}

The circular resonator is formed by bending a waveguide and connecting
its ends. The fundamental disadvantage is that the energy is converted
from propagating modes to free radiation at bending~\cite{wang2011lightpropagation,dikken2010characterization}.
On the other hand these losses decrease exponentially with increasing
the radius of the ring. Moreover it is expected that bending losses
are smaller for higher confined modes.
Circular waveguides can nowadays be made lithographically, but this process leads to polycrystalline structures, with its associated radiative and non-radiative
losses at domain interfaces which impair SPP propagation. Hopefully single-crystal circular waveguides (probably synthesized by chemical means) will be available in the near future.  

\subsubsection{Linear resonator}

Linear resonators can be built by placing reflective mirrors in the
waveguide. Here scattering losses and transmission through these mirrors
have to be avoided for having good resonators. Distributed
Bragg reflectors were predicted to be limited to low reflectivities
for plasmons on 2D metal surfaces\cite{oulton2007scattering}. However, recent
resonator realizations have shown high reflectivities by combining both modes
highly confined to waveguides and Bragg reflectors composed of alternating dielectric layers with
small refractive index differences~\cite{deleon2012tailoring}. In
comparison to optical and microwave cavities, where mirror absorption,
scattering and transmission losses can be reduced to several ppm\cite{vahala2003optical,rempe1992measurement},
plasmon mirrors are expected to exhibit losses in the order of a few percent.

\subsection{Quantum emitters \label{sub:Quantum-TLS/Emitters}}

Emitters should (i) be photostable , (ii) present a large dipole moment and (iii) maintain their properties when either embedded into a solid state substrate or placed on top of a surface. Interesting candidates are color centers in crystals or semiconductor
quantum dots (QDs) grown on surfaces or chemically synthesized as
nano-spheres.

The emission spectrum of single atomic emitters traditionally studied
in quantum optics is simply a Lorentzian with a very narrow \emph{transform
limited linewidth} given by the time-energy uncertainty relation.
In contrast, solid state emitters have higher dipole moments but are
also strongly coupled to their solid state environment. Therefore,
the transform limited line, also called zero-phonon-line (ZPL), is
dominated and covered by phonon sidebands, giving rise to a very broad
non-transform limited spectrum~\cite{laucht2009dephasing,majumdar2010linewidth,auffeves2009pureemitter,auffeves2010controlling}.
As used in section~\ref{sub:Emitter-plasmon-interaction}, this broad
peak can be modeled phenomenological by an additional source of dephasing
in the master equation (\ref{qme}) \cite{auffeves2009pureemitter,gonzalez-tudela2009effectof,auffeves2010controlling,cui2006emission,majumdar2010linewidth}.
At lower temperature the phonon sidebands mostly vanish and the ZPL
can be observed.

The first emitter we consider is the nitrogen-vacancy(NV) center in diamond\cite{aharonovich2011diamondbased}.
Single NV centers can be found embedded in diamond
nanocrystals of sizes down to a few $\unit{nm}$. At room temperature (RT), these centers are highly
stable and present an emission line at $\approx\unit[670]{nm}$, with an spectral width (FWHM) of $\approx \unit[80]{nm}$. 
The strong overall
dipole moment (including RT sidebands) is responsible for a spontaneous
emission rate of $\unit[0.04]{GHz}$ at room temperature. At lower
temperatures ($\sim \unit[2]{K}$) the ZPL prevails at $\unit[638]{nm}$
with an spontaneous emission of $\unit[0.013]{GHz}$. NV-centers are
interesting due to their stability, homogeneous properties and long
spin coherence times, making them ideal for quantum information processing
tasks, as has been recently reviewed in Ref.~\cite{aharonovich2011diamondbased}.

The second emitter we investigate are chemically synthesized CdSe
semiconductor quantum dots. These spherical nanocrystals with diameters of the order of 
several $\unit{nm}$ can also be operated at RT and show a
size-tunable emission wavelength in the red part of the spectrum. An typical example 
is a quantum dot with a spectral width  $\approx\unit[20]{nm}$ at $\unit[650]{nm}$\cite{banin1999evidence,akimov2007generation}
and spontaneous emission rate of $\unit[0.05]{GHz}$ at room temperature\cite{akimov2007generation}
. Finally, we study quantum dots made out of InAs clusters in GaAs\cite{andersen2011strongly,andreani1999strongcoupling}.
They can exhibit very strong dipole moments and spontaneous emission
rates above $\unit[1]{GHz}$ at cryogenic temperatures ($T\sim4K$).

\section{ coupling\label{sec:ws} }

\subsection{Strong coupling condition}

The eigenvalues of Eq.(\ref{HJC}) form the so called JC-ladder. At resonance
$\omega_{r}=\omega_{e}$, the states split in doublets $|\psi_{\pm}\rangle=1/\sqrt{2}(|N,g\rangle\pm|N-1,e\rangle)$
with energies $E_{n,\pm}=\hbar N\omega_{0}\pm\hbar\sqrt{N}g$. The
degeneracy between TLS and resonator is lifted because of the coupling,
yielding an anticrossing splitting of $2g\sqrt{N}$. Considering the smallest
level repulsion with one photon, $N=1$, the SC condition
is usually settled as the parameter range where the emission spectrum
of the cavity-emitter consists on two peaks of different frequencies \cite{auffeves2010controlling}.
This is the case if \cite{laussy2008, delvalle2009luminescence, auffeves2010controlling} 
\begin{equation}
|g|>\frac{1}{4}|\gamma_{r}-\gamma_{e}|\,.\label{eq:SC-condition}
\end{equation}
Here we neglected emitter dephasing. In the opposite case, the losses
dominate over the coupling and a splitting of the lines can not be resolved. This
is the WC  regime.

Reaching SC is \textit{the} objective in many cavity
QED experiments. From the fundamental point of view resolving the
$|\psi_{\pm}\rangle$ states confirms the quantum nature of the light-matter
coupling. Being in the SC regime has multiple practical
applications as well, which we will discuss to some extent in the next
section for the case of plasmons. Nevertheless, the WC 
regime has its own interest, \textit{e.g.} for effective single photon
generation. In both cases the ratio of coupling over losses should
be as large as possible. In the following we compute the coupling
and losses in the case of different plasmonic resonators. The number
of parameters to play is huge so a \textit{brute force} exploration
is unpractical. Therefore we first look at the dependencies
of both the coupling and losses on different parameters, which will help us to find the most promising
configurations for achieving strong coupling.

From Eq.~(\ref{eq:g-res}) we see that $g\sim\sqrt{\Gamma_{0}/(A_{{\rm eff}}L})$.
The coupling strength depends on the emitter via its free space spontaneous
emission and, therefore, its dipole moment $\Gamma_{0}\sim|d|^{2}$.
Larger dipole moments directly translate to higher couplings. The
two remaining dependencies come from the resonator itself. The first
one is the transverse field confinement $\sim1/A_{{\rm eff}}$. The
very small modal area of plasmons was the motivation to investigate
plasmon resonators in the first place. Finally, the coupling depends
on the field strength at the emitter position and consequently $g\sim1/\sqrt{L}$
.

Now we turn to the right side of the SC condition, Eq.~(\ref{eq:SC-condition}),
quantifying decoherence of emitter and resonator. To minimize resonator
losses in Eq. (\ref{eq:gammar}) we must search for long propagation
lengths and highly reflective mirrors. Besides, we see an interesting
dependence with $1/L$ in $\gamma_{leak}$ {[}Cf. Eq. (\ref{eq:gammar})
and discussion below{]}. This must be compared to the $1/\sqrt{L}$
dependence of the coupling strength. Therefore, in a realistic scenario
where the reflectivity is always less than one, these two dependencies
compete and, depending on the other parameters, an optimal length appears.
This discussion is also true for the ring configuration by replacing
leakage through the mirrors by bending losses. The latter also decreases
when increasing the resonator length (but this time exponentially
\cite{dikken2010characterization}) since the curvature is reduced.
Therefore, pretty much like in the linear case, for circular resonators
an optimal length also appears.

\subsection{Temperature and propagation losses}

In the next section we will see that the losses from the plasmon resonators
are often dominated by the small propagation length of the plasmons.
In particular, it may not be sufficient to use sophisticated waveguide
geometries to increase this length. An often overlooked factor affecting
plasmon propagation length is temperature, since usual plasmonic experiments
are operated at room temperature. However, lowering the temperature,
the propagation length of plasmons can be systematically extended
by orders of magnitude\cite{gong2007designof}. This increase is possible
if the conduction losses inside the metal are dominated by scattering
by phonons instead of defects like grain boundaries or impurities.
Furthermore, the metal nanostructure must have a smooth surface or
otherwise electron scattering at the surface will dominate\cite{chang2007strongcoupling,ditlbacher2005silvernanowires}(as
well as radiative losses but they are much smaller\cite{chang2007strongcoupling}).

Using the Drude-Sommerfeld model for free electrons, the imaginary
part of the permittivity, $\epsilon^{\prime\prime}$ is approximately
proportional to the resistivity, $\rho$ (details in~\ref{sec:Temperature-and-propagation}).
Since $\left|\epsilon^{\prime}\right|\gg\left|\epsilon^{\prime\prime}\right|$
the modal properties of the plasmons are not affected and an decrease
in resistivity directly translates into an increase in propagation
length

\begin{equation}
\label{Drude}
\ell\propto\frac{1}{\epsilon^{\prime\prime}}\propto\frac{1}{\rho\left(T\right)}\,.
\end{equation}

Working at lower temperatures is of course an experimental hurdle.
However, many quantum emitters must be operated at low temperatures
anyway. In sufficiently
smooth, pure and single crystalline silver, the propagation length
can be easily enhanced by a factor of about $10$ when using liquid
nitrogen ($77\,{\rm K}$) or even almost $100$ when using liquid helium
($4\,{\rm K}$), see  Appendix ~\ref{sec:Temperature-and-propagation}.

\subsection{Strong and weak coupling in plasmon resonators\label{sub:Strong-and-weak}}

Now we present a systematic study on whether the SC
condition (\ref{eq:SC-condition}) can be fulfilled with combinations
of realistic plasmon waveguides, resonator geometries and emitters.
Since this depends on so many adjustable parameters (different waveguides
each with different geometries, resonator reflectivity and length,
temperature, emitters) it is convenient to have a representation that gives
a broad overview for as many of these parameters as possible. To
this end we rearrange the SC condition {[}Eq.~(\ref{eq:SC-condition}){]}
as 
\begin{equation}
\sqrt{\frac{3}{m}\frac{\Gamma_{0}}{\omega}\frac{v_{p}c}{v_{g}^{2}}\frac{A_{0}}{A_{{\rm eff}}}}>\frac{1}{8}\left(\frac{\ell}{\lambda}\right)^{-1}-\frac{1}{2m}\ln\left|R\right|\,.\label{eq:SC-reformatted}
\end{equation}
We have neglected the emitter losses into non-plasmonic modes and used the relation
$\omega=v_{p}k^{\prime}$. Notice that, the properties of the waveguide are
encoded in only two parameters: (i) the field confinement
$\frac{v_{p}c}{v_{g}^{2}}\frac{A_{0}}{A_{\rm eff}}$ and (ii)
the propagation length normalized to the plasmon wavelength $\ell/\lambda$.
By choosing them to be the axes of a 2D-plot\cite{oulton2008confinement}
in Fig.~\ref{fig:optimize1550nm}, we can represent the boundaries that separate 
strong and weak coupling regimes, which are independent of the actual waveguide used.
Furthermore, this can be done for various resonator
lengths ($L=m\lambda/2$, see Eq.~\ref{eq:L}) and mirror reflectivities
($\left|R\right|$). Notice that the lines with $R=1$ in Fig.~\ref{fig:optimize1550nm} can also represent circular resonators which are long enough to have negligible bending
losses (for instance the line with $\{L=25\lambda, R=1\}$).

We can now overlay in this figure the achievable field confinements and propagation lengths for
different waveguides (nanowire, hybrid,
wedge) as function of the geometrical parameters that define them. Within this representation, 
an emitter and a resonator made of a particular waveguide are in the SC regime 
if the point corresponding to the waveguide is above (meaning higher propagation length
than the minimum required) and to the left (meaning higher confinement than needed) of
the relevant given boundary line.

The calculations of propagation lengths and field confinements were carried
out numerically via a finite element method. The emitter is placed at 10nm from the metal surface. 
In this way, as mentioned above, the coupling into non-radiative channels is strongly 
suppressed while the coupling into plasmons is as large as if the emitter were at the surface.  Beyond this restriction, 
both emitter position and orientation were optimized to provide maximal coupling into surface plasmons.

\begin{figure}
\includegraphics[width=1\columnwidth]{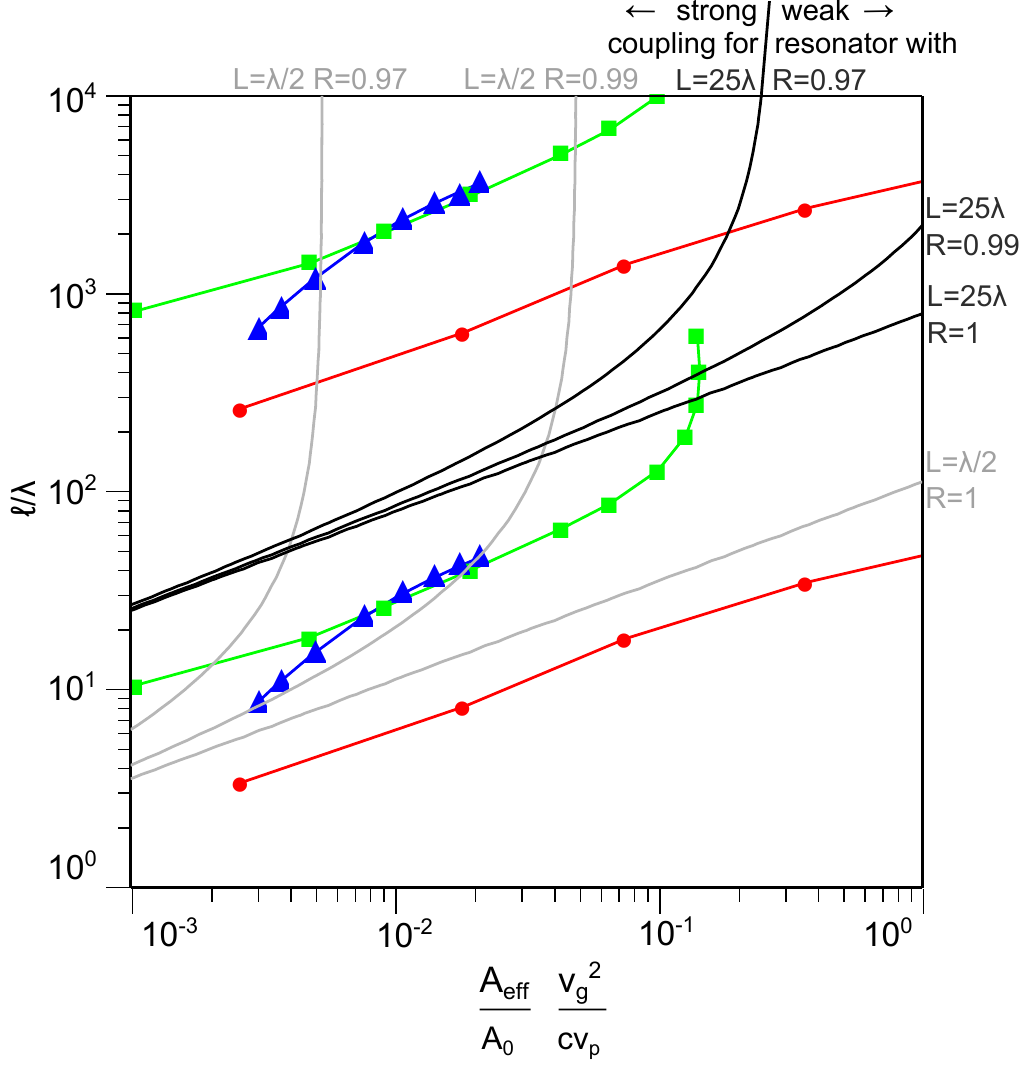}
\caption{Overview of conditions for reaching SC for different resonators, 
made from wedge($\blacktriangle$), hybrid($\blacksquare$) and nanowire($\bullet$) waveguides.
The emitter is assumed to have  $\lambda_{0}=\unit[1550]{nm}$ and $\Gamma_{0}=\unit[1]{GHz}$.
We draw lines separating the region of strong and weak coupling for
multiple resonator realizations with different lengths ($L=\{\lambda/2,25\lambda\}$)
and end reflectivities ($R=\left\{ 1,0.99,0.97\right\} $). The considered waveguides are marked by symbols and are defined by the following parameters (the order in the parameter list corresponds to the order on the waveguide line counted from the leftmost point).  The nanowire waveguide radii are $r_{rad}=\left\{ 25,50,100,250,500,750,1000\right\} \unit{nm}$.
The wedge waveguide angles are $\theta=\left\{ 5,10,20,40,60,80,100,120\right\} ^{\circ}$
and its tip has a radius of $\unit[10]{nm}$. The hybrid waveguide
has separations between the metal and the dielectric nanowire of $h=\left\{ 5,25,50,100,200,300,500,750,1000,1250,1500\right\} \unit{nm}$
and its dielectric nanowire a width of $\unit[200]{nm}$. Furthermore
we show results for waveguides both at room temperature (lower three
lines) and ${\rm {T}}=4{\rm {K}}$ (upper three lines). \label{fig:optimize1550nm}}
\end{figure}

Figure~\ref{fig:optimize1550nm} considers an emitter operating at telecom
frequencies (free space wavelength $\lambda_{0}=\unit[1550]{nm}$
), for instance a self assembled InAs/GaAs quantum dot. As we
can see, SC is hard to achieve at room temperature. With non-perfectly reflecting
mirrors ($R=0.99$) only the hybrid waveguide can lead to resonators in the SC region. The emitted and resonators made of wire or wedge waveguides are always in WC. 
The propagation lengths computed for ${\rm {T}}=4{\rm {K}}$
are two orders of magnitude larger than those computed assuming room temperature. 
Of course, imperfections in the waveguides may limit such
enhancements in propagation length. However, our calculations show that even at the lower
propagation lengths, the improvement when lowering the temperature places the system comfortably in the SC
phase space, specially for long resonators with good reflective ends.

We can conclude that strong
coupling should be reachable at low temperatures between a single high-dipole moment quantum dot (e.g.
InAs/GaAs ) and a plasmonic resonator. This is possible by using (chemically
synthesized) smooth single-crystalline waveguides, realistic DBR mirror
reflectivities above $95\%$ and resonator lengths of several wavelengths.

\subsection{NV-Center or CdSe QDs}

In Fig.~\ref{fig:optimize650nm-0.05GHz} we plot the same as in Fig.~\ref{fig:optimize1550nm}
but for emitters with spontaneous emission rate of $\Gamma_{0}=0.05{\rm {GHz}}$
at $\lambda_{0}=650{\rm {nm}}$. This resembles optimistic values
for CdSe QDs or NV-centers. In this case the SC regime
is harder to reach: the emission rate is smaller and the normalized
propagation length of most of the waveguides is shorter at optical
than at telecom frequencies . Even at lowered temperatures reaching
SC with emitters with such low emission rates presents
an experimental challenge. Especially since at optical frequency interband
transitions, which are independent of temperature, limit the
propagation length increase that can be achieved when lowering the temperature.

\begin{figure}
\includegraphics[width=1\columnwidth]{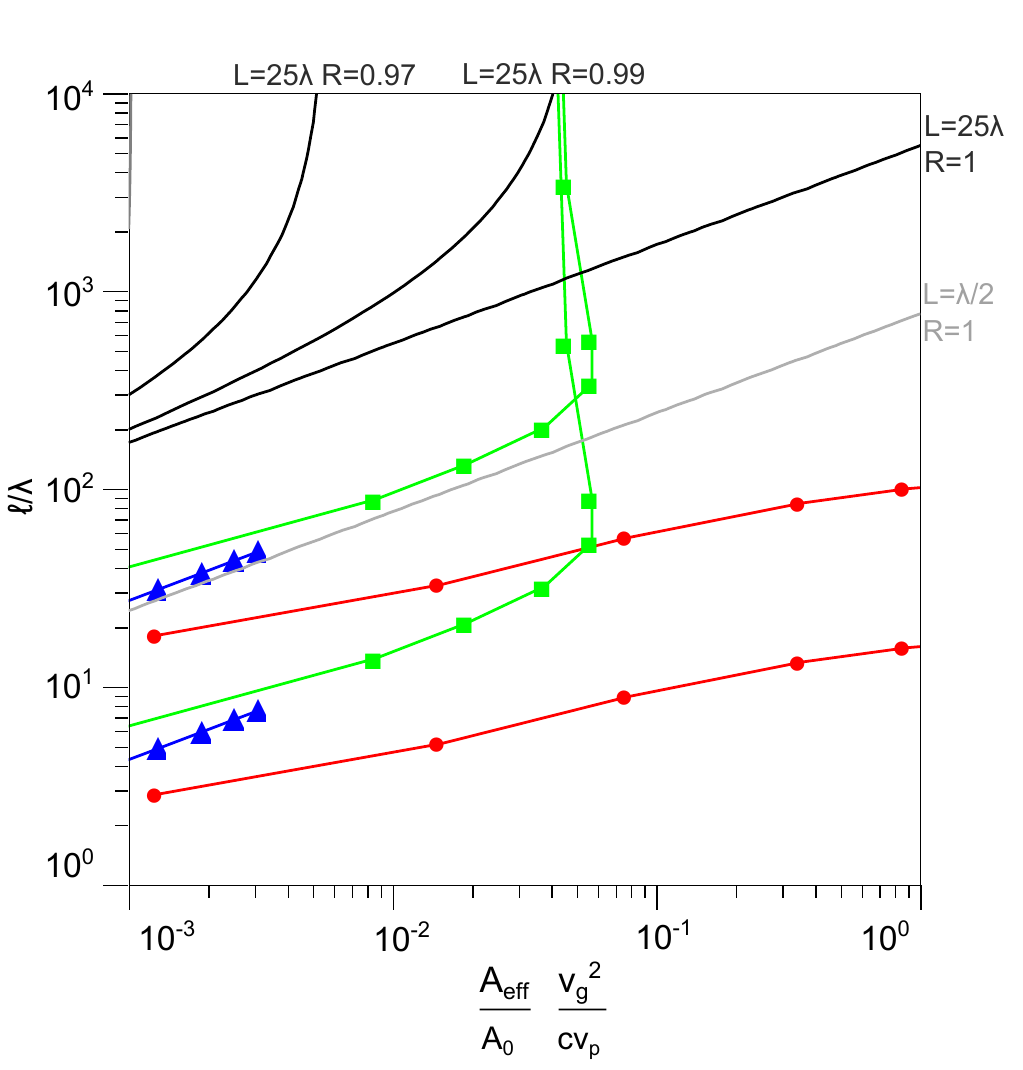}\caption{Same plot as in Fig.~\ref{fig:optimize1550nm} but with an emitter
operating at $\lambda_{0}=\unit[650]{nm}$ and a free space spontaneous
emission rate of $\Gamma_{0}=\unit[0.05]{GHz}$. This roughly corresponds
to CdSe QDs or NV-Centers. The waveguide properties
are all the same as in Fig.~\ref{fig:optimize1550nm} except the
dielectric nanowire that has been adapted to better perform at this
wavelength with a width of $\unit[100]{nm}$.\label{fig:optimize650nm-0.05GHz} }
\end{figure}

\subsection{Tradeoff between confinement and losses}

In Fig.~\ref{fig:optimize1550nm} we first notice the well-known tradeoff between mode confinement and propagation length
for plasmon waveguides \cite{Barnes2003}: The parametric lines for each waveguide run
more or less diagonal from bottom-left to top-right in the 2D parameter-space
plots. However, different waveguide types perform differently, with
the hybrid waveguide offering highest field confinement and propagation
lengths, as noted in Ref.~\cite{oulton2008confinement}. The mode
confinement in the plane perpendicular to the waveguide axis therefore
affects the maximum achievable quality factor of the resonator.

Interestingly, we see that the trend observed for plasmon waveguides
also holds for the resonators as: the stronger the field
confinement in the dimension \emph{along} the waveguide -- e.g. shorter
resonators -- the higher the losses at the ends of the resonators.
This is also true for circular resonators, where building shorter
but more strongly bent resonators results in higher bending losses.

\section{Applications\label{sec:applications}}

Let us discuss some practical applications of the theory presented
so far.

\subsection{Purcell enhancement for single plasmon sources}
\label{sec:purcell}
\begin{figure*}[t]
\includegraphics[width=1\textwidth]{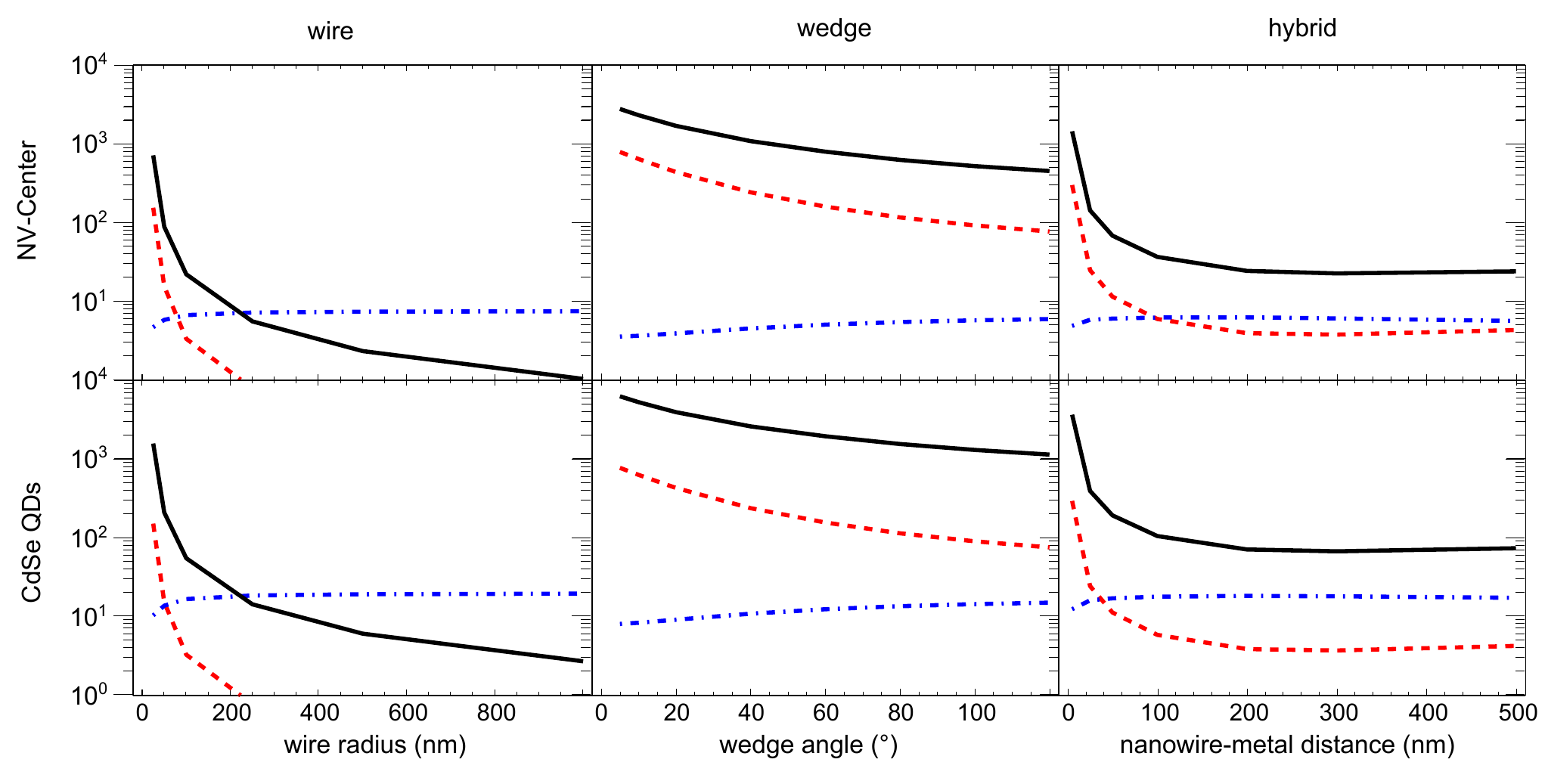}

\caption{Purcell factors for plasmon resonators made of different waveguides.
The emitters used are NV-Centers and CdSe QDs \emph{at room temperature}
with a a free space wavelength of approximately $650{\rm nm}$. The
reflectivity of the resonator ends is $\left|R\right|=0.97$ and its
length $L=\lambda/2$. The red dashed line is the Purcell factor due
to transverse mode confinement of the guided modes ($F_{{\rm guide}}$).
The blue dashed-dotted line is the Purcell factor originating from
the resonator ($F_{{\rm res}}$). The solid black line is the total
Purcell factor. For higher confined modes (which, for each panel, occur at small values in the abscissa),
the resonator Purcell factor decreases since the higher losses decrease
the quality factor of the resonator. For weakly-confined long-propagating
modes (large values in the abscissa) the resonator Purcell factor is limited by the emitter's quality factor. \label{fig:Purcell-factors}}
\end{figure*}

As derived in Appendix\ ~\ref{sub:purcell}, the Purcell factor for plasmon
resonators, \emph{i.e.} emission into surface plasmons compared to
the emission if the emitter would be placed in a homogeneous dielectric
with $\epsilon_{d}$, is 
\begin{equation}
F=F_{{\rm guide}}\times F_{{\rm res}}=\frac{3}{\pi\epsilon_{d}^{3/2}}\frac{c}{v_{g}}\frac{A_{0}}{A_{{\rm eff}}}\times\frac{4}{m\pi}\frac{v_{g}}{v_{p}}\, \frac{1}{\frac{1}{Q_{d}}+\frac{1}{Q_{r}}}\,.
\end{equation}
The first factor is a broadband enhancement due to the strong transversal field confinement
($\propto 1/A_{eff}$) and reduced group velocity for  propagation of EM fields ($\propto 1/v_{g}$) in plasmonic
waveguides. The second part is a resonant (wavelength dependent) enhancement
arising from the longitudinal confinement in the resonator.

At room temperature, the solid state emitters presented in section~\ref{sub:Quantum-TLS/Emitters}
exhibit broad bandwidths due to dephasing. This can be efficiently
expressed in terms of the quality factor of the emitter, $Q=\lambda_{0}/\Delta\lambda$,
where $\Delta\lambda$ is the linewidth of the emitter {\color{blue} almost entirely
attributed to dephasing}. For popular emitters like NV-centers and
CdSe quantum dots the quality factors are $Q_{d,{\rm NV}}\approx670{\rm nm}/80{\rm nm}\approx8$
and $Q_{d,{\rm CdSe}}\approx650{\rm nm}/20{\rm nm}\approx33$, respectively
\cite{aharonovich2011diamondbased,banin1999evidence}. These values
are considerably smaller than those for typical plasmon resonators and therefore
limit the achievable Purcell factors. This can be understood if we
visualize that the linewidth of the resonator is much smaller than
that of the emitter. Therefore the resonator EM modes only resonate with a small
part of the emitter emission spectrum. The advantage of using plasmon resonators
is that the sub-wavelength field confinement related to the underlaying waveguide allows a very high broadband
enhancement of the coupling to the emitter\cite{vesseur2010broadband,deleon2012tailoring,hiscocks2009slotwaveguide}.
Indeed, as we can see in Fig.~\ref{fig:Purcell-factors} the Purcell
factor due to transversal mode confinement is responsible for very high overall Purcell factors. This is even true for very broad
emitters at room temperature where Purcell factors above 1000 are
possible. We especially see that the decrease in propagation length of
more confined modes does not play the dominant role here: the enhancement of the broadband
Purcell factor due to mode confinement amply exceeds the decrease of the resonant
Purcell factor due to the reduction in propagation length.

Let us point out that there are experimental realizations where our theory can be applied. 
In a recent experiment\cite{deleon2012tailoring} Purcell factors
of $75$ have been reported for CdSe QDs when coupled to a $50\,{\rm {nm}}$
radius nanowire embedded in PMMA. Taking the experimental reported
parameters for 
$L=\lambda$ and DBR mirror reflectivity $\approx0.95$, we calculate a Purcell factor 
of $64$ for a distance between the QD and the wire surface of $10\,{\rm {nm}}$. This is a very satisfactory agreement with the reported value,
especially when taking into account that our calculations do not involve any fitting parameter.

Furthermore a look at lowered temperatures is also interesting here.
Higher Purcell factors at lower temperatures can be expected due to higher quality
factors of both resonators and emitters.

Although, in the considered cases, the resonant contribution to the Purcell factor ($F_{{\rm res}}$) 
is smaller than the broadband contribution due to the coupling to the strongly confined waveguide ($F_{{\rm guide}}$), it has several features that
are important for single plasmon sources. First of all, $F_{{\rm res}}$ still has
a value above 5 for the broadband emitters considered in Fig.~\ref{fig:Purcell-factors}
and therefore it is a significant contribution to the high total Purcell
factor. Second, it selectively enhances plasmons with large
propagation lengths, since
it is an effect attributed to cavity resonance. Only the SPP which
match the cavity length are enhanced. This is particular important
when dealing with the ``efficiency'' to emit radiation into this
SPP rather than into other SPPs or non-radiative modes.

\subsection{Strong coupling}

Once we know under which conditions the SC regime is
reachable within plasmonic resonators, we go through some applications.
As anticipated in the introduction, cavity QED systems in the strong
coupling regime are a cornerstone in quantum optics and a huge number of
applications were proposed and implemented in different realizations.
Let us discuss some of them that may have relevance in the manipulation
of light at the nanoscale. All those applications are implicitly or
explicitly related to the coherent coupling between the TLS and the
resonator mode, encapsulated in the ratio, $g/\omega_{r}$ and $g/{\rm max}[\gamma_{r},\gamma_{e},\gamma_{d}]$.
The larger the coupling and smaller the losses, the faster and more
coherent light-matter oscillations are, optimizing the performance
of many of the applications. For later reference, we plot in Fig.~\ref{fig:g-and-gamma} the expected
performance of $g/\gamma_{r}$ as a function of  $g/\omega_{r}$ for several plasmonic resonators
operated at $4{\rm K}$, together with the boundary separating the strong as weak coupling regimes. We consider linear resonators with different lengths and a mirror reflectivity $\left|R\right|=0.99$ but, for the larger resonator length considered the results are also applicable to circular resonators, as in this case bending losses are expected to be negligible.
For ease of comparison to other coupled light-matter systems, Fig.~\ref{fig:g-and-gamma-absolute} reproduces the results rendered in Fig.~\ref{fig:g-and-gamma}, but in terms of explicit values for the coupling and the loss rates.

\begin{figure}
\centering{}\includegraphics[width=0.9\columnwidth]{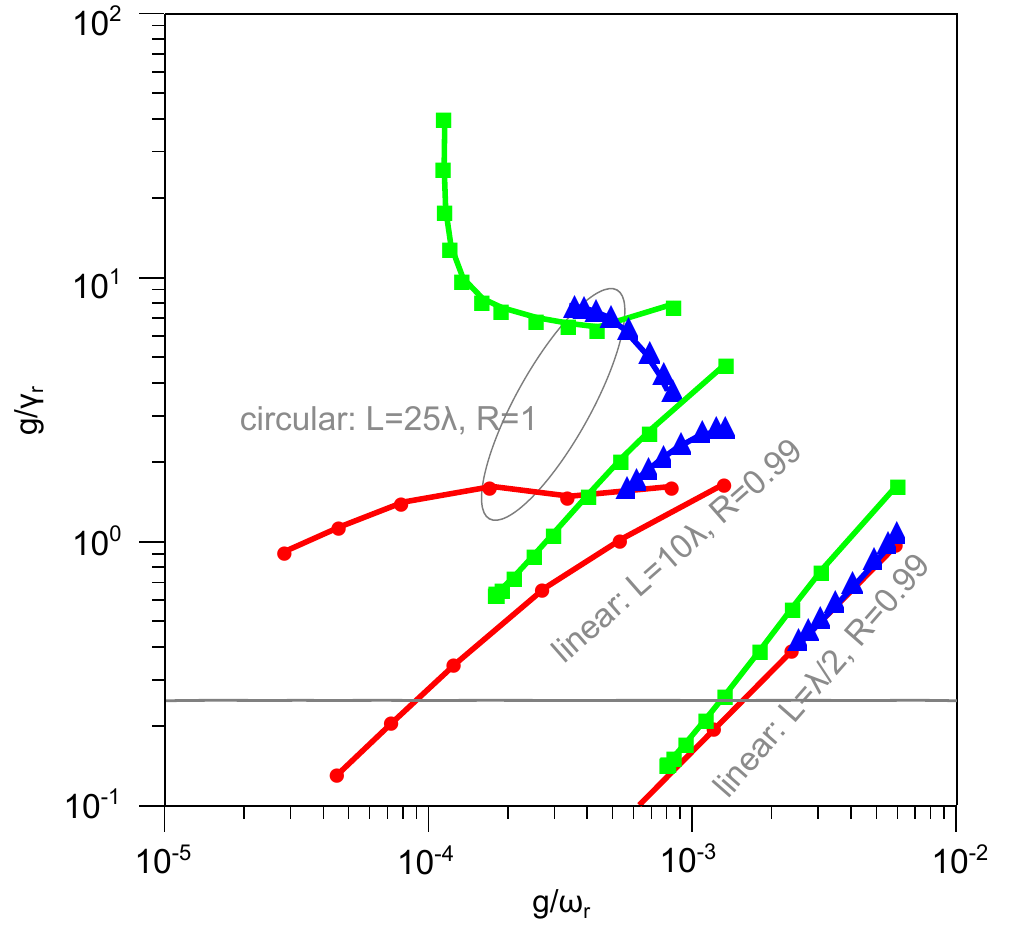}\caption{Coupling strength over losses for an $\Gamma_{0}=\unit[1]{GHz}$ emitter
at $\lambda_{0}=\unit[1550]{nm}$ at $\unit[4]{K}$ for different resonator realizations. The underlying waveguide parameters are varied as in Fig.~\ref{fig:optimize1550nm}. The considered waveguides are marked by symbols and are defined by the following parameters (the order in the parameter list corresponds to the order on the waveguide line counted from the leftmost point).  The nanowire waveguide radii are $r_{rad}=\left\{ 25,50,100,250,500,750,1000\right\} \unit{nm}$.
The wedge waveguide angles are $\theta=\left\{ 5,10,20,40,60,80,100,120\right\} ^{\circ}$
and its tip has a radius of $\unit[10]{nm}$. The hybrid waveguide
has separations between the metal and the dielectric nanowire of $h=\left\{ 5,25,50,100,200,300,500,750,1000,1250,1500\right\} \unit{nm}$
and its dielectric nanowire a width of $\unit[200]{nm}$.  Above the gray line the strong coupling condition is fulfilled.\label{fig:g-and-gamma}}
\end{figure}

\begin{figure}
\centering{}\includegraphics[width=0.9\columnwidth]{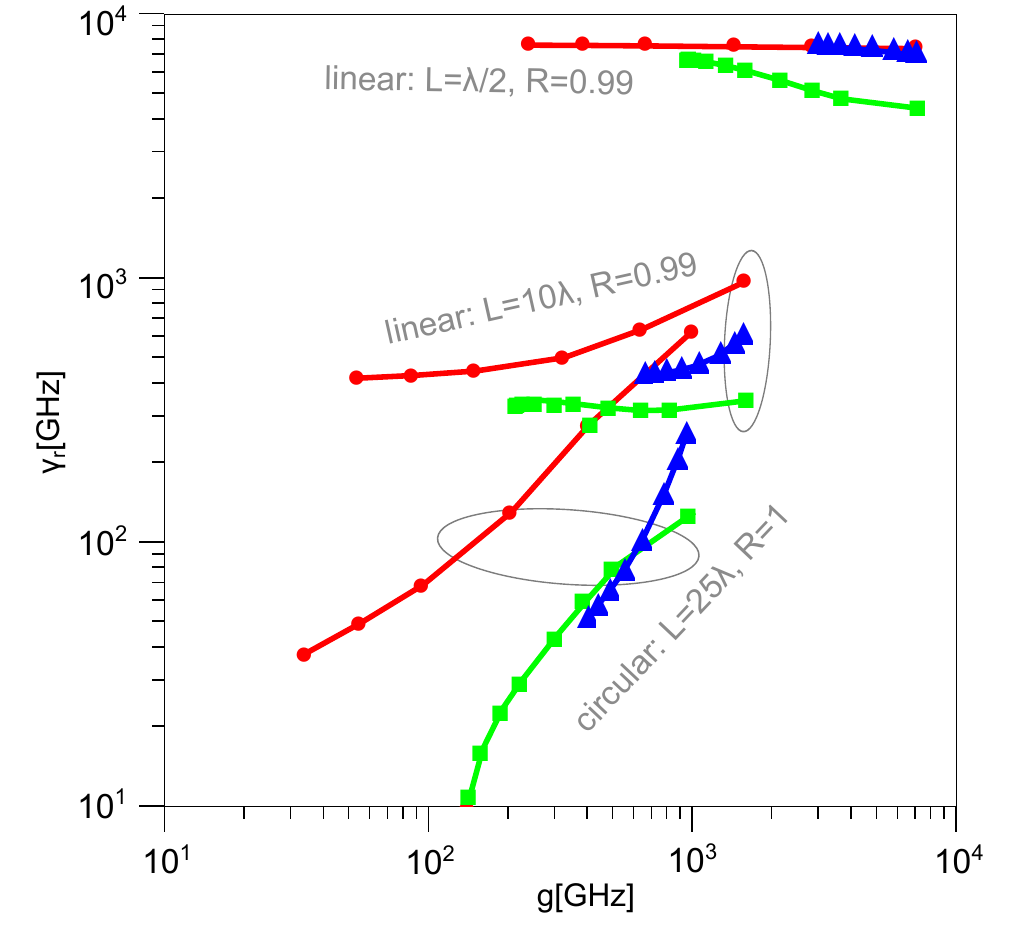}\caption{Absolute values of coupling and resonator loss rates for the same cases considered in Fig.~\ref{fig:g-and-gamma}. \label{fig:g-and-gamma-absolute}}
\end{figure}

\paragraph{Quantum nonlinear optics.}
The JC-model (\ref{HJC}) is nonlinear, the energy levels are not
equally spaced. Therefore, its response to an external stimulus, is
not linear either. In the dispersive regime \cite{Zueco2009}, by
expanding the JC-model in powers of $g/\delta\ll1$ with $\delta=\omega_{r}-\omega_{e}$,
effective Kerr Hamiltonians like $H/\hbar=\omega a^{\dagger}a+\kappa(a^{\dagger}a)^{2}$,
have been proposed \cite{Jacobs2009}. Kerr nonlinearities generate
squeezed states. In a circuit-QED implementation such physics has
been recently reported \cite{Yin2012}. In that work the authors exploited
these nonlinearities to demonstrate, among other things, squeezing.
In this experiment $g/\gamma_{e}\sim10$ and $g/\omega\sim10^{-3}$.
As shown in Fig. \ref{fig:g-and-gamma}, these numbers can be reproduced
with the plasmonic resonators considered in this paper.

\paragraph{Plasmon-plasmon interaction.}
Rooted in the same nonlinearities, the JC-physics can be used to induce
effective photon-photon interactions, as in the so-called photon blockade
phenomenon \cite{Hoffman2011}. Another evidence for photon-photon
interactions in cavity-QED systems have been demonstrated in Ref.
\cite{Englund2012}, where two light beams interact through a QD (InAs)
coupled to a cavity in a photonic crystal. In that experiment, the
reporting numbers are $g/\gamma_{r}\sim1$ and $g/\omega\sim10^{-4}$.
Again, these number are within reach of plasmonic resonators, see
Fig. \ref{fig:g-and-gamma}.

\paragraph{Other.}
Photon-photon interaction allows exploring Bose-Hubbard-like models
in JC-lattices, i.e. arrays of coupled cavity-TLS systems. The non-equally
spaced levels in the JC-model can yield two phases  (localization
and delocalization) depending on the coupling $g$ and the hopping
term between the cavities \cite{Angelakis2007}. These phases survive
even the presence of dissipation \cite{Hummer2012}. Cavity QED is also a building
block for quantum information tasks. On the other hand demonstrations
of quantum computation \cite{Mariantoni2011}, state tomography \cite{Hofheinz2009}
or quantum buses \cite{Majer2007} were done in systems exceeding
the ratios $g/{\rm max}[\gamma_{r},\gamma_{e},\gamma_{d}]$ presented
here. This may be a motivation to further improve current figures, 
for instance the mirror reflectivities, in order to enable these tasks in plasmonic
resonators. Finally, we mention recent advances in doing quantum physics
driven by dissipation \cite{Diehl2008,Verstraete2009}. There, dissipation
is viewed as beneficial for reaching interesting ground states or
doing quantum computation. Because strong dissipation is present in
quantum plasmonics, further investigation in this direction seems
rewarding.

\section{Conclusions}

We have reported a quantum theory for plasmonic resonators coupled
to single quantum emitters. Starting from a \textit{first principle} theory,
and taking into account  the  main losses, we were able to end up in
a master equation for the effective JC model, Cf. Eq.~(\ref{qme}). 
All the coefficients can be obtained via the classical Green's function
together with the emitter characteristics. This allows us to profit
from the studies on plasmonic waveguides. We have studied different
architectures for optimizing the binomia of field enhancement and
losses to reach the SC regime. We have numerically demonstrated
that SC in plasmonics QED is possible at cryogenic temperatures.
Albeit it is demanding at room temperature it is possible if further
improvements are reached, {\it e.g.} higher mirror reflectivities. 
Importantly enough our calculations agree with recent experimental
results in the WC  regime with plasmonic resonators made of
nanowires, see our section \ref{sec:purcell}.  Furthermore, we have shown that other architectures, as
hybrid or wedge waveguides,  can overcome the nanowire implementation and reach even
higher Purcell factors. In the paper we also compare the capabilities of plasmonic
resonators with other technologies.  As demonstrated plasmonic QED can
be used as an effective Kerr media or for generating plasmon-plasmon
interactions, demonstrating its feasibility for controlling the few
plasmon dynamics  at the nanoscale.

\section*{Acknowledgments}

This work was supported by Spanish MICINN projects
FIS2011-25167, MAT2011-28581-C02 and CSD2007-046-Nanolight.es.
FJGV acknowledges financial support by the European Research Council, grant 290981 (PLASMONANOQUANTA).


\appendix

\section{Plamon-Emitter coupling Hamiltonian}
\label{app:A}
In this appendix we justify Hamiltonian (\ref{eq:H}) in the main
text.  First, we quantize the macroscopic Maxwell
equations (taking into account the losses in the metal).  Then, we
provide the emitter-plasmon interaction Hamiltonian.
The material presented here summarizes the formalism used in Refs
\cite{huttner1992quantization,philbin2010canonical} and the book \cite{Vogel2006}.

\subsection{EM quantization in dispersive media: the system-environment approach}

Let us sketch the quantization program
in dispersive (and therefore also lossy) media. We follow an \textit{open
system} approach where the losses are modeled by a reservoir (or ``bath'')
accounting for the irreversible leakage of energy from the system.
We find a \textit{Quantum Langevin Equation}, that in the classical
limit is the Maxwell equation of the EM fields \cite{huttner1992quantization,philbin2010canonical}.

It is convenient to work both in reciprocal space, 
\begin{equation}
\vec{E}(\vec{r},t)=\frac{1}{(2\pi)^{3/2}}\int{\rm d}^{3}k\,\vec{E}(\vec{k},t){\rm e}^{i\vec{k}\cdot\vec{r}}\,,
\end{equation}
and in the Coulomb Gauge, 
\begin{equation}
\vec{A} \cdot \vec k =0\,.
\end{equation}
The quantization is based on a system-bath Lagrangian.  The EM field
is coupled to an infinite collection of resonator.
The latter provide  linear dissipation. 
Being specific we write 
$L_{\rm total}
=
\int {\rm d}^3 k {\mathcal L}_{\rm total}
$
with
\begin{align}
\label{Ltotal}
{\mathcal L}_{{\rm total}}= \, &\epsilon_{0} \left(|\dot{\vec{A}}|^{2}-c\vec{k}^{2}|\vec{A}|^{2}\right)
+\sum_{j}\left(\mu\dot{x}_{j}^{2}-\omega_{j}^{2}x_{j}^{2}\right)
\\ \nonumber
& +\sum_{j}\alpha_{j}\int{\rm d}^{3}k\;\left(\dot{\vec{x}}_{j}\vec{A}^{*}+\dot{\vec{x}}_{j}^{*}\vec{A}\right)\;.
\end{align}
The first line accounts for the EM Lagrangian both in the reciprocal
space and Coulomb gauge and the set of oscillators. The second line
accounts for the interaction.  To alleviate the notation we omit the explicit dependence on $\vec{k}$ and $t$ in $\vec{A}(\vec{k},t)$.
The introduced constants will later on be identified with the system's
material parameters.

With Lagrangian (\ref{Ltotal}) at hand we start the quantization
of $\vec{A}(\vec{k},t)$ and their conjugate momenta 
\begin{equation}
\Pi=\frac{\partial{\mathcal{L}}}{\partial\dot{\vec{A}}^{*}}\;,\quad\Pi^{*}=\frac{\partial{\mathcal{L}}}{\partial\dot{\vec{A}}}
\end{equation}
The quantized fields satisfy the commutation relations\cite{Cohen-Tannoudji1992}
\begin{equation}
[\vec{A}(\vec{k}),\vec{\Pi}(\vec{k}^{\prime})]=0\,,\quad[\vec{A}(\vec{k}),\vec{\Pi}^{\dagger}(\vec{k}^{\prime})]=i\hbar\delta(\vec{k}-\vec{k}^{\prime})
\end{equation}
and the bath's coordinates satisfy 
\begin{equation}
[\vec{x}_{j},\vec{p}_{j^{\prime}}]=i\hbar\vec{\delta}_{jj^{\prime}}
\end{equation}

We write the Heisenberg equations of motion for both \textbf{A} and
the bath operators $x_{j}$. Because of the interaction part, {[}Cf.
third term in (\ref{Ltotal}){]} these equations are coupled 
\begin{eqnarray}
\epsilon_{0}\ddot{\vec{A}} & = & -c^{2}\vec{k}^{2}\vec{A}-\sum_{j}\alpha_{j}\vec{x}_{j}\label{ddotA}\\
\ddot{\vec{x}}_{j} & = & -\omega_{j}^{2}\vec{x}_{j}+\frac{\alpha_{j}}{\mu}\dot{\vec{A}}\,.\label{ddotxj}
\end{eqnarray}
The solution of (\ref{ddotxj}) is given by
\begin{align}
\label{xjsol}
\dot{\vec{x}}_{j}=& i\sqrt{\frac{\hbar\omega_{j}}{2\mu}}\Big(f_{j}^{\dagger}{\rm
  e}^{i\omega_{j}t}-f_{j}{\rm e}^{-i\omega_{j}t}\Big)
\\ \nonumber
&-\frac{\alpha_{j}}{\mu\omega_{j}}\int_{-\infty}^{t}\sin(\omega_{j}(t-t^{\prime}))\ddot{\vec{A}}
\end{align}
with the anhilation/creation operatos:
$[f_i, f^\dagger_j]= \delta_{ij}$.
Inserting the above (\ref{xjsol}) in (\ref{ddotA}) together with
some algebra we end up with an equation for the Fourier components
of the vector potential, 
\begin{equation}
\vec{A}(\vec{k},t)=\int{\rm d}\omega{\rm e}^{-i\omega t}\vec{A}_{\omega}(\vec{k})
\end{equation}
that can be cast to a \emph{Langevin}-like form 
\begin{align}
\label{QLE-M}
-\epsilon(\omega)\omega^{2}\vec{A}_{\omega}= &
-c^{2}\vec{k}^{2}\vec{A}_{\omega}
\\ \nonumber
&-i\sqrt{\frac{\hbar}{\pi}}
\int{\rm d}\nu\,\nu\sqrt{\epsilon^{\prime\prime}(\nu)}\Big(f_{\nu}^{\dagger}{\rm e}^{i\nu t}-f_{\nu}{\rm e}^{-i\nu t}\Big)\,.
\end{align}
The introduced permittivity of the media 
\begin{align}
\label{ew}
\epsilon(\omega)= &
\epsilon^{\prime}(\omega)+i\epsilon^{\prime\prime}(\omega)
\\ \nonumber
= & 1+\frac{1}{\epsilon_{0}}\left({\mathcal{P}}\left[\int{\rm d}\nu\frac{J(\nu)}{\nu-\omega}\right]+i\frac{\pi}{2}J(\omega)\right)
\end{align}
Notice that we have introduced the imaginary part $\epsilon^{\prime\prime}(\omega)$
also appearing in the integrand of (\ref{QLE-M}) and the spectral
density 
\begin{equation}
J(\omega)=\sum_{j}\frac{\alpha_{j}^{2}}{\mu\omega_{j}}\delta(\omega-\omega_{j})\label{Jw}
\end{equation}
Equations (\ref{ew}) and (\ref{Jw}) link the macroscopic complex
permittivity function with a microscopic model accounting for linear
dissipation. 
For example in the Drude-Sommerfeld model, Eq. (\ref{Drude}) we can
write that for frequencies close to the plasma frequency the relation
becomes,
\begin{equation}
J(\omega) \cong \epsilon_0 \frac {\pi}{2} \frac{\Gamma_{\rm el}}{\omega_p^2} \, \omega
\end{equation} 
with $\omega_p$ the plasma frequency and $\Gamma_{\rm el}$ is the damping
parameter.
Finally, and understanding the last term in (\ref{QLE-M}) as {\it the
  source}, rewriting in position-like operators:
$f(\vec r, \omega) = 1/(2 \pi)^{3/2} \int {\rm d}^2 k f (\vec k, t)$
 we express the fields
via the Green function as  expressed in Eq.~(\ref{E}).

\subsection{Emitter-plasmon coupling: some formulae}
As argued in the main text, the emitter-plasmons coupling in the
dipole-dipole approximation is given by:
\begin{equation}
H_{{\rm int}}=-\sigma_{x}\vec{d}\cdot\vec{E}(\vec{r}_{e})\,.\label{dd-app}
\end{equation}
with  $\vec{E}(\vec{r}_{e})=\int_{0}^{\infty}{\rm
  d\omega}\vec{E}(\vec{r}_{e},\omega)$. 
Using (\ref{E}) we can express this interaction Hamiltonian as
\begin{equation}
H_{{\rm int}}=-\sigma_{x} \int_{0}^{\infty}{\rm
  d\omega}
\int{\rm d}^{3}r^{\prime}
g( \omega,r^{\prime},r_{e})
\left(f^{\dagger}(\vec{r}^{\prime},\omega)-f(\vec{r}^{\prime},\omega)\right)
\end{equation}
where we have introduced the shorthand notation
\begin{equation}
g( \omega,r^{\prime},r_{e})
=
i\sqrt{\frac{\hbar}{\pi\epsilon_{0}}}\frac{\omega^{2}}{c^{2}}\sqrt{\epsilon^{\prime\prime}(\vec{r}^{\prime},\omega)} \,
\vec d 
\stackrel
{\leftrightarrow}{G}(\vec{r}_e,\vec{r}^{\prime},\omega)
\,
\,.
\end{equation}
We now define the {\it collective modes} $a(\omega)$
\begin{equation}
\int {\rm d}^{3}r^{\prime}\,
g(\omega,r^{\prime},r_{e})f(r^{\prime},\omega)
\equiv \hbar g(\omega)a(\omega)
\end{equation}
which fullfil the  bosonic conmutation relations
\begin{equation}
[a(\omega), a^\dagger(\omega^\prime) ]
= 
\delta (\omega- \omega^\prime)
\end{equation}
and yield
\begin{widetext}
\begin{equation}
|g(\omega) |^2
= 
\frac{1}{\hbar \pi \epsilon_0} 
\frac{\omega^4}{c^4}
\int {\rm d}^{3}r^{\prime}\,
\epsilon^{\prime\prime}(\vec{r}^{\prime},\omega)
\vec d ^{T}
\stackrel
{\leftrightarrow}{G}
(\vec{r}_e,\vec{r}^{\prime},\omega)
\stackrel
{\leftrightarrow}{G^*}
(\vec{r}_e,\vec{r}^{\prime},\omega)
\vec d\,.
\end{equation}
Using the relation for the Green's tensor~\cite{knoll2000qedin, Dzsotjan2010}
\begin{equation}
\frac{\omega^2}
{c^2} \int {\rm d}^{3}r^{\prime}\,
\epsilon^{\prime\prime}(\vec{r}^{\prime},\omega)
\stackrel
{\leftrightarrow}{G}
(\vec{r}_e,\vec{r}^{\prime},\omega)
\stackrel
{\leftrightarrow}{G^*}
(\vec{r}_e,\vec{r}^{\prime},\omega)
=
{\rm Im} {G}(\vec{r}_e,\vec{r}^{\prime},\omega)
\end{equation}
\end{widetext}

we finally end up with 
\begin{equation}
\left|g(\omega)\right|^{2}=\frac{1}{\hbar\pi\epsilon_{0}}\frac{\omega^{2}}{c^{2}}\vec{d}^{T}{\rm Im}[\stackrel{\leftrightarrow}{G}(\omega,\vec{r}_{e},\vec{r}_{e})]\vec{d}\,.\label{gG1}
\end{equation}
This expression allows the evaluation of the contribution of different electromagnetic modes to the spectral density. In our problem, if the emitter is in an intermediate range of distances to the metal surface (larger than  $\sim 10$ nm to avoid quenching and smaller than the plasmon confinement in the direction normal to the metal surface) the coupling is mainly into plasmons. For this reason, we separate the SPP modes from the rest, treating explicitly the emitter-plasmon coupling via an interacting Hamiltonian (given by Eq.~(\ref{eq:H}) in the main text). The coupling between the emitter and the non-plasmonic electromagnetic modes (considered for this problem as "dissipative channels") are described via Lindblad terms, which affect the non-unitary evolution of the density matrix. With this prescription, the operators 
$a(\omega)$ and $a^\dagger(\omega^\prime)$ are now annihilation and creation operators of {\it plasmon modes} and, correspondingly the spectral density is given by Eq. (\ref{gG}) in the main text. 

\section{Green's function of resonators\label{app:Green}}

\subsection{Linear resonator}

To get the Green's function of a resonator, we first assume that the
system is translational along the z-direction with additional reflections
at the resonator ends, effectively reducing the problem to one dimension.
We further notice that $G_{{\rm 1d}}(\omega,z,z^{\prime})$ can be
obtained by summing all the waves scattered at the mirrors. A resonator
of length $L$ with complex reflection coefficient $R$ ($0\leq\left|R\right|\leq1$)
on the resonator ends located at $x_{l}$ and $x_{r}$ therefore yields
the 1D Green's function 
\begin{widetext}
\begin{eqnarray}
G_{{\rm 1d}}\left(\omega,z,z^{\prime}\right) = \sum_{n=0}^{\infty}\left(e^{ik2L}R^{2}\right)^{n}\frac{i}{2k}\left(e^{ik\left|z-z^{\prime}\right|}+\right.
 \left.Re^{ik\left|2z_{r}-\left(z+z^{\prime}\right)\right|}+Re^{-ik\left|2z_{l}-\left(z+z^{\prime}\right)\right|}+R^{2}e^{ik\left(2L-\left|z-z^{\prime}\right|\right)}\right)\,.\nonumber 
\end{eqnarray}
\end{widetext}
Without loss of generality we set $z_{r}=L/2=-z_{l}$. 

The coupling of an emitter to the resonator, $|g\left(\omega\right)|^{2}\propto{\rm Im}G_{{\rm 1d}}\left(\omega,z,z\right)$,
depends on the position $z$. To maximize the coupling we place the emitter along the resonator axis at an antinode of the electric field.

\subsection{Circular resonator}

T the boundary condition in the circular resonator configuration presents
$2\pi$-periodicity. In a similar way as for the linear case, summing
all the different partial waves yields the Green's function 
\begin{equation}
G_{{\rm 1d}}\left(\omega,z,z^{\prime}\right)=\sum_{n=0}^{\infty}\left(e^{ik2}\right)^{n}\frac{i}{2k}\left(e^{ik\left|z-z^{\prime}\right|}+e^{ik\left(L-\left|z-z^{\prime}\right|\right)}\right)\,.
\end{equation}
Notice that, as expected, in this case the coupling between emitter and resonator does not depend on the
emitter position.

\subsection{Approximating resonances}

The 1D-Green's function evaluated at a field antinode and
$z=z^{\prime}$ can be written, for both linear and circular resonators, as
\begin{equation}
G_{{\rm 1d}}\left(\omega,z,z\right)=\frac{1}{2k}\frac{i\,\sinh\theta^{\prime\prime}+\sin\theta^{\prime}}{\cosh\theta^{\prime\prime}-\cos\theta^{\prime}}\label{eq:app-1Dgreen-theta},
\end{equation}
where we have defined
\begin{equation}
\theta=\theta^{\prime}+i\theta^{\prime\prime}=k^{\prime}L+\varphi+i\left(k^{\prime\prime}L-\ln\left|R\right|\right)
\end{equation}
and  $\varphi=\arg\left(R\right)$ is the phase picked up by reflection at each resonator end, which
is $\approx \pi$ for the linear resonator and zero in the circular configuration,

We see that $\theta^{\prime\prime}$ quantifies the losses, both due to
propagation (via the imaginary part of the propagation constant, $k^{\prime\prime}$),
and the losses through the mirrors (via $|R|<1$). For radiation losses
due to bending in the circular resonator we can phenomenologically
add a term $k_{{\rm bend}}^{\prime\prime}$ to the imaginary part
of the propagation constant, $k^{\prime\prime}$.

The condition for resonances is 
\begin{equation}
L=\frac{\pi}{k_{r}^{\prime}}m=\frac{\lambda}{2}m
\end{equation}
where $m$ is the number of antinodes in the resonator and has to
be an even integer for circular resonators and any integer for the
linear configuration, In both configurations the coupling 
\begin{equation}
\left|g(\omega)\right|^{2}=\Gamma_{{\rm guide}}\frac{1}{\pi}{\rm Im}\left\{ k\, G_{{\rm 1d}}\left(\omega,z,z\right)\right\} \,,
\end{equation}
 can be approximated by a Lorentzian near a resonance. We approximate
the cosine around the center of the resonance peaks 
\begin{equation}
\cos(\theta^{\prime})\cong1-\frac{1}{2}L^{2}(k^{\prime}-k_{0}^{\prime})^{2}=1-\frac{L^{2}}{2v_{g}^{2}}\left(\omega-\omega_{r}\right)
\end{equation}
Therefore we can write, 
\begin{equation}
\left|g(\omega)\right|^{2}\cong g^{2}\frac{1}{\pi}\,\frac{\gamma_{r}/2}{(\omega-\omega_{r})^{2}+\left(\gamma_{r}/2\right)^{2}}\,.\label{eq:app-gw}
\end{equation}
with the width of the Lorentzian (FWHM) $\gamma_{r}$ being the resonator
decay rate 
\begin{equation}
\gamma_{r}=2\frac{v_{g}}{L}\sqrt{2\left(\cosh\theta^{\prime\prime}-1\right)}\approx2\frac{v_{g}}{L}\theta^{\prime\prime}\label{eq:app-gamma}
\end{equation}
and the coupling defined as 
\begin{equation}
g=\sqrt{\Gamma_{{\rm guide}}\frac{v_{g}}{L}}\,\sqrt{\frac{\sinh\theta^{\prime\prime}}{\sqrt{2\left(\cosh\theta^{\prime\prime}-1\right)}}}\approx\sqrt{\Gamma_{{\rm guide}}\frac{v_{g}}{L}}\,.\label{eq:app-g}
\end{equation}
The approximated results were obtained by assuming weak losses, $\theta^{\prime\prime}\ll1$,
and consequently Taylor expanding $\sinh(\theta^{\prime\prime})$
and $\cosh(\theta^{\prime\prime})$.

For small enough $\gamma_{r}$ the Lorentzian in Eq.~(\ref{eq:app-gw})
can be approximated by the expression in Eq.~(\ref{gwmap}), which is used in the
mapping to the Jaynes-Cummings model.

\section{Purcell factors\label{sub:purcell} }

Applying the Markov approximation, the emitter coupled to a plasmon
resonator undergoes exponential decay into surface plasmons, with a
rate given by the spectral density at the frequency of the emitter,
$\omega_{e}$\cite{lekien2009cavityenhanced}(Fermi's golden rule)
\begin{equation}
\Gamma_{{\rm res}}=2\pi\left|g(\omega_{e})\right|^{2}\,.\label{eq:Emission-rate-markov}
\end{equation}
We define the Purcell factor as the ratio of this emission into surface
plasmons compared to $\Gamma_{0}$, the emission if the emitter would
be placed in a homogeneous medium characterized by $\epsilon_{d}$,
\begin{equation}
F=\frac{\Gamma_{{\rm {\rm res}}}}{\Gamma_{0}}\;,\label{eq:PF}
\end{equation}
If emission into other channels is negligible small compared to the
emission into surface plasmons, this Purcell factor measures the decrease
of the emitter lifetime.

In the case of plasmonic resonators the Purcell factor can be written
as the product of two different contributions
\begin{equation}
F=\frac{\Gamma_{{\rm res}}}{\Gamma_{0}}=F_{{\rm guide}}\times F_{{\rm res}}\,.
\end{equation}
The waveguide Purcell Factor $F_{{\rm guide}}$ exists even without
a cavity. It is a result of the small mode area and higher density
of states ($\partial k/\partial\omega=v_{g}^{-1}$) of guided surface
plasmons

\begin{equation}
F_{{\rm guide}}\equiv\frac{\Gamma_{{\rm guide}}}{\Gamma_{0}}\equiv\frac{3}{\pi\epsilon_{d}^{3/2}}\frac{c}{v_{g}}\frac{A_{{\rm 0}}}{A_{{\rm eff}}}\,.
\end{equation}
Notice that $F_{{\rm guide}}$ has no resonance origin, so it is broadband. The additional contribution to the
Purcell factor ($F_{{\rm res}}$)
is, for $g\ll\max\left\{ \gamma_{r},\gamma_{p},\gamma_{e}\right\} $,
\begin{equation}
F_{{\rm res}}=\frac{\Gamma_{{\rm res}}}{\Gamma_{{\rm guide}}}=\frac{4g^{2}}{\gamma_{r}+\gamma_{d}}/\Gamma_{{\rm guide}}\,,
\end{equation}
where $\gamma_{d}$ is the linewidth due to emitter dephasing (spectral
diffusion). Using $g^{2}=\frac{v_{g}}{L}\Gamma_{{\rm guide}}$, $\omega=v_{p}2\pi/\lambda$,  $L=m \lambda/2$, $w_e=w_r$ 
and the quality factors $Q_{d}=\omega_{e}/\gamma_{d}$(emitter dephasing)
and $Q_{r}=\omega_{r}/\gamma_{r}$ (resonator) we can rewrite the
resonator Purcell factor in terms of

\begin{equation}
F_{{\rm res}}=\frac{4}{m\pi}\frac{v_{g}}{v_{p}}\frac{1}{\frac{1}{Q_{p}}+\frac{1}{Q_{r}}}\,.
\end{equation}
The fraction of emission guided into surface plasmons is given by

\begin{equation}
\beta=\frac{\Gamma_{{\rm res}}}{\Gamma_{{\rm res}}+\gamma_{e}}
\end{equation}
In normal cavity QED the emitter decay rates to channels not in the cavity ($\gamma_{e}$)
stay approximately the same with and without the cavity since the
cavity only affects the modes in a small spatial angle. In plasmonics
the presence of the metal surface leaves the emission into radiation modes practically unaltered. However,
an absorbing metal surface also introduces decay into non-radiative
modes that dissipate in the metal. The decay into non-radiative modes is dominant at distances below $\sim10-20$ nm. Here, even at these distances, the plasmonic cavity is quite
useful as it increases the decay into surface plasmons and not the
decay into non-radiative modes. This is an advantage of the ``resonator
Purcell factor'' over the ``broadband Purcell factor''.

\section{Temperature and propagation losses\label{sec:Temperature-and-propagation}}

Since the phonon population is strongly temperature dependent, we
can significantly reduce scattering of electrons and increase the
plasmon propagation lengths.

To get the permittivity as a function of temperature, we recourse
to the Drude-Sommerfeld model for free electrons which is well-applicable
and sufficient for near infrared to telecom ($\lambda_{0}\approx\unit[1550]{nm}$)
wavelengths, where interband transitions can be safely neglected in
silver. Here, the permittivity is given by the plasma frequency of
the free electrons $\omega_{p}$ and the damping rate of the electrons
$\Gamma_{{\rm el}}$

\begin{equation}
\label{Drude}
\epsilon_{{\rm Drude}}(\omega)=1-\frac{\omega_{p}^{2}}{\omega^{2}+i\Gamma_{{\rm el}}\omega}\,.
\end{equation}
The damping is a function of Fermi velocity $v_{F}$ and the mean
free path of the electrons $l_{{\rm el}}$, $\Gamma_{{\rm el}}=\frac{v_{F}}{l_{{\rm el}}}$.
The mean free path is in turn proportional to the resistivity $\rho$,
$l_{{\rm el}}\propto1/\rho$. Using $\omega\gg\Gamma_{{\rm el}}$
the imaginary part of the permittivity is thus approximately proportional
to the resistivity

\begin{equation}
\epsilon_{{\rm Drude}}^{\prime\prime}(\omega)\propto\Gamma_{{\rm el}}\propto\rho\,.
\end{equation}
By using tabulated data for the resistivity of silver\cite{smith1995lowtemperature}
at different temperatures and scaling $\Gamma_{{\rm el}}$ accordingly,
we get the imaginary part of the permittivity at different temperatures.
The real part stays approximately constant.

With $\left|\epsilon^{\prime}\right|\gg\left|\epsilon^{\prime\prime}\right|$
the modal shape of the propagating plasmons is not affected and the
reduced imaginary part of the permittivity directly translates into
increased propagation length. The imaginary part of the permittivity
is plotted in Fig.~\ref{fig:Temperature}. It translates to a propagation
length increase that is universal for all waveguides analyzed in this
paper.

\begin{figure}
\includegraphics[width=0.8\columnwidth]{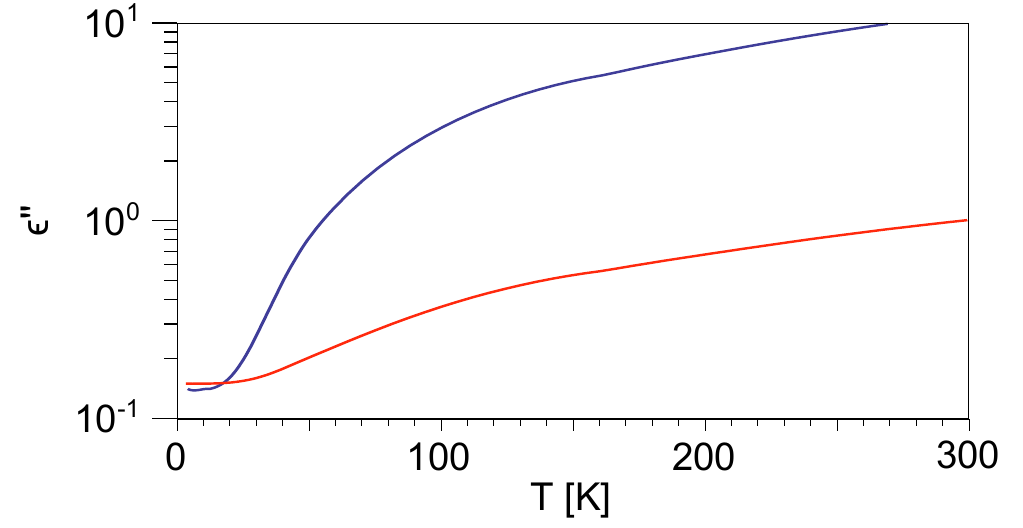}\caption{Permittivity of silver as a function of temperature for\textcolor{red}{{}
}$\lambda_{0}=\unit[1550]{nm}$ (blue line) and $\lambda_{0}=\unit[650]{nm}$
(red line). We see that an increase of about 100 is possible in the
first case, while an increase of about 10 is possible for optical
frequencies.\label{fig:Temperature}}
\end{figure}

At low temperatures $\epsilon^{\prime\prime}$ saturates since the
dominant electron scattering happens at lattice impurities. Furthermore,
when the scattering due to the Drude-Sommerfield model vanishes, 
small but maybe finite interband transitions may play a role. Note
that at optical frequencies they may even dominate. That is why the
change of $\epsilon^{\prime\prime}$ is less pronounced for $\lambda_{0}=\unit[650]{nm}$
in Fig.~\ref{fig:Temperature}. However, the data used for interband
transitions\cite{rodrigo2008influence} may be very vague at lower temperatures since it is obtained
from a fit at room temperature. A more close experimental analysis
is needed here.

Of special interest are the properties at $\unit[77]{K}$ and $\unit[4]{K}$
reachable with liquid nitrogen and liquid helium, respectively. For $\lambda=1550{\rm nm}$,  the plasmon propagation length at 
$\unit[77]{K}$  is $\sim 5$ times larger than the room temperature value,  while at $\unit[4]{K}$ this factor is $80$. Notice that these are estimations obtained from experimental values of the resistivity of silver. Single crystal resonators may present even larger enhancements. 


\bibliographystyle{apsrev4-1}
\bibliography{paper-plasmon}

\begin{thebibliography}{88}%
\makeatletter
\providecommand \@ifxundefined [1]{%
 \@ifx{#1\undefined}
}%
\providecommand \@ifnum [1]{%
 \ifnum #1\expandafter \@firstoftwo
 \else \expandafter \@secondoftwo
 \fi
}%
\providecommand \@ifx [1]{%
 \ifx #1\expandafter \@firstoftwo
 \else \expandafter \@secondoftwo
 \fi
}%
\providecommand \natexlab [1]{#1}%
\providecommand \enquote  [1]{``#1''}%
\providecommand \bibnamefont  [1]{#1}%
\providecommand \bibfnamefont [1]{#1}%
\providecommand \citenamefont [1]{#1}%
\providecommand \href@noop [0]{\@secondoftwo}%
\providecommand \href [0]{\begingroup \@sanitize@url \@href}%
\providecommand \@href[1]{\@@startlink{#1}\@@href}%
\providecommand \@@href[1]{\endgroup#1\@@endlink}%
\providecommand \@sanitize@url [0]{\catcode `\\12\catcode `\$12\catcode
  `\&12\catcode `\#12\catcode `\^12\catcode `\_12\catcode `\%12\relax}%
\providecommand \@@startlink[1]{}%
\providecommand \@@endlink[0]{}%
\providecommand \url  [0]{\begingroup\@sanitize@url \@url }%
\providecommand \@url [1]{\endgroup\@href {#1}{\urlprefix }}%
\providecommand \urlprefix  [0]{URL }%
\providecommand \Eprint [0]{\href }%
\providecommand \doibase [0]{http://dx.doi.org/}%
\providecommand \selectlanguage [0]{\@gobble}%
\providecommand \bibinfo  [0]{\@secondoftwo}%
\providecommand \bibfield  [0]{\@secondoftwo}%
\providecommand \translation [1]{[#1]}%
\providecommand \BibitemOpen [0]{}%
\providecommand \bibitemStop [0]{}%
\providecommand \bibitemNoStop [0]{.\EOS\space}%
\providecommand \EOS [0]{\spacefactor3000\relax}%
\providecommand \BibitemShut  [1]{\csname bibitem#1\endcsname}%
\let\auto@bib@innerbib\@empty
\bibitem [{\citenamefont {Haroche}\ and\ \citenamefont
  {Raimond}(1993)}]{Haroche1993}%
  \BibitemOpen
  \bibfield  {author} {\bibinfo {author} {\bibfnamefont {S.}~\bibnamefont
  {Haroche}}\ and\ \bibinfo {author} {\bibfnamefont {J.}~\bibnamefont
  {Raimond}},\ }\href
  {http://www.osti.gov/energycitations/product.biblio.jsp?osti\_id=121765}
  {\bibfield  {journal} {\bibinfo  {journal} {Scientific American}\ }\textbf
  {\bibinfo {volume} {54}},\ \bibinfo {pages} {26} (\bibinfo {year}
  {1993})}\BibitemShut {NoStop}%
\bibitem [{\citenamefont {Raimond}\ \emph {et~al.}(2001)\citenamefont
  {Raimond}, \citenamefont {Brune},\ and\ \citenamefont
  {Haroche}}]{Raimond2001}%
  \BibitemOpen
  \bibfield  {author} {\bibinfo {author} {\bibfnamefont {J.}~\bibnamefont
  {Raimond}}, \bibinfo {author} {\bibfnamefont {M.}~\bibnamefont {Brune}}, \
  and\ \bibinfo {author} {\bibfnamefont {S.}~\bibnamefont {Haroche}},\ }\href
  {\doibase 10.1103/RevModPhys.73.565} {\bibfield  {journal} {\bibinfo
  {journal} {Reviews of Modern Physics}\ }\textbf {\bibinfo {volume} {73}},\
  \bibinfo {pages} {565} (\bibinfo {year} {2001})}\BibitemShut {NoStop}%
\bibitem [{\citenamefont {Wallraff}\ \emph {et~al.}(2004)\citenamefont
  {Wallraff}, \citenamefont {Schuster}, \citenamefont {Blais}, \citenamefont
  {Frunzio}, \citenamefont {Huang}, \citenamefont {Majer}, \citenamefont
  {Kumar}, \citenamefont {Girvin},\ and\ \citenamefont
  {Schoelkopf}}]{Wallraff2004}%
  \BibitemOpen
  \bibfield  {author} {\bibinfo {author} {\bibfnamefont {A.}~\bibnamefont
  {Wallraff}}, \bibinfo {author} {\bibfnamefont {D.~I.}\ \bibnamefont
  {Schuster}}, \bibinfo {author} {\bibfnamefont {A.}~\bibnamefont {Blais}},
  \bibinfo {author} {\bibfnamefont {L.}~\bibnamefont {Frunzio}}, \bibinfo
  {author} {\bibfnamefont {R.-S.}\ \bibnamefont {Huang}}, \bibinfo {author}
  {\bibfnamefont {J.}~\bibnamefont {Majer}}, \bibinfo {author} {\bibfnamefont
  {S.}~\bibnamefont {Kumar}}, \bibinfo {author} {\bibfnamefont {S.~M.}\
  \bibnamefont {Girvin}}, \ and\ \bibinfo {author} {\bibfnamefont {R.~J.}\
  \bibnamefont {Schoelkopf}},\ }\href {\doibase 10.1038/nature02851} {\bibfield
   {journal} {\bibinfo  {journal} {Nature}\ }\textbf {\bibinfo {volume}
  {431}},\ \bibinfo {pages} {162} (\bibinfo {year} {2004})}\BibitemShut
  {NoStop}%
\bibitem [{\citenamefont {LaHaye}\ \emph {et~al.}(2009)\citenamefont {LaHaye},
  \citenamefont {Suh}, \citenamefont {Echternach}, \citenamefont {Schwab},\
  and\ \citenamefont {Roukes}}]{LaHaye2009}%
  \BibitemOpen
  \bibfield  {author} {\bibinfo {author} {\bibfnamefont {M.~D.}\ \bibnamefont
  {LaHaye}}, \bibinfo {author} {\bibfnamefont {J.}~\bibnamefont {Suh}},
  \bibinfo {author} {\bibfnamefont {P.~M.}\ \bibnamefont {Echternach}},
  \bibinfo {author} {\bibfnamefont {K.~C.}\ \bibnamefont {Schwab}}, \ and\
  \bibinfo {author} {\bibfnamefont {M.~L.}\ \bibnamefont {Roukes}},\ }\href
  {\doibase 10.1038/nature08093} {\bibfield  {journal} {\bibinfo  {journal}
  {Nature}\ }\textbf {\bibinfo {volume} {459}},\ \bibinfo {pages} {960}
  (\bibinfo {year} {2009})}\BibitemShut {NoStop}%
\bibitem [{\citenamefont {Steele}\ \emph {et~al.}(2009)\citenamefont {Steele},
  \citenamefont {H\"{u}ttel}, \citenamefont {Witkamp}, \citenamefont {Poot},
  \citenamefont {Meerwaldt}, \citenamefont {Kouwenhoven},\ and\ \citenamefont
  {van~der Zant}}]{Steele2009}%
  \BibitemOpen
  \bibfield  {author} {\bibinfo {author} {\bibfnamefont {G.~A.}\ \bibnamefont
  {Steele}}, \bibinfo {author} {\bibfnamefont {A.~K.}\ \bibnamefont
  {H\"{u}ttel}}, \bibinfo {author} {\bibfnamefont {B.}~\bibnamefont {Witkamp}},
  \bibinfo {author} {\bibfnamefont {M.}~\bibnamefont {Poot}}, \bibinfo {author}
  {\bibfnamefont {H.~B.}\ \bibnamefont {Meerwaldt}}, \bibinfo {author}
  {\bibfnamefont {L.~P.}\ \bibnamefont {Kouwenhoven}}, \ and\ \bibinfo {author}
  {\bibfnamefont {H.~S.~J.}\ \bibnamefont {van~der Zant}},\ }\href {\doibase
  10.1126/science.1176076} {\bibfield  {journal} {\bibinfo  {journal} {Science
  (New York, N.Y.)}\ }\textbf {\bibinfo {volume} {325}},\ \bibinfo {pages}
  {1103} (\bibinfo {year} {2009})}\BibitemShut {NoStop}%
\bibitem [{\citenamefont {Hennessy}\ \emph {et~al.}(2007)\citenamefont
  {Hennessy}, \citenamefont {Badolato}, \citenamefont {Winger}, \citenamefont
  {Gerace}, \citenamefont {Atat\"{u}re}, \citenamefont {Gulde}, \citenamefont
  {F\"{a}lt}, \citenamefont {Hu},\ and\ \citenamefont
  {Imamoglu}}]{hennessy2007quantum}%
  \BibitemOpen
  \bibfield  {author} {\bibinfo {author} {\bibfnamefont {K.}~\bibnamefont
  {Hennessy}}, \bibinfo {author} {\bibfnamefont {A.}~\bibnamefont {Badolato}},
  \bibinfo {author} {\bibfnamefont {M.}~\bibnamefont {Winger}}, \bibinfo
  {author} {\bibfnamefont {D.}~\bibnamefont {Gerace}}, \bibinfo {author}
  {\bibfnamefont {M.}~\bibnamefont {Atat\"{u}re}}, \bibinfo {author}
  {\bibfnamefont {S.}~\bibnamefont {Gulde}}, \bibinfo {author} {\bibfnamefont
  {S.}~\bibnamefont {F\"{a}lt}}, \bibinfo {author} {\bibfnamefont
  {E.}~\bibnamefont {Hu}}, \ and\ \bibinfo {author} {\bibfnamefont
  {A.}~\bibnamefont {Imamoglu}},\ }\href
  {http://www.nature.com/nature/journal/vaop/ncurrent/full/nature05586.html}
  {\bibfield  {journal} {\bibinfo  {journal} {Nature}\ }\textbf {\bibinfo
  {volume} {445}},\ \bibinfo {pages} {896{\textendash}899} (\bibinfo {year}
  {2007})}\BibitemShut {NoStop}%
\bibitem [{\citenamefont {Zhu}\ \emph {et~al.}(2011)\citenamefont {Zhu},
  \citenamefont {Saito}, \citenamefont {Kemp}, \citenamefont {Kakuyanagi},
  \citenamefont {Karimoto}, \citenamefont {Nakano}, \citenamefont {Munro},
  \citenamefont {Tokura}, \citenamefont {Everitt}, \citenamefont {Nemoto},
  \citenamefont {Kasu}, \citenamefont {Mizuochi},\ and\ \citenamefont
  {Semba}}]{Zhu2011}%
  \BibitemOpen
  \bibfield  {author} {\bibinfo {author} {\bibfnamefont {X.}~\bibnamefont
  {Zhu}}, \bibinfo {author} {\bibfnamefont {S.}~\bibnamefont {Saito}}, \bibinfo
  {author} {\bibfnamefont {A.}~\bibnamefont {Kemp}}, \bibinfo {author}
  {\bibfnamefont {K.}~\bibnamefont {Kakuyanagi}}, \bibinfo {author}
  {\bibfnamefont {S.-i.}\ \bibnamefont {Karimoto}}, \bibinfo {author}
  {\bibfnamefont {H.}~\bibnamefont {Nakano}}, \bibinfo {author} {\bibfnamefont
  {W.~J.}\ \bibnamefont {Munro}}, \bibinfo {author} {\bibfnamefont
  {Y.}~\bibnamefont {Tokura}}, \bibinfo {author} {\bibfnamefont {M.~S.}\
  \bibnamefont {Everitt}}, \bibinfo {author} {\bibfnamefont {K.}~\bibnamefont
  {Nemoto}}, \bibinfo {author} {\bibfnamefont {M.}~\bibnamefont {Kasu}},
  \bibinfo {author} {\bibfnamefont {N.}~\bibnamefont {Mizuochi}}, \ and\
  \bibinfo {author} {\bibfnamefont {K.}~\bibnamefont {Semba}},\ }\href
  {\doibase 10.1038/nature10462} {\bibfield  {journal} {\bibinfo  {journal}
  {Nature}\ }\textbf {\bibinfo {volume} {478}},\ \bibinfo {pages} {221}
  (\bibinfo {year} {2011})}\BibitemShut {NoStop}%
\bibitem [{\citenamefont {Romero-Isart}\ \emph {et~al.}(2010)\citenamefont
  {Romero-Isart}, \citenamefont {Juan}, \citenamefont {Quidant},\ and\
  \citenamefont {Cirac}}]{Romero-Isart2010}%
  \BibitemOpen
  \bibfield  {author} {\bibinfo {author} {\bibfnamefont {O.}~\bibnamefont
  {Romero-Isart}}, \bibinfo {author} {\bibfnamefont {M.~L.}\ \bibnamefont
  {Juan}}, \bibinfo {author} {\bibfnamefont {R.}~\bibnamefont {Quidant}}, \
  and\ \bibinfo {author} {\bibfnamefont {J.~I.}\ \bibnamefont {Cirac}},\ }\href
  {\doibase 10.1088/1367-2630/12/3/033015} {\bibfield  {journal} {\bibinfo
  {journal} {New Journal of Physics}\ }\textbf {\bibinfo {volume} {12}},\
  \bibinfo {pages} {033015} (\bibinfo {year} {2010})}\BibitemShut {NoStop}%
\bibitem [{\citenamefont {de~Leon}\ \emph {et~al.}(2012)\citenamefont
  {de~Leon}, \citenamefont {Shields}, \citenamefont {Yu}, \citenamefont
  {Englund}, \citenamefont {Akimov}, \citenamefont {Lukin},\ and\ \citenamefont
  {Park}}]{deleon2012tailoring}%
  \BibitemOpen
  \bibfield  {author} {\bibinfo {author} {\bibfnamefont {N.~P.}\ \bibnamefont
  {de~Leon}}, \bibinfo {author} {\bibfnamefont {B.~J.}\ \bibnamefont
  {Shields}}, \bibinfo {author} {\bibfnamefont {C.~L.}\ \bibnamefont {Yu}},
  \bibinfo {author} {\bibfnamefont {D.~E.}\ \bibnamefont {Englund}}, \bibinfo
  {author} {\bibfnamefont {A.~V.}\ \bibnamefont {Akimov}}, \bibinfo {author}
  {\bibfnamefont {M.~D.}\ \bibnamefont {Lukin}}, \ and\ \bibinfo {author}
  {\bibfnamefont {H.}~\bibnamefont {Park}},\ }\href {\doibase
  10.1103/PhysRevLett.108.226803} {\bibfield  {journal} {\bibinfo  {journal}
  {Physical Review Letters}\ }\textbf {\bibinfo {volume} {108}},\ \bibinfo
  {pages} {226803} (\bibinfo {year} {2012})}\BibitemShut {NoStop}%
\bibitem [{\citenamefont {Mu}\ and\ \citenamefont
  {Savage}(1992)}]{mu1992oneatom}%
  \BibitemOpen
  \bibfield  {author} {\bibinfo {author} {\bibfnamefont {Y.}~\bibnamefont
  {Mu}}\ and\ \bibinfo {author} {\bibfnamefont {C.~M.}\ \bibnamefont
  {Savage}},\ }\href {\doibase 10.1103/PhysRevA.46.5944} {\bibfield  {journal}
  {\bibinfo  {journal} {Physical Review A}\ }\textbf {\bibinfo {volume} {46}},\
  \bibinfo {pages} {5944} (\bibinfo {year} {1992})}\BibitemShut {NoStop}%
\bibitem [{\citenamefont {Yin}\ \emph {et~al.}(2012)\citenamefont {Yin},
  \citenamefont {Wang}, \citenamefont {Mariantoni}, \citenamefont {Bialczak},
  \citenamefont {Barends}, \citenamefont {Chen}, \citenamefont {Lenander},
  \citenamefont {Lucero}, \citenamefont {Neeley}, \citenamefont {O'Connell},
  \citenamefont {Sank}, \citenamefont {Weides}, \citenamefont {Wenner},
  \citenamefont {Yamamoto}, \citenamefont {Zhao}, \citenamefont {Cleland},\
  and\ \citenamefont {Martinis}}]{Yin2012}%
  \BibitemOpen
  \bibfield  {author} {\bibinfo {author} {\bibfnamefont {Y.}~\bibnamefont
  {Yin}}, \bibinfo {author} {\bibfnamefont {H.}~\bibnamefont {Wang}}, \bibinfo
  {author} {\bibfnamefont {M.}~\bibnamefont {Mariantoni}}, \bibinfo {author}
  {\bibfnamefont {R.}~\bibnamefont {Bialczak}}, \bibinfo {author}
  {\bibfnamefont {R.}~\bibnamefont {Barends}}, \bibinfo {author} {\bibfnamefont
  {Y.}~\bibnamefont {Chen}}, \bibinfo {author} {\bibfnamefont {M.}~\bibnamefont
  {Lenander}}, \bibinfo {author} {\bibfnamefont {E.}~\bibnamefont {Lucero}},
  \bibinfo {author} {\bibfnamefont {M.}~\bibnamefont {Neeley}}, \bibinfo
  {author} {\bibfnamefont {A.}~\bibnamefont {O'Connell}}, \bibinfo {author}
  {\bibfnamefont {D.}~\bibnamefont {Sank}}, \bibinfo {author} {\bibfnamefont
  {M.}~\bibnamefont {Weides}}, \bibinfo {author} {\bibfnamefont
  {J.}~\bibnamefont {Wenner}}, \bibinfo {author} {\bibfnamefont
  {T.}~\bibnamefont {Yamamoto}}, \bibinfo {author} {\bibfnamefont
  {J.}~\bibnamefont {Zhao}}, \bibinfo {author} {\bibfnamefont {A.}~\bibnamefont
  {Cleland}}, \ and\ \bibinfo {author} {\bibfnamefont {J.}~\bibnamefont
  {Martinis}},\ }\href {\doibase 10.1103/PhysRevA.85.023826} {\bibfield
  {journal} {\bibinfo  {journal} {Physical Review A}\ }\textbf {\bibinfo
  {volume} {85}} (\bibinfo {year} {2012}),\
  10.1103/PhysRevA.85.023826}\BibitemShut {NoStop}%
\bibitem [{\citenamefont {Englund}\ \emph {et~al.}(2012)\citenamefont
  {Englund}, \citenamefont {Majumdar}, \citenamefont {Bajcsy}, \citenamefont
  {Faraon}, \citenamefont {Petroff},\ and\ \citenamefont
  {Vu\v{c}kovi\'{c}}}]{Englund2012}%
  \BibitemOpen
  \bibfield  {author} {\bibinfo {author} {\bibfnamefont {D.}~\bibnamefont
  {Englund}}, \bibinfo {author} {\bibfnamefont {A.}~\bibnamefont {Majumdar}},
  \bibinfo {author} {\bibfnamefont {M.}~\bibnamefont {Bajcsy}}, \bibinfo
  {author} {\bibfnamefont {A.}~\bibnamefont {Faraon}}, \bibinfo {author}
  {\bibfnamefont {P.}~\bibnamefont {Petroff}}, \ and\ \bibinfo {author}
  {\bibfnamefont {J.}~\bibnamefont {Vu\v{c}kovi\'{c}}},\ }\href {\doibase
  10.1103/PhysRevLett.108.093604} {\bibfield  {journal} {\bibinfo  {journal}
  {Physical Review Letters}\ }\textbf {\bibinfo {volume} {108}} (\bibinfo
  {year} {2012}),\ 10.1103/PhysRevLett.108.093604}\BibitemShut {NoStop}%
\bibitem [{\citenamefont {Ladd}\ \emph {et~al.}(2010)\citenamefont {Ladd},
  \citenamefont {Jelezko}, \citenamefont {Laflamme}, \citenamefont {Nakamura},
  \citenamefont {Monroe},\ and\ \citenamefont {O'Brien}}]{Ladd2010}%
  \BibitemOpen
  \bibfield  {author} {\bibinfo {author} {\bibfnamefont {T.~D.}\ \bibnamefont
  {Ladd}}, \bibinfo {author} {\bibfnamefont {F.}~\bibnamefont {Jelezko}},
  \bibinfo {author} {\bibfnamefont {R.}~\bibnamefont {Laflamme}}, \bibinfo
  {author} {\bibfnamefont {Y.}~\bibnamefont {Nakamura}}, \bibinfo {author}
  {\bibfnamefont {C.}~\bibnamefont {Monroe}}, \ and\ \bibinfo {author}
  {\bibfnamefont {J.~L.}\ \bibnamefont {O'Brien}},\ }\href {\doibase
  10.1038/nature08812} {\bibfield  {journal} {\bibinfo  {journal} {Nature}\
  }\textbf {\bibinfo {volume} {464}},\ \bibinfo {pages} {45} (\bibinfo {year}
  {2010})}\BibitemShut {NoStop}%
\bibitem [{\citenamefont {Niemczyk}\ \emph {et~al.}(2010)\citenamefont
  {Niemczyk}, \citenamefont {Deppe}, \citenamefont {Huebl}, \citenamefont
  {Menzel}, \citenamefont {Hocke}, \citenamefont {Schwarz}, \citenamefont
  {Garcia-Ripoll}, \citenamefont {Zueco}, \citenamefont {H\"{u}mmer},
  \citenamefont {Solano}, \citenamefont {Marx},\ and\ \citenamefont
  {Gross}}]{Niemczyk2010}%
  \BibitemOpen
  \bibfield  {author} {\bibinfo {author} {\bibfnamefont {T.}~\bibnamefont
  {Niemczyk}}, \bibinfo {author} {\bibfnamefont {F.}~\bibnamefont {Deppe}},
  \bibinfo {author} {\bibfnamefont {H.}~\bibnamefont {Huebl}}, \bibinfo
  {author} {\bibfnamefont {E.~P.}\ \bibnamefont {Menzel}}, \bibinfo {author}
  {\bibfnamefont {F.}~\bibnamefont {Hocke}}, \bibinfo {author} {\bibfnamefont
  {M.~J.}\ \bibnamefont {Schwarz}}, \bibinfo {author} {\bibfnamefont {J.~J.}\
  \bibnamefont {Garcia-Ripoll}}, \bibinfo {author} {\bibfnamefont
  {D.}~\bibnamefont {Zueco}}, \bibinfo {author} {\bibfnamefont
  {T.}~\bibnamefont {H\"{u}mmer}}, \bibinfo {author} {\bibfnamefont
  {E.}~\bibnamefont {Solano}}, \bibinfo {author} {\bibfnamefont
  {A.}~\bibnamefont {Marx}}, \ and\ \bibinfo {author} {\bibfnamefont
  {R.}~\bibnamefont {Gross}},\ }\href {\doibase 10.1038/nphys1730} {\bibfield
  {journal} {\bibinfo  {journal} {Nature Physics}\ }\textbf {\bibinfo {volume}
  {6}},\ \bibinfo {pages} {772} (\bibinfo {year} {2010})}\BibitemShut {NoStop}%
\bibitem [{\citenamefont {Forn-D\'{\i}az}\ \emph {et~al.}(2010)\citenamefont
  {Forn-D\'{\i}az}, \citenamefont {Lisenfeld}, \citenamefont {Marcos},
  \citenamefont {Garc\'{\i}a-Ripoll}, \citenamefont {Solano}, \citenamefont
  {Harmans},\ and\ \citenamefont {Mooij}}]{Forn-Diaz2010}%
  \BibitemOpen
  \bibfield  {author} {\bibinfo {author} {\bibfnamefont {P.}~\bibnamefont
  {Forn-D\'{\i}az}}, \bibinfo {author} {\bibfnamefont {J.}~\bibnamefont
  {Lisenfeld}}, \bibinfo {author} {\bibfnamefont {D.}~\bibnamefont {Marcos}},
  \bibinfo {author} {\bibfnamefont {J.}~\bibnamefont {Garc\'{\i}a-Ripoll}},
  \bibinfo {author} {\bibfnamefont {E.}~\bibnamefont {Solano}}, \bibinfo
  {author} {\bibfnamefont {C.}~\bibnamefont {Harmans}}, \ and\ \bibinfo
  {author} {\bibfnamefont {J.}~\bibnamefont {Mooij}},\ }\href
  {http://prl.aps.org/abstract/PRL/v105/i23/e237001} {\bibfield  {journal}
  {\bibinfo  {journal} {Physical Review Letters}\ }\textbf {\bibinfo {volume}
  {105}},\ \bibinfo {pages} {237001} (\bibinfo {year} {2010})}\BibitemShut
  {NoStop}%
\bibitem [{\citenamefont {Chang}\ \emph {et~al.}(2007)\citenamefont {Chang},
  \citenamefont {S{\o}rensen}, \citenamefont {Hemmer},\ and\ \citenamefont
  {Lukin}}]{chang2007strongcoupling}%
  \BibitemOpen
  \bibfield  {author} {\bibinfo {author} {\bibfnamefont {D.~E.}\ \bibnamefont
  {Chang}}, \bibinfo {author} {\bibfnamefont {A.~S.}\ \bibnamefont
  {S{\o}rensen}}, \bibinfo {author} {\bibfnamefont {P.~R.}\ \bibnamefont
  {Hemmer}}, \ and\ \bibinfo {author} {\bibfnamefont {M.~D.}\ \bibnamefont
  {Lukin}},\ }\href {\doibase 10.1103/PhysRevB.76.035420} {\bibfield  {journal}
  {\bibinfo  {journal} {Physical Review B}\ }\textbf {\bibinfo {volume} {76}},\
  \bibinfo {pages} {035420} (\bibinfo {year} {2007})}\BibitemShut {NoStop}%
\bibitem [{\citenamefont {Oulton}\ \emph
  {et~al.}(2008{\natexlab{a}})\citenamefont {Oulton}, \citenamefont {Bartal},
  \citenamefont {Pile},\ and\ \citenamefont {Zhang}}]{oulton2008confinement}%
  \BibitemOpen
  \bibfield  {author} {\bibinfo {author} {\bibfnamefont {R.~F.}\ \bibnamefont
  {Oulton}}, \bibinfo {author} {\bibfnamefont {G.}~\bibnamefont {Bartal}},
  \bibinfo {author} {\bibfnamefont {D.~F.~P.}\ \bibnamefont {Pile}}, \ and\
  \bibinfo {author} {\bibfnamefont {X.}~\bibnamefont {Zhang}},\ }\href
  {\doibase 10.1088/1367-2630/10/10/105018} {\bibfield  {journal} {\bibinfo
  {journal} {New Journal of Physics}\ }\textbf {\bibinfo {volume} {10}},\
  \bibinfo {pages} {105018} (\bibinfo {year} {2008}{\natexlab{a}})}\BibitemShut
  {NoStop}%
\bibitem [{\citenamefont {Oulton}\ \emph
  {et~al.}(2008{\natexlab{b}})\citenamefont {Oulton}, \citenamefont {Sorger},
  \citenamefont {Genov}, \citenamefont {Pile},\ and\ \citenamefont
  {Zhang}}]{oulton2008ahybrid}%
  \BibitemOpen
  \bibfield  {author} {\bibinfo {author} {\bibfnamefont {R.~F.}\ \bibnamefont
  {Oulton}}, \bibinfo {author} {\bibfnamefont {V.~J.}\ \bibnamefont {Sorger}},
  \bibinfo {author} {\bibfnamefont {D.~A.}\ \bibnamefont {Genov}}, \bibinfo
  {author} {\bibfnamefont {D.~F.~P.}\ \bibnamefont {Pile}}, \ and\ \bibinfo
  {author} {\bibfnamefont {X.}~\bibnamefont {Zhang}},\ }\href {\doibase
  10.1038/nphoton.2008.131} {\bibfield  {journal} {\bibinfo  {journal} {Nature
  Photonics}\ }\textbf {\bibinfo {volume} {2}},\ \bibinfo {pages} {496}
  (\bibinfo {year} {2008}{\natexlab{b}})}\BibitemShut {NoStop}%
\bibitem [{\citenamefont {Boltasseva}\ \emph {et~al.}(2008)\citenamefont
  {Boltasseva}, \citenamefont {Volkov}, \citenamefont {Nielsen}, \citenamefont
  {Moreno}, \citenamefont {Rodrigo},\ and\ \citenamefont
  {Bozhevolnyi}}]{boltasseva2008triangular}%
  \BibitemOpen
  \bibfield  {author} {\bibinfo {author} {\bibfnamefont {A.}~\bibnamefont
  {Boltasseva}}, \bibinfo {author} {\bibfnamefont {V.~S.}\ \bibnamefont
  {Volkov}}, \bibinfo {author} {\bibfnamefont {R.~B.}\ \bibnamefont {Nielsen}},
  \bibinfo {author} {\bibfnamefont {E.}~\bibnamefont {Moreno}}, \bibinfo
  {author} {\bibfnamefont {S.~G.}\ \bibnamefont {Rodrigo}}, \ and\ \bibinfo
  {author} {\bibfnamefont {S.~I.}\ \bibnamefont {Bozhevolnyi}},\ }\href
  {\doibase 10.1364/OE.16.005252} {\bibfield  {journal} {\bibinfo  {journal}
  {Optics Express}\ }\textbf {\bibinfo {volume} {16}},\ \bibinfo {pages} {5252}
  (\bibinfo {year} {2008})}\BibitemShut {NoStop}%
\bibitem [{\citenamefont {Bozhevolnyi}\ \emph {et~al.}(2006)\citenamefont
  {Bozhevolnyi}, \citenamefont {Volkov}, \citenamefont {Devaux}, \citenamefont
  {Laluet},\ and\ \citenamefont {Ebbesen}}]{bozhevolnyi2006channel}%
  \BibitemOpen
  \bibfield  {author} {\bibinfo {author} {\bibfnamefont {S.~I.}\ \bibnamefont
  {Bozhevolnyi}}, \bibinfo {author} {\bibfnamefont {V.~S.}\ \bibnamefont
  {Volkov}}, \bibinfo {author} {\bibfnamefont {E.}~\bibnamefont {Devaux}},
  \bibinfo {author} {\bibfnamefont {J.}~\bibnamefont {Laluet}}, \ and\ \bibinfo
  {author} {\bibfnamefont {T.~W.}\ \bibnamefont {Ebbesen}},\ }\href {\doibase
  10.1038/nature04594} {\bibfield  {journal} {\bibinfo  {journal} {Nature}\
  }\textbf {\bibinfo {volume} {440}},\ \bibinfo {pages} {508} (\bibinfo {year}
  {2006})}\BibitemShut {NoStop}%
\bibitem [{\citenamefont {Stockman}(2004)}]{Stockman04}%
  \BibitemOpen
  \bibfield  {author} {\bibinfo {author} {\bibfnamefont {M.~I.}\ \bibnamefont
  {Stockman}},\ }\href {\doibase 10.1103/PhysRevLett.93.137404} {\bibfield
  {journal} {\bibinfo  {journal} {Phys. Rev. Lett.}\ }\textbf {\bibinfo
  {volume} {93}},\ \bibinfo {pages} {137404} (\bibinfo {year}
  {2004})}\BibitemShut {NoStop}%
\bibitem [{\citenamefont {Volkov}\ \emph {et~al.}(2009)\citenamefont {Volkov},
  \citenamefont {Bozhevolnyi}, \citenamefont {Rodrigo}, \citenamefont
  {Martin-Moreno}, \citenamefont {Garcia-Vidal}, \citenamefont {Devaux},\ and\
  \citenamefont {Ebbesen}}]{Volkov09}%
  \BibitemOpen
  \bibfield  {author} {\bibinfo {author} {\bibfnamefont {V.~S.}\ \bibnamefont
  {Volkov}}, \bibinfo {author} {\bibfnamefont {S.~I.}\ \bibnamefont
  {Bozhevolnyi}}, \bibinfo {author} {\bibfnamefont {S.~G.}\ \bibnamefont
  {Rodrigo}}, \bibinfo {author} {\bibfnamefont {L.}~\bibnamefont
  {Martin-Moreno}}, \bibinfo {author} {\bibfnamefont {F.~J.}\ \bibnamefont
  {Garcia-Vidal}}, \bibinfo {author} {\bibfnamefont {E.}~\bibnamefont
  {Devaux}}, \ and\ \bibinfo {author} {\bibfnamefont {T.~W.}\ \bibnamefont
  {Ebbesen}},\ }\href {\doibase 10.1021/nl900268v} {\bibfield  {journal}
  {\bibinfo  {journal} {Nano Letters}\ }\textbf {\bibinfo {volume} {9}},\
  \bibinfo {pages} {1278} (\bibinfo {year} {2009})},\ \Eprint
  {http://arxiv.org/abs/http://pubs.acs.org/doi/pdf/10.1021/nl900268v}
  {http://pubs.acs.org/doi/pdf/10.1021/nl900268v} \BibitemShut {NoStop}%
\bibitem [{\citenamefont {Oulton}\ \emph {et~al.}(2009)\citenamefont {Oulton},
  \citenamefont {Sorger}, \citenamefont {Zentgraf}, \citenamefont {Ma},
  \citenamefont {Gladden}, \citenamefont {Dai}, \citenamefont {Bartal},\ and\
  \citenamefont {Zhang}}]{oulton2009plasmon}%
  \BibitemOpen
  \bibfield  {author} {\bibinfo {author} {\bibfnamefont {R.~F.}\ \bibnamefont
  {Oulton}}, \bibinfo {author} {\bibfnamefont {V.~J.}\ \bibnamefont {Sorger}},
  \bibinfo {author} {\bibfnamefont {T.}~\bibnamefont {Zentgraf}}, \bibinfo
  {author} {\bibfnamefont {R.}~\bibnamefont {Ma}}, \bibinfo {author}
  {\bibfnamefont {C.}~\bibnamefont {Gladden}}, \bibinfo {author} {\bibfnamefont
  {L.}~\bibnamefont {Dai}}, \bibinfo {author} {\bibfnamefont {G.}~\bibnamefont
  {Bartal}}, \ and\ \bibinfo {author} {\bibfnamefont {X.}~\bibnamefont
  {Zhang}},\ }\href {\doibase 10.1038/nature08364} {\bibfield  {journal}
  {\bibinfo  {journal} {Nature}\ }\textbf {\bibinfo {volume} {461}},\ \bibinfo
  {pages} {629} (\bibinfo {year} {2009})}\BibitemShut {NoStop}%
\bibitem [{\citenamefont {Martin-Cano}\ \emph {et~al.}(2010)\citenamefont
  {Martin-Cano}, \citenamefont {Martin-Moreno}, \citenamefont {Garcia-Vidal},\
  and\ \citenamefont {Moreno}}]{MartinCano10}%
  \BibitemOpen
  \bibfield  {author} {\bibinfo {author} {\bibfnamefont {D.}~\bibnamefont
  {Martin-Cano}}, \bibinfo {author} {\bibfnamefont {L.}~\bibnamefont
  {Martin-Moreno}}, \bibinfo {author} {\bibfnamefont {F.~J.}\ \bibnamefont
  {Garcia-Vidal}}, \ and\ \bibinfo {author} {\bibfnamefont {E.}~\bibnamefont
  {Moreno}},\ }\href {\doibase 10.1021/nl101876f} {\bibfield  {journal}
  {\bibinfo  {journal} {Nano Letters}\ }\textbf {\bibinfo {volume} {10}},\
  \bibinfo {pages} {3129} (\bibinfo {year} {2010})},\ \Eprint
  {http://arxiv.org/abs/http://pubs.acs.org/doi/pdf/10.1021/nl101876f}
  {http://pubs.acs.org/doi/pdf/10.1021/nl101876f} \BibitemShut {NoStop}%
\bibitem [{\citenamefont {Gonzalez-Tudela}\ \emph {et~al.}(2011)\citenamefont
  {Gonzalez-Tudela}, \citenamefont {Martin-Cano}, \citenamefont {Moreno},
  \citenamefont {Martin-Moreno}, \citenamefont {Tejedor},\ and\ \citenamefont
  {Garcia-Vidal}}]{Tudela11}%
  \BibitemOpen
  \bibfield  {author} {\bibinfo {author} {\bibfnamefont {A.}~\bibnamefont
  {Gonzalez-Tudela}}, \bibinfo {author} {\bibfnamefont {D.}~\bibnamefont
  {Martin-Cano}}, \bibinfo {author} {\bibfnamefont {E.}~\bibnamefont {Moreno}},
  \bibinfo {author} {\bibfnamefont {L.}~\bibnamefont {Martin-Moreno}}, \bibinfo
  {author} {\bibfnamefont {C.}~\bibnamefont {Tejedor}}, \ and\ \bibinfo
  {author} {\bibfnamefont {F.~J.}\ \bibnamefont {Garcia-Vidal}},\ }\href
  {\doibase 10.1103/PhysRevLett.106.020501} {\bibfield  {journal} {\bibinfo
  {journal} {Phys. Rev. Lett.}\ }\textbf {\bibinfo {volume} {106}},\ \bibinfo
  {pages} {020501} (\bibinfo {year} {2011})}\BibitemShut {NoStop}%
\bibitem [{\citenamefont {Russell}\ \emph {et~al.}(2012)\citenamefont
  {Russell}, \citenamefont {Liu}, \citenamefont {Cui},\ and\ \citenamefont
  {Hu}}]{Russell2012}%
  \BibitemOpen
  \bibfield  {author} {\bibinfo {author} {\bibfnamefont {K.~J.}\ \bibnamefont
  {Russell}}, \bibinfo {author} {\bibfnamefont {T.-L.}\ \bibnamefont {Liu}},
  \bibinfo {author} {\bibfnamefont {S.}~\bibnamefont {Cui}}, \ and\ \bibinfo
  {author} {\bibfnamefont {E.~L.}\ \bibnamefont {Hu}},\ }\href {\doibase
  10.1038/nphoton.2012.112} {\bibfield  {journal} {\bibinfo  {journal} {Nature
  Photonics}\ }\textbf {\bibinfo {volume} {6}},\ \bibinfo {pages} {459}
  (\bibinfo {year} {2012})}\BibitemShut {NoStop}%
\bibitem [{\citenamefont {Gong}\ and\ \citenamefont
  {Vu\v{c}kovi\'{c}}(2007)}]{gong2007designof}%
  \BibitemOpen
  \bibfield  {author} {\bibinfo {author} {\bibfnamefont {Y.}~\bibnamefont
  {Gong}}\ and\ \bibinfo {author} {\bibfnamefont {J.}~\bibnamefont
  {Vu\v{c}kovi\'{c}}},\ }\href {\doibase 10.1063/1.2431450} {\bibfield
  {journal} {\bibinfo  {journal} {Applied Physics Letters}\ }\textbf {\bibinfo
  {volume} {90}},\ \bibinfo {pages} {033113} (\bibinfo {year}
  {2007})}\BibitemShut {NoStop}%
\bibitem [{\citenamefont {Novotny}\ and\ \citenamefont
  {Hecht}(2006)}]{novotny2006principles}%
  \BibitemOpen
  \bibfield  {author} {\bibinfo {author} {\bibfnamefont {L.}~\bibnamefont
  {Novotny}}\ and\ \bibinfo {author} {\bibfnamefont {B.}~\bibnamefont
  {Hecht}},\ }\href@noop {} {\emph {\bibinfo {title} {Principles of
  {Nano-Optics}}}}\ (\bibinfo  {publisher} {Cambridge University Press},\
  \bibinfo {year} {2006})\BibitemShut {NoStop}%
\bibitem [{\citenamefont {Weiss}(2008)}]{weiss2008quantum}%
  \BibitemOpen
  \bibfield  {author} {\bibinfo {author} {\bibfnamefont {U.}~\bibnamefont
  {Weiss}},\ }\href@noop {} {\emph {\bibinfo {title} {Quantum dissipative
  systems}}}\ (\bibinfo  {publisher} {World Scientific},\ \bibinfo {year}
  {2008})\BibitemShut {NoStop}%
\bibitem [{\citenamefont {Huttner}\ and\ \citenamefont
  {Barnett}(1992)}]{huttner1992quantization}%
  \BibitemOpen
  \bibfield  {author} {\bibinfo {author} {\bibfnamefont {B.}~\bibnamefont
  {Huttner}}\ and\ \bibinfo {author} {\bibfnamefont {S.~M.}\ \bibnamefont
  {Barnett}},\ }\href {\doibase 10.1103/PhysRevA.46.4306} {\bibfield  {journal}
  {\bibinfo  {journal} {Physical Review A}\ }\textbf {\bibinfo {volume} {46}},\
  \bibinfo {pages} {4306} (\bibinfo {year} {1992})}\BibitemShut {NoStop}%
\bibitem [{\citenamefont {Philbin}(2010)}]{philbin2010canonical}%
  \BibitemOpen
  \bibfield  {author} {\bibinfo {author} {\bibfnamefont {T.~G.}\ \bibnamefont
  {Philbin}},\ }\href {http://iopscience.iop.org/1367-2630/12/12/123008}
  {\bibfield  {journal} {\bibinfo  {journal} {New Journal of Physics}\ }\textbf
  {\bibinfo {volume} {12}},\ \bibinfo {pages} {123008} (\bibinfo {year}
  {2010})}\BibitemShut {NoStop}%
\bibitem [{\citenamefont {Kn\"oll}\ \emph {et~al.}(2001)\citenamefont
  {Kn\"oll}, \citenamefont {Scheel},\ and\ \citenamefont
  {Welsch}}]{knoll2000qedin}%
  \BibitemOpen
  \bibfield  {author} {\bibinfo {author} {\bibfnamefont {L.}~\bibnamefont
  {Kn\"oll}}, \bibinfo {author} {\bibfnamefont {S.}~\bibnamefont {Scheel}}, \
  and\ \bibinfo {author} {\bibfnamefont {D.}~\bibnamefont {Welsch}},\ }in\
  \href@noop {} {\emph {\bibinfo {booktitle} {in Coherence and Statistics of
  Photons and Atoms}}}\ (\bibinfo {year} {2001})\BibitemShut {NoStop}%
\bibitem [{\citenamefont {Dung}\ \emph {et~al.}(1998)\citenamefont {Dung},
  \citenamefont {Kn\"{o}ll},\ and\ \citenamefont
  {Welsch}}]{dung1998threedimensional}%
  \BibitemOpen
  \bibfield  {author} {\bibinfo {author} {\bibfnamefont {H.~T.}\ \bibnamefont
  {Dung}}, \bibinfo {author} {\bibfnamefont {L.}~\bibnamefont {Kn\"{o}ll}}, \
  and\ \bibinfo {author} {\bibfnamefont {D.}~\bibnamefont {Welsch}},\ }\href
  {\doibase 10.1103/PhysRevA.57.3931} {\bibfield  {journal} {\bibinfo
  {journal} {Physical Review A}\ }\textbf {\bibinfo {volume} {57}},\ \bibinfo
  {pages} {3931} (\bibinfo {year} {1998})}\BibitemShut {NoStop}%
\bibitem [{\citenamefont {Dung}\ \emph {et~al.}(2003)\citenamefont {Dung},
  \citenamefont {Buhmann}, \citenamefont {Kn\"{o}ll}, \citenamefont {Welsch},
  \citenamefont {Scheel},\ and\ \citenamefont
  {K\"{a}stel}}]{dung2003electromagneticfield}%
  \BibitemOpen
  \bibfield  {author} {\bibinfo {author} {\bibfnamefont {H.~T.}\ \bibnamefont
  {Dung}}, \bibinfo {author} {\bibfnamefont {S.~Y.}\ \bibnamefont {Buhmann}},
  \bibinfo {author} {\bibfnamefont {L.}~\bibnamefont {Kn\"{o}ll}}, \bibinfo
  {author} {\bibfnamefont {D.}~\bibnamefont {Welsch}}, \bibinfo {author}
  {\bibfnamefont {S.}~\bibnamefont {Scheel}}, \ and\ \bibinfo {author}
  {\bibfnamefont {J.}~\bibnamefont {K\"{a}stel}},\ }\href {\doibase
  10.1103/PhysRevA.68.043816} {\bibfield  {journal} {\bibinfo  {journal}
  {Physical Review A}\ }\textbf {\bibinfo {volume} {68}},\ \bibinfo {pages}
  {043816} (\bibinfo {year} {2003})}\BibitemShut {NoStop}%
\bibitem [{\citenamefont {Gruner}\ and\ \citenamefont
  {Welsch}(1996)}]{gruner1996greenfunction}%
  \BibitemOpen
  \bibfield  {author} {\bibinfo {author} {\bibfnamefont {T.}~\bibnamefont
  {Gruner}}\ and\ \bibinfo {author} {\bibfnamefont {D.~G.}\ \bibnamefont
  {Welsch}},\ }\href@noop {} {\bibfield  {journal} {\bibinfo  {journal}
  {Physical Review A}\ }\textbf {\bibinfo {volume} {53}},\ \bibinfo {pages}
  {1818} (\bibinfo {year} {1996})}\BibitemShut {NoStop}%
\bibitem [{\citenamefont {Schleich}(2001)}]{schleich2001quantum}%
  \BibitemOpen
  \bibfield  {author} {\bibinfo {author} {\bibfnamefont {W.~P.}\ \bibnamefont
  {Schleich}},\ }\href@noop {} {\emph {\bibinfo {title} {Quantum Optics in
  Phase Space}}},\ \bibinfo {edition} {1st}\ ed.\ (\bibinfo  {publisher}
  {{Wiley-VCH}},\ \bibinfo {year} {2001})\BibitemShut {NoStop}%
\bibitem [{\citenamefont {Dzsotjan}\ \emph {et~al.}(2010)\citenamefont
  {Dzsotjan}, \citenamefont {S{\o}rensen},\ and\ \citenamefont
  {Fleischhauer}}]{Dzsotjan2010}%
  \BibitemOpen
  \bibfield  {author} {\bibinfo {author} {\bibfnamefont {D.}~\bibnamefont
  {Dzsotjan}}, \bibinfo {author} {\bibfnamefont {A.}~\bibnamefont
  {S{\o}rensen}}, \ and\ \bibinfo {author} {\bibfnamefont {M.}~\bibnamefont
  {Fleischhauer}},\ }\href {\doibase 10.1103/PhysRevB.82.075427} {\bibfield
  {journal} {\bibinfo  {journal} {Physical Review B}\ }\textbf {\bibinfo
  {volume} {82}},\ \bibinfo {pages} {075427} (\bibinfo {year}
  {2010})}\BibitemShut {NoStop}%
\bibitem [{\citenamefont {Chen}\ \emph {et~al.}(2010)\citenamefont {Chen},
  \citenamefont {Nielsen}, \citenamefont {Gregersen}, \citenamefont {Lodahl},\
  and\ \citenamefont {M{\o}rk}}]{chen2010finiteelement}%
  \BibitemOpen
  \bibfield  {author} {\bibinfo {author} {\bibfnamefont {Y.}~\bibnamefont
  {Chen}}, \bibinfo {author} {\bibfnamefont {T.~R.}\ \bibnamefont {Nielsen}},
  \bibinfo {author} {\bibfnamefont {N.}~\bibnamefont {Gregersen}}, \bibinfo
  {author} {\bibfnamefont {P.}~\bibnamefont {Lodahl}}, \ and\ \bibinfo {author}
  {\bibfnamefont {J.}~\bibnamefont {M{\o}rk}},\ }\href {\doibase
  10.1103/PhysRevB.81.125431} {\bibfield  {journal} {\bibinfo  {journal}
  {Physical Review B}\ }\textbf {\bibinfo {volume} {81}},\ \bibinfo {pages}
  {125431} (\bibinfo {year} {2010})}\BibitemShut {NoStop}%
\bibitem [{\citenamefont {S{\o}ndergaard}\ and\ \citenamefont
  {Tromborg}(2001)}]{sondergaard2001general}%
  \BibitemOpen
  \bibfield  {author} {\bibinfo {author} {\bibfnamefont {T.}~\bibnamefont
  {S{\o}ndergaard}}\ and\ \bibinfo {author} {\bibfnamefont {B.}~\bibnamefont
  {Tromborg}},\ }\href {\doibase 10.1103/PhysRevA.64.033812} {\bibfield
  {journal} {\bibinfo  {journal} {Physical Review A}\ }\textbf {\bibinfo
  {volume} {64}},\ \bibinfo {pages} {033812} (\bibinfo {year}
  {2001})}\BibitemShut {NoStop}%
\bibitem [{\citenamefont {Rodrigo}\ \emph {et~al.}(2008)\citenamefont
  {Rodrigo}, \citenamefont {{Garcia-Vidal}},\ and\ \citenamefont
  {{Martin-Moreno}}}]{rodrigo2008influence}%
  \BibitemOpen
  \bibfield  {author} {\bibinfo {author} {\bibfnamefont {S.~G.}\ \bibnamefont
  {Rodrigo}}, \bibinfo {author} {\bibfnamefont {F.~J.}\ \bibnamefont
  {{Garcia-Vidal}}}, \ and\ \bibinfo {author} {\bibfnamefont {L.}~\bibnamefont
  {{Martin-Moreno}}},\ }\href {\doibase 10.1103/PhysRevB.77.075401} {\bibfield
  {journal} {\bibinfo  {journal} {Physical Review B}\ }\textbf {\bibinfo
  {volume} {77}},\ \bibinfo {pages} {075401} (\bibinfo {year}
  {2008})}\BibitemShut {NoStop}%
\bibitem [{\citenamefont {Tai}(1996)}]{tai1996dyadicgreen}%
  \BibitemOpen
  \bibfield  {author} {\bibinfo {author} {\bibfnamefont {C.}~\bibnamefont
  {Tai}},\ }\href@noop {} {\emph {\bibinfo {title} {Dyadic Green Functions in
  Electromagnetic Theory}}}\ (\bibinfo  {publisher} {Oxford University Press,
  {USA}},\ \bibinfo {year} {1996})\BibitemShut {NoStop}%
\bibitem [{\citenamefont {Hanson}\ and\ \citenamefont
  {Yakovlev}(2002)}]{hanson2002operator}%
  \BibitemOpen
  \bibfield  {author} {\bibinfo {author} {\bibfnamefont {G.~W.}\ \bibnamefont
  {Hanson}}\ and\ \bibinfo {author} {\bibfnamefont {A.~B.}\ \bibnamefont
  {Yakovlev}},\ }\href@noop {} {\emph {\bibinfo {title} {Operator Theory for
  Electromagnetics: An Introduction}}}\ (\bibinfo  {publisher} {Springer},\
  \bibinfo {year} {2002})\BibitemShut {NoStop}%
\bibitem [{\citenamefont {Leggett}(1984)}]{leggett1984quantum}%
  \BibitemOpen
  \bibfield  {author} {\bibinfo {author} {\bibfnamefont {A.~J.}\ \bibnamefont
  {Leggett}},\ }\href {\doibase 10.1103/PhysRevB.30.1208} {\bibfield  {journal}
  {\bibinfo  {journal} {Physical Review B}\ }\textbf {\bibinfo {volume} {30}},\
  \bibinfo {pages} {1208} (\bibinfo {year} {1984})}\BibitemShut {NoStop}%
\bibitem [{\citenamefont {Garg}\ \emph {et~al.}(1985)\citenamefont {Garg},
  \citenamefont {Onuchic},\ and\ \citenamefont
  {Ambegaokar}}]{garg1985effectof}%
  \BibitemOpen
  \bibfield  {author} {\bibinfo {author} {\bibfnamefont {A.}~\bibnamefont
  {Garg}}, \bibinfo {author} {\bibfnamefont {J.~N.}\ \bibnamefont {Onuchic}}, \
  and\ \bibinfo {author} {\bibfnamefont {V.}~\bibnamefont {Ambegaokar}},\
  }\href {\doibase 10.1063/1.449017} {\bibfield  {journal} {\bibinfo  {journal}
  {The Journal of Chemical Physics}\ }\textbf {\bibinfo {volume} {83}},\
  \bibinfo {pages} {4491} (\bibinfo {year} {1985})}\BibitemShut {NoStop}%
\bibitem [{\citenamefont {Goorden}\ \emph {et~al.}(2004)\citenamefont
  {Goorden}, \citenamefont {Thorwart},\ and\ \citenamefont
  {Grifoni}}]{goorden2004entanglement}%
  \BibitemOpen
  \bibfield  {author} {\bibinfo {author} {\bibfnamefont {M.~C.}\ \bibnamefont
  {Goorden}}, \bibinfo {author} {\bibfnamefont {M.}~\bibnamefont {Thorwart}}, \
  and\ \bibinfo {author} {\bibfnamefont {M.}~\bibnamefont {Grifoni}},\ }\href
  {\doibase 10.1103/PhysRevLett.93.267005} {\bibfield  {journal} {\bibinfo
  {journal} {Physical Review Letters}\ }\textbf {\bibinfo {volume} {93}},\
  \bibinfo {pages} {267005} (\bibinfo {year} {2004})}\BibitemShut {NoStop}%
\bibitem [{\citenamefont {{Xian-Ting}}(2007)}]{xian-ting2007decoherence}%
  \BibitemOpen
  \bibfield  {author} {\bibinfo {author} {\bibfnamefont {L.}~\bibnamefont
  {{Xian-Ting}}},\ }\href {\doibase 10.1016/j.cplett.2007.10.074} {\bibfield
  {journal} {\bibinfo  {journal} {Chemical Physics Letters}\ }\textbf {\bibinfo
  {volume} {449}},\ \bibinfo {pages} {296} (\bibinfo {year}
  {2007})}\BibitemShut {NoStop}%
\bibitem [{\citenamefont {Breuer}\ and\ \citenamefont
  {Petruccione}(2002)}]{breuer2002thetheory}%
  \BibitemOpen
  \bibfield  {author} {\bibinfo {author} {\bibfnamefont {H.}~\bibnamefont
  {Breuer}}\ and\ \bibinfo {author} {\bibfnamefont {F.}~\bibnamefont
  {Petruccione}},\ }\href@noop {} {\emph {\bibinfo {title} {The theory of open
  quantum systems}}}\ (\bibinfo  {publisher} {Oxford University Press},\
  \bibinfo {year} {2002})\BibitemShut {NoStop}%
\bibitem [{\citenamefont {Cohen-Tannoudji}\ \emph {et~al.}(1992)\citenamefont
  {Cohen-Tannoudji}, \citenamefont {Dupont-Roc},\ and\ \citenamefont
  {Grynberg}}]{Cohen-Tannoudji1992}%
  \BibitemOpen
  \bibfield  {author} {\bibinfo {author} {\bibfnamefont {C.}~\bibnamefont
  {Cohen-Tannoudji}}, \bibinfo {author} {\bibfnamefont {J.}~\bibnamefont
  {Dupont-Roc}}, \ and\ \bibinfo {author} {\bibfnamefont {G.}~\bibnamefont
  {Grynberg}},\ }\href
  {http://www.amazon.com/Atom-Photon-Interactions-Basic-Processes-Applications/dp/0471625566}
  {\emph {\bibinfo {title} {{Atom-Photon Interactions: Basic Processes and
  Applications}}}}\ (\bibinfo  {publisher} {Wiley-Interscience},\ \bibinfo
  {year} {1992})\ p.\ \bibinfo {pages} {680}\BibitemShut {NoStop}%
\bibitem [{\citenamefont {{Martin-Cano}}\ \emph {et~al.}(2011)\citenamefont
  {{Martin-Cano}}, \citenamefont {{Gonz\'{a}lez-Tudela}}, \citenamefont
  {{Martin-Moreno}}, \citenamefont {{Garcia-Vidal}}, \citenamefont {Tejedor},\
  and\ \citenamefont {Moreno}}]{martin-cano2011dissipationdriven}%
  \BibitemOpen
  \bibfield  {author} {\bibinfo {author} {\bibfnamefont {D.}~\bibnamefont
  {{Martin-Cano}}}, \bibinfo {author} {\bibfnamefont {A.}~\bibnamefont
  {{Gonz\'{a}lez-Tudela}}}, \bibinfo {author} {\bibfnamefont {L.}~\bibnamefont
  {{Martin-Moreno}}}, \bibinfo {author} {\bibfnamefont {F.~J.}\ \bibnamefont
  {{Garcia-Vidal}}}, \bibinfo {author} {\bibfnamefont {C.}~\bibnamefont
  {Tejedor}}, \ and\ \bibinfo {author} {\bibfnamefont {E.}~\bibnamefont
  {Moreno}},\ }\href {\doibase 10.1103/PhysRevB.84.235306} {\bibfield
  {journal} {\bibinfo  {journal} {Physical Review B}\ }\textbf {\bibinfo
  {volume} {84}},\ \bibinfo {pages} {235306} (\bibinfo {year}
  {2011})}\BibitemShut {NoStop}%
\bibitem [{\citenamefont {Gonzalez-Tudela}\ \emph {et~al.}(2012)\citenamefont
  {Gonzalez-Tudela}, \citenamefont {Huidobro}, \citenamefont {Martin-Moreno},
  \citenamefont {Tejedor},\ and\ \citenamefont
  {Garcia-Vidal}}]{Gonzalez-Tudela2012a}%
  \BibitemOpen
  \bibfield  {author} {\bibinfo {author} {\bibfnamefont {A.}~\bibnamefont
  {Gonzalez-Tudela}}, \bibinfo {author} {\bibfnamefont {P.~A.}\ \bibnamefont
  {Huidobro}}, \bibinfo {author} {\bibfnamefont {L.}~\bibnamefont
  {Martin-Moreno}}, \bibinfo {author} {\bibfnamefont {C.}~\bibnamefont
  {Tejedor}}, \ and\ \bibinfo {author} {\bibfnamefont {F.~J.}\ \bibnamefont
  {Garcia-Vidal}},\ }\href {http://arxiv.org/abs/1205.3938} {\  (\bibinfo
  {year} {2012})},\ \Eprint {http://arxiv.org/abs/1205.3938} {arXiv:1205.3938}
  \BibitemShut {NoStop}%
\bibitem [{\citenamefont {Chance}\ \emph {et~al.}(1978)\citenamefont {Chance},
  \citenamefont {Prock},\ and\ \citenamefont {Silbey}}]{Chance1978}%
  \BibitemOpen
  \bibfield  {author} {\bibinfo {author} {\bibfnamefont {R.~R.}\ \bibnamefont
  {Chance}}, \bibinfo {author} {\bibfnamefont {A.}~\bibnamefont {Prock}}, \
  and\ \bibinfo {author} {\bibfnamefont {R.}~\bibnamefont {Silbey}},\
  }\href@noop {} {\bibfield  {journal} {\bibinfo  {journal} {Adv. chem. Phys.}\
  }\textbf {\bibinfo {volume} {37}},\ \bibinfo {pages} {1} (\bibinfo {year}
  {1978})}\BibitemShut {NoStop}%
\bibitem [{\citenamefont {Barnes}(1998)}]{Barnes1998}%
  \BibitemOpen
  \bibfield  {author} {\bibinfo {author} {\bibfnamefont {W.~L.}\ \bibnamefont
  {Barnes}},\ }\href@noop {} {\bibfield  {journal} {\bibinfo  {journal}
  {Journal of Modern Optics}\ }\textbf {\bibinfo {volume} {45}},\ \bibinfo
  {pages} {661} (\bibinfo {year} {1998})}\BibitemShut {NoStop}%
\bibitem [{\citenamefont {Auff\`{e}ves}\ \emph {et~al.}(2009)\citenamefont
  {Auff\`{e}ves}, \citenamefont {G\'{e}rard},\ and\ \citenamefont
  {Poizat}}]{auffeves2009pureemitter}%
  \BibitemOpen
  \bibfield  {author} {\bibinfo {author} {\bibfnamefont {A.}~\bibnamefont
  {Auff\`{e}ves}}, \bibinfo {author} {\bibfnamefont {J.~M.}\ \bibnamefont
  {G\'{e}rard}}, \ and\ \bibinfo {author} {\bibfnamefont {J.~P.}\ \bibnamefont
  {Poizat}},\ }\href@noop {} {\bibfield  {journal} {\bibinfo  {journal}
  {Physical Review A}\ }\textbf {\bibinfo {volume} {79}},\ \bibinfo {pages}
  {053838} (\bibinfo {year} {2009})}\BibitemShut {NoStop}%
\bibitem [{\citenamefont {{Gonzalez-Tudela}}\ \emph {et~al.}(2009)\citenamefont
  {{Gonzalez-Tudela}}, \citenamefont {del Valle}, \citenamefont {Cancellieri},
  \citenamefont {Tejedor}, \citenamefont {Sanvitto},\ and\ \citenamefont
  {Laussy}}]{gonzalez-tudela2009effectof}%
  \BibitemOpen
  \bibfield  {author} {\bibinfo {author} {\bibfnamefont {A.}~\bibnamefont
  {{Gonzalez-Tudela}}}, \bibinfo {author} {\bibfnamefont {E.}~\bibnamefont {del
  Valle}}, \bibinfo {author} {\bibfnamefont {E.}~\bibnamefont {Cancellieri}},
  \bibinfo {author} {\bibfnamefont {C.}~\bibnamefont {Tejedor}}, \bibinfo
  {author} {\bibfnamefont {D.}~\bibnamefont {Sanvitto}}, \ and\ \bibinfo
  {author} {\bibfnamefont {F.~P.}\ \bibnamefont {Laussy}},\ }\href@noop {} {\
  (\bibinfo {year} {2009})},\ \bibinfo {note} {optics Express, Vol. 18, Issue
  7, pp. 7002-7009 (2010)}\BibitemShut {NoStop}%
\bibitem [{\citenamefont {Auff\`{e}ves}\ \emph {et~al.}(2010)\citenamefont
  {Auff\`{e}ves}, \citenamefont {Gerace}, \citenamefont {G\'{e}rard},
  \citenamefont {Santos}, \citenamefont {Andreani},\ and\ \citenamefont
  {Poizat}}]{auffeves2010controlling}%
  \BibitemOpen
  \bibfield  {author} {\bibinfo {author} {\bibfnamefont {A.}~\bibnamefont
  {Auff\`{e}ves}}, \bibinfo {author} {\bibfnamefont {D.}~\bibnamefont
  {Gerace}}, \bibinfo {author} {\bibfnamefont {J.}~\bibnamefont {G\'{e}rard}},
  \bibinfo {author} {\bibfnamefont {M.~F.}\ \bibnamefont {Santos}}, \bibinfo
  {author} {\bibfnamefont {L.~C.}\ \bibnamefont {Andreani}}, \ and\ \bibinfo
  {author} {\bibfnamefont {J.}~\bibnamefont {Poizat}},\ }\href@noop {}
  {\bibfield  {journal} {\bibinfo  {journal} {Physical Review B}\ }\textbf
  {\bibinfo {volume} {81}},\ \bibinfo {pages} {245419} (\bibinfo {year}
  {2010})}\BibitemShut {NoStop}%
\bibitem [{\citenamefont {Cui}\ and\ \citenamefont
  {Raymer}(2006)}]{cui2006emission}%
  \BibitemOpen
  \bibfield  {author} {\bibinfo {author} {\bibfnamefont {G.}~\bibnamefont
  {Cui}}\ and\ \bibinfo {author} {\bibfnamefont {M.~G.}\ \bibnamefont
  {Raymer}},\ }\href {\doibase 10.1103/PhysRevA.73.053807} {\bibfield
  {journal} {\bibinfo  {journal} {Physical Review A}\ }\textbf {\bibinfo
  {volume} {73}},\ \bibinfo {pages} {053807} (\bibinfo {year}
  {2006})}\BibitemShut {NoStop}%
\bibitem [{\citenamefont {Majumdar}\ \emph {et~al.}(2010)\citenamefont
  {Majumdar}, \citenamefont {Faraon}, \citenamefont {Kim}, \citenamefont
  {Englund}, \citenamefont {Kim}, \citenamefont {Petroff},\ and\ \citenamefont
  {Vu\v{c}kovi\'{c}}}]{majumdar2010linewidth}%
  \BibitemOpen
  \bibfield  {author} {\bibinfo {author} {\bibfnamefont {A.}~\bibnamefont
  {Majumdar}}, \bibinfo {author} {\bibfnamefont {A.}~\bibnamefont {Faraon}},
  \bibinfo {author} {\bibfnamefont {E.~D.}\ \bibnamefont {Kim}}, \bibinfo
  {author} {\bibfnamefont {D.}~\bibnamefont {Englund}}, \bibinfo {author}
  {\bibfnamefont {H.}~\bibnamefont {Kim}}, \bibinfo {author} {\bibfnamefont
  {P.}~\bibnamefont {Petroff}}, \ and\ \bibinfo {author} {\bibfnamefont
  {J.}~\bibnamefont {Vu\v{c}kovi\'{c}}},\ }\href {\doibase
  10.1103/PhysRevB.82.045306} {\bibfield  {journal} {\bibinfo  {journal}
  {Physical Review B}\ }\textbf {\bibinfo {volume} {82}},\ \bibinfo {pages}
  {045306} (\bibinfo {year} {2010})}\BibitemShut {NoStop}%
\bibitem [{\citenamefont {Moreno}\ \emph {et~al.}(2008)\citenamefont {Moreno},
  \citenamefont {Rodrigo}, \citenamefont {Bozhevolnyi}, \citenamefont
  {Martin-Moreno},\ and\ \citenamefont {Garcia-Vidal}}]{moreno2008}%
  \BibitemOpen
  \bibfield  {author} {\bibinfo {author} {\bibfnamefont {E.}~\bibnamefont
  {Moreno}}, \bibinfo {author} {\bibfnamefont {S.~G.}\ \bibnamefont {Rodrigo}},
  \bibinfo {author} {\bibfnamefont {S.~I.}\ \bibnamefont {Bozhevolnyi}},
  \bibinfo {author} {\bibfnamefont {L.}~\bibnamefont {Martin-Moreno}}, \ and\
  \bibinfo {author} {\bibfnamefont {F.~J.}\ \bibnamefont {Garcia-Vidal}},\
  }\href {\doibase 10.1103/PhysRevLett.100.023901} {\bibfield  {journal}
  {\bibinfo  {journal} {Phys. Rev. Lett.}\ }\textbf {\bibinfo {volume} {100}},\
  \bibinfo {pages} {023901} (\bibinfo {year} {2008})}\BibitemShut {NoStop}%
\bibitem [{\citenamefont {Wang}\ \emph {et~al.}(2011)\citenamefont {Wang},
  \citenamefont {Yang}, \citenamefont {Fan}, \citenamefont {Xu},\ and\
  \citenamefont {Wang}}]{wang2011lightpropagation}%
  \BibitemOpen
  \bibfield  {author} {\bibinfo {author} {\bibfnamefont {W.}~\bibnamefont
  {Wang}}, \bibinfo {author} {\bibfnamefont {Q.}~\bibnamefont {Yang}}, \bibinfo
  {author} {\bibfnamefont {F.}~\bibnamefont {Fan}}, \bibinfo {author}
  {\bibfnamefont {H.}~\bibnamefont {Xu}}, \ and\ \bibinfo {author}
  {\bibfnamefont {Z.~L.}\ \bibnamefont {Wang}},\ }\href {\doibase
  10.1021/nl104514m} {\bibfield  {journal} {\bibinfo  {journal} {Nano Lett.}\
  }\textbf {\bibinfo {volume} {11}},\ \bibinfo {pages} {1603} (\bibinfo {year}
  {2011})}\BibitemShut {NoStop}%
\bibitem [{\citenamefont {Dikken}\ \emph {et~al.}(2010)\citenamefont {Dikken},
  \citenamefont {Spasenovi}, \citenamefont {Verhagen}, \citenamefont {van
  Oosten},\ and\ \citenamefont
  {{(Kobus)}~Kuipers}}]{dikken2010characterization}%
  \BibitemOpen
  \bibfield  {author} {\bibinfo {author} {\bibfnamefont {D.~J.}\ \bibnamefont
  {Dikken}}, \bibinfo {author} {\bibfnamefont {M.}~\bibnamefont {Spasenovi}},
  \bibinfo {author} {\bibfnamefont {E.}~\bibnamefont {Verhagen}}, \bibinfo
  {author} {\bibfnamefont {D.}~\bibnamefont {van Oosten}}, \ and\ \bibinfo
  {author} {\bibfnamefont {L.}~\bibnamefont {{(Kobus)}~Kuipers}},\ }\href
  {\doibase 10.1364/OE.18.016112} {\bibfield  {journal} {\bibinfo  {journal}
  {Optics Express}\ }\textbf {\bibinfo {volume} {18}},\ \bibinfo {pages}
  {16112} (\bibinfo {year} {2010})}\BibitemShut {NoStop}%
\bibitem [{\citenamefont {Oulton}\ \emph {et~al.}(2007)\citenamefont {Oulton},
  \citenamefont {Pile}, \citenamefont {Liu},\ and\ \citenamefont
  {Zhang}}]{oulton2007scattering}%
  \BibitemOpen
  \bibfield  {author} {\bibinfo {author} {\bibfnamefont {R.~F.}\ \bibnamefont
  {Oulton}}, \bibinfo {author} {\bibfnamefont {D.~F.~P.}\ \bibnamefont {Pile}},
  \bibinfo {author} {\bibfnamefont {Y.}~\bibnamefont {Liu}}, \ and\ \bibinfo
  {author} {\bibfnamefont {X.}~\bibnamefont {Zhang}},\ }\href {\doibase
  10.1103/PhysRevB.76.035408} {\bibfield  {journal} {\bibinfo  {journal}
  {Physical Review B}\ }\textbf {\bibinfo {volume} {76}},\ \bibinfo {pages}
  {035408} (\bibinfo {year} {2007})}\BibitemShut {NoStop}%
\bibitem [{\citenamefont {Vahala}(2003)}]{vahala2003optical}%
  \BibitemOpen
  \bibfield  {author} {\bibinfo {author} {\bibfnamefont {K.~J.}\ \bibnamefont
  {Vahala}},\ }\href {\doibase 10.1038/nature01939} {\bibfield  {journal}
  {\bibinfo  {journal} {Nature}\ }\textbf {\bibinfo {volume} {424}},\ \bibinfo
  {pages} {839} (\bibinfo {year} {2003})}\BibitemShut {NoStop}%
\bibitem [{\citenamefont {Rempe}\ \emph {et~al.}(1992)\citenamefont {Rempe},
  \citenamefont {Thompson}, \citenamefont {Kimble},\ and\ \citenamefont
  {Lalezari}}]{rempe1992measurement}%
  \BibitemOpen
  \bibfield  {author} {\bibinfo {author} {\bibfnamefont {G.}~\bibnamefont
  {Rempe}}, \bibinfo {author} {\bibfnamefont {R.~J.}\ \bibnamefont {Thompson}},
  \bibinfo {author} {\bibfnamefont {H.~J.}\ \bibnamefont {Kimble}}, \ and\
  \bibinfo {author} {\bibfnamefont {R.}~\bibnamefont {Lalezari}},\ }\href
  {http://www.opticsinfobase.org/abstract.cfm?id=11045} {\bibfield  {journal}
  {\bibinfo  {journal} {Optics letters}\ }\textbf {\bibinfo {volume} {17}},\
  \bibinfo {pages} {363{\textendash}365} (\bibinfo {year} {1992})}\BibitemShut
  {NoStop}%
\bibitem [{\citenamefont {Laucht}\ \emph {et~al.}(2009)\citenamefont {Laucht},
  \citenamefont {Hauke}, \citenamefont {{Villas-B\^{o}as}}, \citenamefont
  {Hofbauer}, \citenamefont {B\"{o}hm}, \citenamefont {Kaniber},\ and\
  \citenamefont {Finley}}]{laucht2009dephasing}%
  \BibitemOpen
  \bibfield  {author} {\bibinfo {author} {\bibfnamefont {A.}~\bibnamefont
  {Laucht}}, \bibinfo {author} {\bibfnamefont {N.}~\bibnamefont {Hauke}},
  \bibinfo {author} {\bibfnamefont {J.~M.}\ \bibnamefont {{Villas-B\^{o}as}}},
  \bibinfo {author} {\bibfnamefont {F.}~\bibnamefont {Hofbauer}}, \bibinfo
  {author} {\bibfnamefont {G.}~\bibnamefont {B\"{o}hm}}, \bibinfo {author}
  {\bibfnamefont {M.}~\bibnamefont {Kaniber}}, \ and\ \bibinfo {author}
  {\bibfnamefont {J.~J.}\ \bibnamefont {Finley}},\ }\href {\doibase
  10.1103/PhysRevLett.103.087405} {\bibfield  {journal} {\bibinfo  {journal}
  {Physical Review Letters}\ }\textbf {\bibinfo {volume} {103}},\ \bibinfo
  {pages} {087405} (\bibinfo {year} {2009})}\BibitemShut {NoStop}%
\bibitem [{\citenamefont {Aharonovich}\ \emph {et~al.}(2011)\citenamefont
  {Aharonovich}, \citenamefont {Castelletto}, \citenamefont {Simpson},
  \citenamefont {Su}, \citenamefont {Greentree},\ and\ \citenamefont
  {Prawer}}]{aharonovich2011diamondbased}%
  \BibitemOpen
  \bibfield  {author} {\bibinfo {author} {\bibfnamefont {I.}~\bibnamefont
  {Aharonovich}}, \bibinfo {author} {\bibfnamefont {S.}~\bibnamefont
  {Castelletto}}, \bibinfo {author} {\bibfnamefont {D.~A.}\ \bibnamefont
  {Simpson}}, \bibinfo {author} {\bibfnamefont {C.~H.}\ \bibnamefont {Su}},
  \bibinfo {author} {\bibfnamefont {A.~D.}\ \bibnamefont {Greentree}}, \ and\
  \bibinfo {author} {\bibfnamefont {S.}~\bibnamefont {Prawer}},\ }\href@noop {}
  {\bibfield  {journal} {\bibinfo  {journal} {Reports on Progress in Physics}\
  }\textbf {\bibinfo {volume} {74}},\ \bibinfo {pages} {076501} (\bibinfo
  {year} {2011})}\BibitemShut {NoStop}%
\bibitem [{\citenamefont {Banin}\ \emph {et~al.}(1999)\citenamefont {Banin},
  \citenamefont {Bruchez}, \citenamefont {Alivisatos}, \citenamefont {Ha},
  \citenamefont {Weiss},\ and\ \citenamefont {Chemla}}]{banin1999evidence}%
  \BibitemOpen
  \bibfield  {author} {\bibinfo {author} {\bibfnamefont {U.}~\bibnamefont
  {Banin}}, \bibinfo {author} {\bibfnamefont {M.}~\bibnamefont {Bruchez}},
  \bibinfo {author} {\bibfnamefont {A.~P.}\ \bibnamefont {Alivisatos}},
  \bibinfo {author} {\bibfnamefont {T.}~\bibnamefont {Ha}}, \bibinfo {author}
  {\bibfnamefont {S.}~\bibnamefont {Weiss}}, \ and\ \bibinfo {author}
  {\bibfnamefont {D.~S.}\ \bibnamefont {Chemla}},\ }\href {\doibase
  doi:10.1063/1.478161} {\bibfield  {journal} {\bibinfo  {journal} {The Journal
  of Chemical Physics}\ }\textbf {\bibinfo {volume} {110}},\ \bibinfo {pages}
  {1195} (\bibinfo {year} {1999})}\BibitemShut {NoStop}%
\bibitem [{\citenamefont {Akimov}\ \emph {et~al.}(2007)\citenamefont {Akimov},
  \citenamefont {Mukherjee}, \citenamefont {Yu}, \citenamefont {Chang},
  \citenamefont {Zibrov}, \citenamefont {Hemmer}, \citenamefont {Park},\ and\
  \citenamefont {Lukin}}]{akimov2007generation}%
  \BibitemOpen
  \bibfield  {author} {\bibinfo {author} {\bibfnamefont {A.~V.}\ \bibnamefont
  {Akimov}}, \bibinfo {author} {\bibfnamefont {A.}~\bibnamefont {Mukherjee}},
  \bibinfo {author} {\bibfnamefont {C.~L.}\ \bibnamefont {Yu}}, \bibinfo
  {author} {\bibfnamefont {D.~E.}\ \bibnamefont {Chang}}, \bibinfo {author}
  {\bibfnamefont {A.~S.}\ \bibnamefont {Zibrov}}, \bibinfo {author}
  {\bibfnamefont {P.~R.}\ \bibnamefont {Hemmer}}, \bibinfo {author}
  {\bibfnamefont {H.}~\bibnamefont {Park}}, \ and\ \bibinfo {author}
  {\bibfnamefont {M.~D.}\ \bibnamefont {Lukin}},\ }\href {\doibase
  10.1038/nature06230} {\bibfield  {journal} {\bibinfo  {journal} {Nature}\
  }\textbf {\bibinfo {volume} {450}},\ \bibinfo {pages} {402} (\bibinfo {year}
  {2007})}\BibitemShut {NoStop}%
\bibitem [{\citenamefont {Andersen}\ \emph {et~al.}(2011)\citenamefont
  {Andersen}, \citenamefont {Stobbe}, \citenamefont {Sorensen},\ and\
  \citenamefont {Lodahl}}]{andersen2011strongly}%
  \BibitemOpen
  \bibfield  {author} {\bibinfo {author} {\bibfnamefont {M.~L.}\ \bibnamefont
  {Andersen}}, \bibinfo {author} {\bibfnamefont {S.}~\bibnamefont {Stobbe}},
  \bibinfo {author} {\bibfnamefont {A.~S.}\ \bibnamefont {Sorensen}}, \ and\
  \bibinfo {author} {\bibfnamefont {P.}~\bibnamefont {Lodahl}},\ }\href
  {\doibase 10.1038/nphys1870} {\bibfield  {journal} {\bibinfo  {journal} {Nat
  Phys}\ }\textbf {\bibinfo {volume} {7}},\ \bibinfo {pages} {215} (\bibinfo
  {year} {2011})}\BibitemShut {NoStop}%
\bibitem [{\citenamefont {Andreani}\ \emph {et~al.}(1999)\citenamefont
  {Andreani}, \citenamefont {Panzarini},\ and\ \citenamefont
  {G\'{e}rard}}]{andreani1999strongcoupling}%
  \BibitemOpen
  \bibfield  {author} {\bibinfo {author} {\bibfnamefont {L.~C.}\ \bibnamefont
  {Andreani}}, \bibinfo {author} {\bibfnamefont {G.}~\bibnamefont {Panzarini}},
  \ and\ \bibinfo {author} {\bibfnamefont {J.}~\bibnamefont {G\'{e}rard}},\
  }\href {\doibase 10.1103/PhysRevB.60.13276} {\bibfield  {journal} {\bibinfo
  {journal} {Physical Review B}\ }\textbf {\bibinfo {volume} {60}},\ \bibinfo
  {pages} {13276} (\bibinfo {year} {1999})}\BibitemShut {NoStop}%
\bibitem [{\citenamefont {Laussy}\ \emph {et~al.}(2008)\citenamefont {Laussy},
  \citenamefont {del Valle},\ and\ \citenamefont {Tejedor}}]{laussy2008}%
  \BibitemOpen
  \bibfield  {author} {\bibinfo {author} {\bibfnamefont {F.}~\bibnamefont
  {Laussy}}, \bibinfo {author} {\bibfnamefont {E.}~\bibnamefont {del Valle}}, \
  and\ \bibinfo {author} {\bibfnamefont {C.}~\bibnamefont {Tejedor}},\ }\href
  {http://prl.aps.org/abstract/PRL/v101/i8/e083601} {\bibfield  {journal}
  {\bibinfo  {journal} {Physical Review Letters}\ }\textbf {\bibinfo {volume}
  {101}},\ \bibinfo {pages} {083601} (\bibinfo {year} {2008})}\BibitemShut
  {NoStop}%
\bibitem [{\citenamefont {del Valle}\ \emph {et~al.}(2009)\citenamefont {del
  Valle}, \citenamefont {Laussy},\ and\ \citenamefont
  {Tejedor}}]{delvalle2009luminescence}%
  \BibitemOpen
  \bibfield  {author} {\bibinfo {author} {\bibfnamefont {E.}~\bibnamefont {del
  Valle}}, \bibinfo {author} {\bibfnamefont {F.~P.}\ \bibnamefont {Laussy}}, \
  and\ \bibinfo {author} {\bibfnamefont {C.}~\bibnamefont {Tejedor}},\ }\href
  {\doibase 10.1103/PhysRevB.79.235326} {\bibfield  {journal} {\bibinfo
  {journal} {Physical Review B}\ }\textbf {\bibinfo {volume} {79}},\ \bibinfo
  {pages} {235326} (\bibinfo {year} {2009})}\BibitemShut {NoStop}%
\bibitem [{\citenamefont {Ditlbacher}\ \emph {et~al.}(2005)\citenamefont
  {Ditlbacher}, \citenamefont {Hohenau}, \citenamefont {Wagner}, \citenamefont
  {Kreibig}, \citenamefont {Rogers}, \citenamefont {Hofer}, \citenamefont
  {Aussenegg},\ and\ \citenamefont {Krenn}}]{ditlbacher2005silvernanowires}%
  \BibitemOpen
  \bibfield  {author} {\bibinfo {author} {\bibfnamefont {H.}~\bibnamefont
  {Ditlbacher}}, \bibinfo {author} {\bibfnamefont {A.}~\bibnamefont {Hohenau}},
  \bibinfo {author} {\bibfnamefont {D.}~\bibnamefont {Wagner}}, \bibinfo
  {author} {\bibfnamefont {U.}~\bibnamefont {Kreibig}}, \bibinfo {author}
  {\bibfnamefont {M.}~\bibnamefont {Rogers}}, \bibinfo {author} {\bibfnamefont
  {F.}~\bibnamefont {Hofer}}, \bibinfo {author} {\bibfnamefont
  {F.}~\bibnamefont {Aussenegg}}, \ and\ \bibinfo {author} {\bibfnamefont
  {J.}~\bibnamefont {Krenn}},\ }\href@noop {} {\bibfield  {journal} {\bibinfo
  {journal} {Physical review letters}\ }\textbf {\bibinfo {volume} {95}},\
  \bibinfo {pages} {257403} (\bibinfo {year} {2005})}\BibitemShut {NoStop}%
\bibitem [{\citenamefont {Barnes}\ \emph {et~al.}(2003)\citenamefont {Barnes},
  \citenamefont {Dereux},\ and\ \citenamefont {Ebbesen}}]{Barnes2003}%
  \BibitemOpen
  \bibfield  {author} {\bibinfo {author} {\bibfnamefont {W.~L.}\ \bibnamefont
  {Barnes}}, \bibinfo {author} {\bibfnamefont {A.}~\bibnamefont {Dereux}}, \
  and\ \bibinfo {author} {\bibfnamefont {T.~W.}\ \bibnamefont {Ebbesen}},\
  }\href@noop {} {\bibfield  {journal} {\bibinfo  {journal} {Nature}\ }\textbf
  {\bibinfo {volume} {424}},\ \bibinfo {pages} {824} (\bibinfo {year}
  {2003})}\BibitemShut {NoStop}%
\bibitem [{\citenamefont {Vesseur}\ \emph {et~al.}(2010)\citenamefont
  {Vesseur}, \citenamefont {de~Abajo},\ and\ \citenamefont
  {Polman}}]{vesseur2010broadband}%
  \BibitemOpen
  \bibfield  {author} {\bibinfo {author} {\bibfnamefont {E.~J.~R.}\
  \bibnamefont {Vesseur}}, \bibinfo {author} {\bibfnamefont {F.~J.~G.}\
  \bibnamefont {de~Abajo}}, \ and\ \bibinfo {author} {\bibfnamefont
  {A.}~\bibnamefont {Polman}},\ }\href {\doibase 10.1103/PhysRevB.82.165419}
  {\bibfield  {journal} {\bibinfo  {journal} {Physical Review B}\ }\textbf
  {\bibinfo {volume} {82}},\ \bibinfo {pages} {165419} (\bibinfo {year}
  {2010})}\BibitemShut {NoStop}%
\bibitem [{\citenamefont {Hiscocks}\ \emph {et~al.}(2009)\citenamefont
  {Hiscocks}, \citenamefont {Su}, \citenamefont {Gibson}, \citenamefont
  {Greentree}, \citenamefont {Hollenberg},\ and\ \citenamefont
  {Ladouceur}}]{hiscocks2009slotwaveguide}%
  \BibitemOpen
  \bibfield  {author} {\bibinfo {author} {\bibfnamefont {M.~P.}\ \bibnamefont
  {Hiscocks}}, \bibinfo {author} {\bibfnamefont {C.}~\bibnamefont {Su}},
  \bibinfo {author} {\bibfnamefont {B.~C.}\ \bibnamefont {Gibson}}, \bibinfo
  {author} {\bibfnamefont {A.~D.}\ \bibnamefont {Greentree}}, \bibinfo {author}
  {\bibfnamefont {L.~C.~L.}\ \bibnamefont {Hollenberg}}, \ and\ \bibinfo
  {author} {\bibfnamefont {F.}~\bibnamefont {Ladouceur}},\ }\href {\doibase
  10.1364/OE.17.007295} {\bibfield  {journal} {\bibinfo  {journal} {Optics
  Express}\ }\textbf {\bibinfo {volume} {17}},\ \bibinfo {pages} {7295}
  (\bibinfo {year} {2009})}\BibitemShut {NoStop}%
\bibitem [{\citenamefont {Zueco}\ \emph {et~al.}(2009)\citenamefont {Zueco},
  \citenamefont {Reuther}, \citenamefont {Kohler},\ and\ \citenamefont
  {H\"{a}nggi}}]{Zueco2009}%
  \BibitemOpen
  \bibfield  {author} {\bibinfo {author} {\bibfnamefont {D.}~\bibnamefont
  {Zueco}}, \bibinfo {author} {\bibfnamefont {G.}~\bibnamefont {Reuther}},
  \bibinfo {author} {\bibfnamefont {S.}~\bibnamefont {Kohler}}, \ and\ \bibinfo
  {author} {\bibfnamefont {P.}~\bibnamefont {H\"{a}nggi}},\ }\href {\doibase
  10.1103/PhysRevA.80.033846} {\bibfield  {journal} {\bibinfo  {journal}
  {Physical Review A}\ }\textbf {\bibinfo {volume} {80}} (\bibinfo {year}
  {2009}),\ 10.1103/PhysRevA.80.033846}\BibitemShut {NoStop}%
\bibitem [{\citenamefont {Jacobs}\ and\ \citenamefont
  {Landahl}(2009)}]{Jacobs2009}%
  \BibitemOpen
  \bibfield  {author} {\bibinfo {author} {\bibfnamefont {K.}~\bibnamefont
  {Jacobs}}\ and\ \bibinfo {author} {\bibfnamefont {A.}~\bibnamefont
  {Landahl}},\ }\href {\doibase 10.1103/PhysRevLett.103.067201} {\bibfield
  {journal} {\bibinfo  {journal} {Physical Review Letters}\ }\textbf {\bibinfo
  {volume} {103}} (\bibinfo {year} {2009}),\
  10.1103/PhysRevLett.103.067201}\BibitemShut {NoStop}%
\bibitem [{\citenamefont {Hoffman}\ \emph {et~al.}(2011)\citenamefont
  {Hoffman}, \citenamefont {Srinivasan}, \citenamefont {Schmidt}, \citenamefont
  {Spietz}, \citenamefont {Aumentado}, \citenamefont {T\"{u}reci},\ and\
  \citenamefont {Houck}}]{Hoffman2011}%
  \BibitemOpen
  \bibfield  {author} {\bibinfo {author} {\bibfnamefont {A.}~\bibnamefont
  {Hoffman}}, \bibinfo {author} {\bibfnamefont {S.}~\bibnamefont {Srinivasan}},
  \bibinfo {author} {\bibfnamefont {S.}~\bibnamefont {Schmidt}}, \bibinfo
  {author} {\bibfnamefont {L.}~\bibnamefont {Spietz}}, \bibinfo {author}
  {\bibfnamefont {J.}~\bibnamefont {Aumentado}}, \bibinfo {author}
  {\bibfnamefont {H.}~\bibnamefont {T\"{u}reci}}, \ and\ \bibinfo {author}
  {\bibfnamefont {A.}~\bibnamefont {Houck}},\ }\href {\doibase
  10.1103/PhysRevLett.107.053602} {\bibfield  {journal} {\bibinfo  {journal}
  {Physical Review Letters}\ }\textbf {\bibinfo {volume} {107}} (\bibinfo
  {year} {2011}),\ 10.1103/PhysRevLett.107.053602}\BibitemShut {NoStop}%
\bibitem [{\citenamefont {Angelakis}\ \emph {et~al.}(2007)\citenamefont
  {Angelakis}, \citenamefont {Santos},\ and\ \citenamefont
  {Bose}}]{Angelakis2007}%
  \BibitemOpen
  \bibfield  {author} {\bibinfo {author} {\bibfnamefont {D.}~\bibnamefont
  {Angelakis}}, \bibinfo {author} {\bibfnamefont {M.}~\bibnamefont {Santos}}, \
  and\ \bibinfo {author} {\bibfnamefont {S.}~\bibnamefont {Bose}},\ }\href
  {\doibase 10.1103/PhysRevA.76.031805} {\bibfield  {journal} {\bibinfo
  {journal} {Physical Review A}\ }\textbf {\bibinfo {volume} {76}} (\bibinfo
  {year} {2007}),\ 10.1103/PhysRevA.76.031805}\BibitemShut {NoStop}%
\bibitem [{\citenamefont {H\"{u}mmer}\ \emph {et~al.}(2012)\citenamefont
  {H\"{u}mmer}, \citenamefont {Reuther}, \citenamefont {H\"{a}nggi},\ and\
  \citenamefont {Zueco}}]{Hummer2012}%
  \BibitemOpen
  \bibfield  {author} {\bibinfo {author} {\bibfnamefont {T.}~\bibnamefont
  {H\"{u}mmer}}, \bibinfo {author} {\bibfnamefont {G.}~\bibnamefont {Reuther}},
  \bibinfo {author} {\bibfnamefont {P.}~\bibnamefont {H\"{a}nggi}}, \ and\
  \bibinfo {author} {\bibfnamefont {D.}~\bibnamefont {Zueco}},\ }\href
  {\doibase 10.1103/PhysRevA.85.052320} {\bibfield  {journal} {\bibinfo
  {journal} {Physical Review A}\ }\textbf {\bibinfo {volume} {85}} (\bibinfo
  {year} {2012}),\ 10.1103/PhysRevA.85.052320}\BibitemShut {NoStop}%
\bibitem [{\citenamefont {Mariantoni}\ \emph {et~al.}(2011)\citenamefont
  {Mariantoni}, \citenamefont {Wang}, \citenamefont {Yamamoto}, \citenamefont
  {Neeley}, \citenamefont {Bialczak}, \citenamefont {Chen}, \citenamefont
  {Lenander}, \citenamefont {Lucero}, \citenamefont {O'Connell}, \citenamefont
  {Sank}, \citenamefont {Weides}, \citenamefont {Wenner}, \citenamefont {Yin},
  \citenamefont {Zhao}, \citenamefont {Korotkov}, \citenamefont {Cleland},\
  and\ \citenamefont {Martinis}}]{Mariantoni2011}%
  \BibitemOpen
  \bibfield  {author} {\bibinfo {author} {\bibfnamefont {M.}~\bibnamefont
  {Mariantoni}}, \bibinfo {author} {\bibfnamefont {H.}~\bibnamefont {Wang}},
  \bibinfo {author} {\bibfnamefont {T.}~\bibnamefont {Yamamoto}}, \bibinfo
  {author} {\bibfnamefont {M.}~\bibnamefont {Neeley}}, \bibinfo {author}
  {\bibfnamefont {R.~C.}\ \bibnamefont {Bialczak}}, \bibinfo {author}
  {\bibfnamefont {Y.}~\bibnamefont {Chen}}, \bibinfo {author} {\bibfnamefont
  {M.}~\bibnamefont {Lenander}}, \bibinfo {author} {\bibfnamefont
  {E.}~\bibnamefont {Lucero}}, \bibinfo {author} {\bibfnamefont {A.~D.}\
  \bibnamefont {O'Connell}}, \bibinfo {author} {\bibfnamefont {D.}~\bibnamefont
  {Sank}}, \bibinfo {author} {\bibfnamefont {M.}~\bibnamefont {Weides}},
  \bibinfo {author} {\bibfnamefont {J.}~\bibnamefont {Wenner}}, \bibinfo
  {author} {\bibfnamefont {Y.}~\bibnamefont {Yin}}, \bibinfo {author}
  {\bibfnamefont {J.}~\bibnamefont {Zhao}}, \bibinfo {author} {\bibfnamefont
  {A.~N.}\ \bibnamefont {Korotkov}}, \bibinfo {author} {\bibfnamefont {A.~N.}\
  \bibnamefont {Cleland}}, \ and\ \bibinfo {author} {\bibfnamefont {J.~M.}\
  \bibnamefont {Martinis}},\ }\href {\doibase 10.1126/science.1208517}
  {\bibfield  {journal} {\bibinfo  {journal} {Science (New York, N.Y.)}\
  }\textbf {\bibinfo {volume} {334}},\ \bibinfo {pages} {61} (\bibinfo {year}
  {2011})}\BibitemShut {NoStop}%
\bibitem [{\citenamefont {Hofheinz}\ \emph {et~al.}(2009)\citenamefont
  {Hofheinz}, \citenamefont {Wang}, \citenamefont {Ansmann}, \citenamefont
  {Bialczak}, \citenamefont {Lucero}, \citenamefont {Neeley}, \citenamefont
  {O'Connell}, \citenamefont {Sank}, \citenamefont {Wenner}, \citenamefont
  {Martinis},\ and\ \citenamefont {Cleland}}]{Hofheinz2009}%
  \BibitemOpen
  \bibfield  {author} {\bibinfo {author} {\bibfnamefont {M.}~\bibnamefont
  {Hofheinz}}, \bibinfo {author} {\bibfnamefont {H.}~\bibnamefont {Wang}},
  \bibinfo {author} {\bibfnamefont {M.}~\bibnamefont {Ansmann}}, \bibinfo
  {author} {\bibfnamefont {R.~C.}\ \bibnamefont {Bialczak}}, \bibinfo {author}
  {\bibfnamefont {E.}~\bibnamefont {Lucero}}, \bibinfo {author} {\bibfnamefont
  {M.}~\bibnamefont {Neeley}}, \bibinfo {author} {\bibfnamefont {A.~D.}\
  \bibnamefont {O'Connell}}, \bibinfo {author} {\bibfnamefont {D.}~\bibnamefont
  {Sank}}, \bibinfo {author} {\bibfnamefont {J.}~\bibnamefont {Wenner}},
  \bibinfo {author} {\bibfnamefont {J.~M.}\ \bibnamefont {Martinis}}, \ and\
  \bibinfo {author} {\bibfnamefont {A.~N.}\ \bibnamefont {Cleland}},\ }\href
  {\doibase 10.1038/nature08005} {\bibfield  {journal} {\bibinfo  {journal}
  {Nature}\ }\textbf {\bibinfo {volume} {459}},\ \bibinfo {pages} {546}
  (\bibinfo {year} {2009})}\BibitemShut {NoStop}%
\bibitem [{\citenamefont {Majer}\ \emph {et~al.}(2007)\citenamefont {Majer},
  \citenamefont {Chow}, \citenamefont {Gambetta}, \citenamefont {Koch},
  \citenamefont {Johnson}, \citenamefont {Schreier}, \citenamefont {Frunzio},
  \citenamefont {Schuster}, \citenamefont {Houck}, \citenamefont {Wallraff},
  \citenamefont {Blais}, \citenamefont {Devoret}, \citenamefont {Girvin},\ and\
  \citenamefont {Schoelkopf}}]{Majer2007}%
  \BibitemOpen
  \bibfield  {author} {\bibinfo {author} {\bibfnamefont {J.}~\bibnamefont
  {Majer}}, \bibinfo {author} {\bibfnamefont {J.~M.}\ \bibnamefont {Chow}},
  \bibinfo {author} {\bibfnamefont {J.~M.}\ \bibnamefont {Gambetta}}, \bibinfo
  {author} {\bibfnamefont {J.}~\bibnamefont {Koch}}, \bibinfo {author}
  {\bibfnamefont {B.~R.}\ \bibnamefont {Johnson}}, \bibinfo {author}
  {\bibfnamefont {J.~A.}\ \bibnamefont {Schreier}}, \bibinfo {author}
  {\bibfnamefont {L.}~\bibnamefont {Frunzio}}, \bibinfo {author} {\bibfnamefont
  {D.~I.}\ \bibnamefont {Schuster}}, \bibinfo {author} {\bibfnamefont {A.~A.}\
  \bibnamefont {Houck}}, \bibinfo {author} {\bibfnamefont {A.}~\bibnamefont
  {Wallraff}}, \bibinfo {author} {\bibfnamefont {A.}~\bibnamefont {Blais}},
  \bibinfo {author} {\bibfnamefont {M.~H.}\ \bibnamefont {Devoret}}, \bibinfo
  {author} {\bibfnamefont {S.~M.}\ \bibnamefont {Girvin}}, \ and\ \bibinfo
  {author} {\bibfnamefont {R.~J.}\ \bibnamefont {Schoelkopf}},\ }\href
  {\doibase 10.1038/nature06184} {\bibfield  {journal} {\bibinfo  {journal}
  {Nature}\ }\textbf {\bibinfo {volume} {449}},\ \bibinfo {pages} {443}
  (\bibinfo {year} {2007})}\BibitemShut {NoStop}%
\bibitem [{\citenamefont {Diehl}\ \emph {et~al.}(2008)\citenamefont {Diehl},
  \citenamefont {Micheli}, \citenamefont {Kantian}, \citenamefont {Kraus},
  \citenamefont {B\"{u}chler},\ and\ \citenamefont {Zoller}}]{Diehl2008}%
  \BibitemOpen
  \bibfield  {author} {\bibinfo {author} {\bibfnamefont {S.}~\bibnamefont
  {Diehl}}, \bibinfo {author} {\bibfnamefont {A.}~\bibnamefont {Micheli}},
  \bibinfo {author} {\bibfnamefont {A.}~\bibnamefont {Kantian}}, \bibinfo
  {author} {\bibfnamefont {B.}~\bibnamefont {Kraus}}, \bibinfo {author}
  {\bibfnamefont {H.~P.}\ \bibnamefont {B\"{u}chler}}, \ and\ \bibinfo {author}
  {\bibfnamefont {P.}~\bibnamefont {Zoller}},\ }\href {\doibase
  10.1038/nphys1073} {\bibfield  {journal} {\bibinfo  {journal} {Nature
  Physics}\ }\textbf {\bibinfo {volume} {4}},\ \bibinfo {pages} {878} (\bibinfo
  {year} {2008})}\BibitemShut {NoStop}%
\bibitem [{\citenamefont {Verstraete}\ \emph {et~al.}(2009)\citenamefont
  {Verstraete}, \citenamefont {Wolf},\ and\ \citenamefont {{Ignacio
  Cirac}}}]{Verstraete2009}%
  \BibitemOpen
  \bibfield  {author} {\bibinfo {author} {\bibfnamefont {F.}~\bibnamefont
  {Verstraete}}, \bibinfo {author} {\bibfnamefont {M.~M.}\ \bibnamefont
  {Wolf}}, \ and\ \bibinfo {author} {\bibfnamefont {J.}~\bibnamefont {{Ignacio
  Cirac}}},\ }\href {\doibase 10.1038/nphys1342} {\bibfield  {journal}
  {\bibinfo  {journal} {Nature Physics}\ }\textbf {\bibinfo {volume} {5}},\
  \bibinfo {pages} {633} (\bibinfo {year} {2009})}\BibitemShut {NoStop}%
\bibitem [{\citenamefont {Vogel}\ and\ \citenamefont
  {Welsch}(2006)}]{Vogel2006}%
  \BibitemOpen
  \bibfield  {author} {\bibinfo {author} {\bibfnamefont {W.}~\bibnamefont
  {Vogel}}\ and\ \bibinfo {author} {\bibfnamefont {D.-G.}\ \bibnamefont
  {Welsch}},\ }\href@noop {} {\emph {\bibinfo {title} {{Quantum Optics}}}}\
  (\bibinfo  {publisher} {Wiley-VCH},\ \bibinfo {year} {2006})\ p.\ \bibinfo
  {pages} {520}\BibitemShut {NoStop}%
\bibitem [{\citenamefont {Le~Kien}\ and\ \citenamefont
  {Hakuta}(2009)}]{lekien2009cavityenhanced}%
  \BibitemOpen
  \bibfield  {author} {\bibinfo {author} {\bibfnamefont {F.}~\bibnamefont
  {Le~Kien}}\ and\ \bibinfo {author} {\bibfnamefont {K.}~\bibnamefont
  {Hakuta}},\ }\href {\doibase 10.1103/PhysRevA.80.053826} {\bibfield
  {journal} {\bibinfo  {journal} {Physical Review A}\ }\textbf {\bibinfo
  {volume} {80}},\ \bibinfo {pages} {053826} (\bibinfo {year}
  {2009})}\BibitemShut {NoStop}%
\bibitem [{\citenamefont {Smith}\ and\ \citenamefont
  {Fickett}(1995)}]{smith1995lowtemperature}%
  \BibitemOpen
  \bibfield  {author} {\bibinfo {author} {\bibfnamefont {D.}~\bibnamefont
  {Smith}}\ and\ \bibinfo {author} {\bibfnamefont {F.}~\bibnamefont
  {Fickett}},\ }\href@noop {} {\bibfield  {journal} {\bibinfo  {journal} {J.
  Res. Natl. Inst. Stand. Technol.}\ }\textbf {\bibinfo {volume} {100}},\
  \bibinfo {pages} {119{\textendash}119} (\bibinfo {year} {1995})}\BibitemShut
  {NoStop}%
\end{thebibliography}%
\end{document}